# Perovskite-Inspired Materials for Photovoltaics – From Design to Devices

*Yi-Teng Huang,[1] Seán R. Kavanagh,[2,3,4] David O. Scanlon,[2,4,5] Aron Walsh,[3,6] and Robert L. Z. Hoye[3,\*]*

[1]*Department of Physics, University of Cambridge, JJ Thomson Ave, Cambridge CB3 0HE, UK*
[2]*Department of Chemistry, University College London, 20 Gordon Street, London WC1H 0AJ, UK*
[3]*Department of Materials, Imperial College London, Exhibition Road, London SW7 2AZ, UK*
[4]*Thomas Young Centre, University College London, Gower Street, London WC1E 6BT, UK*
[5]*Diamond Light Source Ltd., Diamond House, Harwell Science and Innovation Campus, Didcot, Oxfordshire OX11 0DE, UK*
[6] *Department of Materials Science and Engineering, Yonsei University, Seoul 120-749, South Korea*

**\* Corresponding author**
*Email:* r.hoye@imperial.ac.uk (R.L.Z.H.)

**Abstract**

Lead-halide perovskites have demonstrated astonishing increases in power conversion efficiency in photovoltaics over the last decade. The most efficient perovskite devices now outperform industry-standard multi-crystalline silicon solar cells, despite the fact that perovskites are typically grown at low temperature using simple solution-based methods. However, the toxicity of lead and its ready solubility in water are concerns for widespread implementation. These challenges, alongside the many successes of the perovskites, have motivated significant efforts across multiple disciplines to find lead-free and stable alternatives which could mimic the ability of the perovskites to achieve high performance with low temperature, facile fabrication methods. This Review discusses the computational and experimental approaches that have been taken to discover lead-free perovskite-inspired materials, and the recent successes and challenges in synthesizing these compounds. The atomistic origins of the extraordinary performance exhibited by lead-halide perovskites in photovoltaic devices is discussed, alongside the key challenges in engineering such high-performance in alternative, next-generation materials. Beyond photovoltaics, this Review discusses the impact perovskite-inspired materials have had in spurring efforts to apply new materials in other optoelectronic applications, namely light-emitting diodes, photocatalysts, radiation detectors, thin film transistors and memristors. Finally, the prospects and key challenges faced by the field in advancing the development of perovskite-inspired materials towards realization in commercial devices is discussed.

## 1. Introduction

Sunlight is a vast, reliable and ubiquitous source of renewable energy [1]. Each year, terrestrial solar energy amounts to 1500 exawatt-hours [2], whereas global electrical energy demand was only 0.03 exawatt-hours in 2018. Arguably the most promising method to convert solar energy to electricity with no carbon-based by-products is through photovoltaics (PVs) [1]. However, solar energy production only reached 440 terawatt-hours in 2018, with 500 gigawatts of PV capacity installed worldwide by the end of 2018 [1]. Although PV deployment is projected to soon reach the terawatt level [1], it is predicted that 37–180 TW of PV capacity would be needed by 2050 in order to fulfil the ambitions of over 60 nations to achieve net-zero carbon emissions [3].

Numerous technoeconomic analyses have examined the challenges that must be addressed to enable the required scaling of PV capacity, some of which are covered in Ref. [1, 3, 4]. Notably, this will require reductions in the levelized cost of electricity (LCOE) and capital-intensity of manufacturing [4], which could be achieved through the use of thin film solar cells, particularly in tandem devices. Tandem solar cells combine two sub-cells which absorb in complementary parts of the solar spectrum, thus surpassing the theoretical limit of a single-junction device. Thin film solar cells are also suitable for integration in buildings (*e.g.*, solar windows), or use indoors to power small standalone gadgets, forming part of the Internet of Things [6].

While thin film PVs have been developed for decades, lead-halide perovskites (LHPs) have recently emerged, demonstrating higher device learning-rates than any other PV material [7]. Since their first report in PVs in 2009 (with 3.8% efficiency [8]), LHP PVs have now reached a certified efficiency of >25% [8] for small area cells, exceeding the efficiency of multi-crystalline silicon and approaching the certified record efficiency of crystalline silicon solar cells (26.7% [9]). Tandems between LHP top-cells and silicon bottom-cells have already reached a certified efficiency of 29.1% after only 6 years of development [10, 11]. The LHP band gap can be easily tuned over a wide range through manipulation of the composition. As a result, perovskite-perovskite tandems have been realized, with 25% efficiency already achieved [11]. Both tandem structures are projected to achieve lower LCOEs than single-junction silicon solar cells [5]. Single-junction perovskite devices have also demonstrated 35.2% efficiency under indoor lighting [12]. Critically, these high efficiencies can be achieved through low-temperature and facile processing, namely solution-processing and thermal evaporation. LHPs therefore hold significant potential for reducing the LCOE and capital-intensity of PVs.

However, there remains debate over whether the toxicity of lead (present in a soluble form in LHPs) could present a barrier to terawatt-scale deployment [13]. In addition, most LHPs are unstable in ambient air. While there has been significant effort to address this, with encapsulated perovskite solar cells now having met industry stability protocols (IEC 61215-1-3:2016) [14], LHPs used on the utility scale would ultimately need to have a field stability exceeding 25 years. More broadly, the success of LHPs defies conventional wisdom regarding the role of defects, in which high efficiencies are achieved despite thin films exhibiting large defect concentrations, several orders of magnitude greater than in traditional silicon or gallium arsenide materials [15]. This prompts the question of how such tolerance to defects arises, and whether this property could be generalized and designed in other classes of materials, especially those composed of abundant, non-toxic elements, and which are stable in air.

Broadly, three strategies have been used to identify perovskite-inspired materials (PIMs): 1) searching for chemically-analogous materials, 2) investigating materials with a perovskite crystal structure, and 3) inverse design of materials based on defect tolerance [18–21]. This effort spans across multiple academic disciplines. The key questions focus not only on structure-mechanism-property relationships of the materials and their performance in devices, but also how computational tools are being developed to accelerate the design and discovery of novel materials for high-performance optoelectronic devices.

In this Topical Review, we bring computational and theoretical research together with experimental efforts to engineer the properties and performance of these materials (Figure 1). Our Review begins with a discussion of the high-throughput materials discovery process, alongside the challenges of this research method. Following this, the chemical origins of defect tolerance are explored in detail, before relating these properties to the remarkable photovoltaic performance of LHPs and the inverse design of advanced perovskite-inspired materials (PIMs) that may also achieve defect tolerance. Additionally, a brief introduction to the practical implementation, challenges and frontiers of computational defect investigations is provided. We next discuss the classes of materials that have been considered, their properties, and synthesis methods developed. This is followed by a discussion on the performance of these materials in photovoltaic devices, as well as the opportunities found in using the materials for alternative applications (including light-emitting diodes and radiation detectors). Finally, we discuss the key frontiers and challenges in discovering high-performance PIMs, and the important questions that urgently need to be answered in order to achieve commercial application.

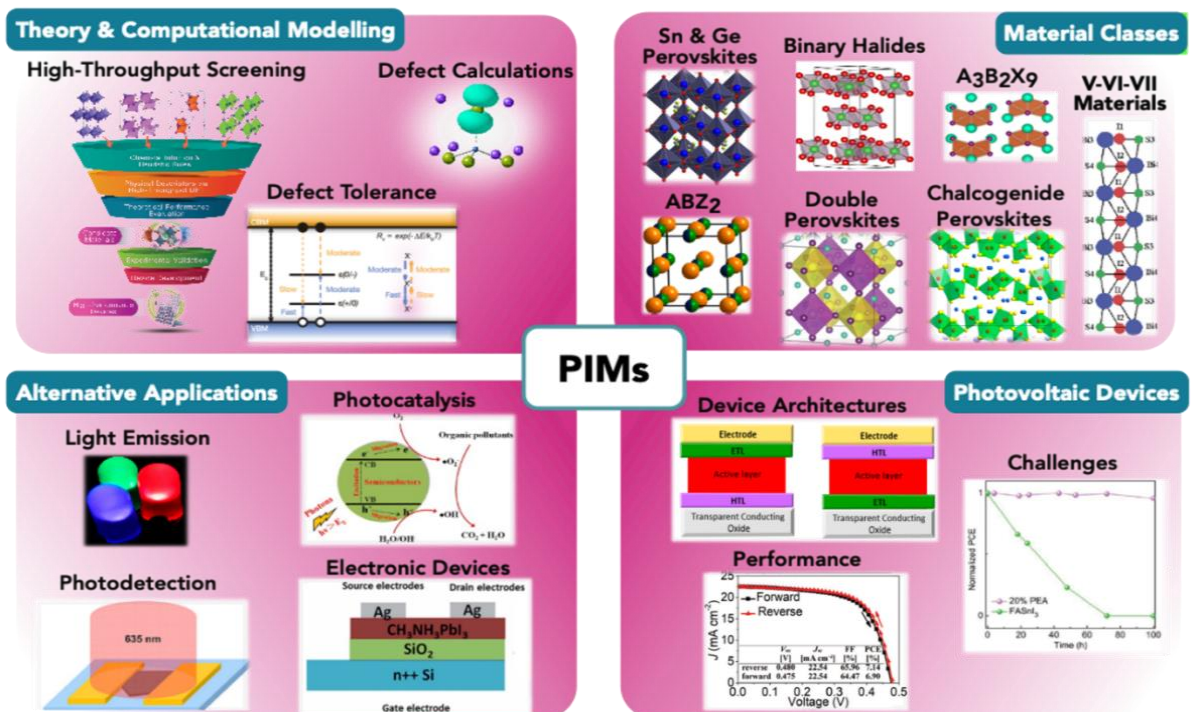

**Figure 1.** Overview of this Topical Review, which covers computational investigations into defect tolerance and tools developed for the screening of perovskite-inspired materials (PIMs), the properties of the materials developed and how they can be grown, and their applications in photovoltaics and beyond. Reprinted with permission from Ref. [22] (Copyright 2014 Royal Society of Chemisitry), [23] (Copyright 2019 Wiley), [24] (Copyright 2018 American Chemical Society), [25] (Copyright 2018 American Chemical Society), [26] (Copyright 2019 American Chemical Society), [27] (Copyright 2016 Royal Society of Chemistry), [28] (Copyright 2007 Elsevier), [29] (Copyright 2019 Wiley), [30] (Copyright 2018 American

Chemical Society), [31] (Copyright 2015 Elsevier), [32] (Copyright 2016 Royal Society of Chemistry).

## 2. Computational and Theoretical Approaches to Accelerate the Discovery of Perovskite-Inspired Materials

### 2.1 Computational Materials Discovery

Materials modelling is a rapidly expanding scientific field, most notably in the realm of materials discovery and design, where computational methods have significantly accelerated research efforts [33]. While computational materials science has flourished as a fertile research field since the 1970s, with a rich history of Nobel laureates, rapid technological advancements alongside financial investment have supported a renaissance in this area in recent times. Long-established approximations are no longer necessary and simulation size/time constraints can be overcome, providing the capability to perform increasingly quantitative and realistic calculations. Historically, computer simulations have played a responsive, augmentative role in relation to experiment. However, these recent advances have facilitated the ability of computational predictions to guide experiment, representing a critical component in the efficient discovery of novel high-performance materials.

The fundamental challenge of materials discovery is to identify promising candidate materials (for the specified application) from the near-infinite chemical space of possible compounds. To illustrate the scale of this physical search space, one can simply look at the Inorganic Crystal Structure Database (ICSD; icsd.cds.rsc.org), which contains over 150,000 entries for unique materials. Taking a more fundamental combinatorial approach, the number of possible two-, three- and four-element combinations, each with specific oxidation states and stoichiometries, in fact exceeds the trillions ($10^{12}$ possibilities). This indicates the presence of a vast unexplored compositional space, which may very well contain hidden 'wonder materials', patiently waiting for the opportunity to revolutionize modern technologies.

The exploration of such a broad physical space is evidently intractable to experiment or *ab-initio* calculations. However, high-throughput screening procedures involving a tiered, 'funnel' approach (as depicted schematically in Figure 2) offer a promising pathway to efficiently probe this raw chemical landscape.

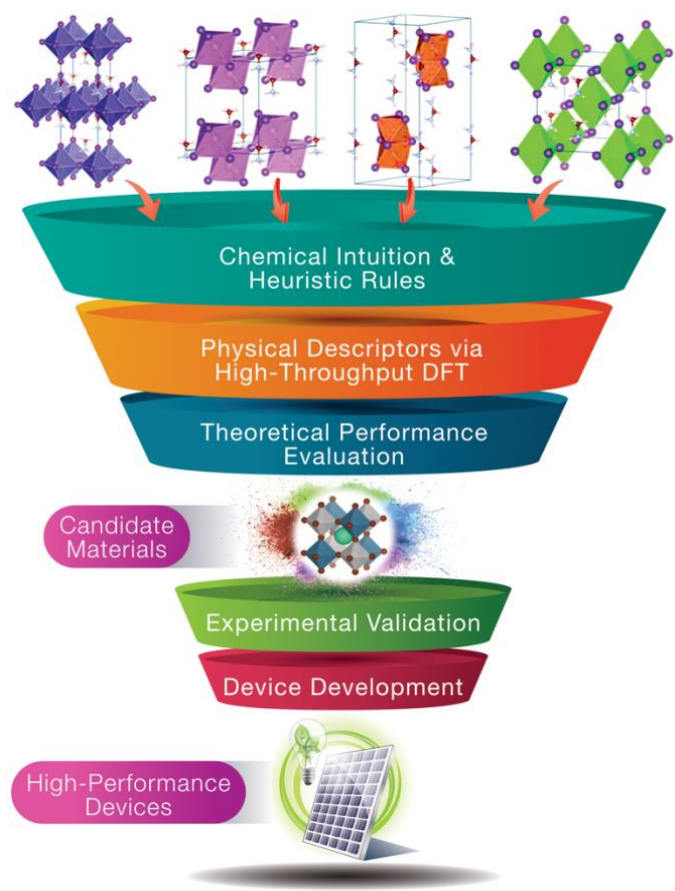

**Figure 2.** Simplified procedural outline of a typical high-throughput screening investigation, with a large input material space reduced to a small selection of promising candidates for further experimental verification and device development.

### 2.1.1 Reducing the Search Space

The first step in a materials discovery process is to apply certain universal constraints to immediately filter out unrealistic and irrelevant candidates. It is common practice to restrict the search space to specific material classes (compositions, structures, symmetries, motifs, etc.) such as, *inter alia*, layered compounds, perovskite structures or metal-organic frameworks, based on the intended application and chemical intuition. A pragmatic choice is to further restrict the search to abundant non-toxic elements, which are suitable candidates for wide-scale deployment in devices. Moreover, application of inexpensive charge neutrality, valency and electronegativity filters to remove chemically-implausible materials can reduce the combinatorial search space by two orders of magnitude [34].

A prerequisite for a viable candidate material is that it be synthesizable and (meta-)stable. Hence, this constraint is commonly applied at an early stage of the screening process, before any predictions of material properties and/or performance metrics are made. For instance, in a search for photoelectrochemically-active semiconductors, Davies *et al.* employed Shannon ionic radii and Goldschmidt tolerance factors to predict the stability of inorganic elemental combinations in the cubic or orthorhombic perovskite structure, reducing their input dataset from 176,851 to 26,567 possible candidates [34].

### 2.1.2    Electronic Structure Theory

After reducing the material search space to a computationally-tractable subset, electronic structure techniques may be invoked to apply a more rigorous evaluation of the remaining candidate materials. A widespread choice is Density Functional Theory (DFT), due to its well-established accuracy (in predicting ground-state structures and formation energies) at relatively moderate computational cost. From high-throughput DFT calculations, theoretical predictions of relevant material properties and structure are performed, alongside validation of stability/synthesizability. While certain higher levels of theory, such as the inclusion of many-body effects within the Random Phase Approximation (RPA), yield improved predictions for the ground-state structural polymorphs of materials (relative to semi-local DFT) [20], the computational expense of such approaches typically render them infeasible for high-throughput investigations, for which speed must be prioritized over accuracy. This becomes increasingly apparent when one considers the success of semi-local DFT in predicting ground-state structures of typical chemical compounds, with only a handful of specific material classes requiring higher levels of theory.

Within DFT, there remains the choice of the vital ingredient: the exchange-correlation functional, which describes the quantum mechanical electron-electron interactions. Two common forms are the Local Density Approximation (LDA) and the Generalized Gradient Approximations (GGA). Decades of practical experience have revealed clear favorites for various material types or structural motifs, such as PBEsol (GGA) for standard inorganic solid compounds [35], DFT+U for materials with magnetic interactions or localized *d* orbitals [36], and dispersion-corrected functionals such as optB86b-vdW or Grimme's D3 for layered structures [37–39].

Databases of material simulations, such as the Materials Project (materialsproject.org), NOMAD repository (nomad-repository.eu) or MatNavi (mits.nims.go.jp), can often be of benefit in accelerating these initial screening steps and reducing the computational demand by providing calculated material structures and properties (*e.g.* formation energy). However, care must be exercised in these cases, as standardized calculation parameters (*e.g.* the choice of exchange-correlation functional) may not be suitable for certain material classes (*e.g.* layered compounds with van der Waals interactions). Thus, knowledge and experience of material simulations is required to judge the applicability of database information for a given screening investigation.

### 2.1.3   Chemical Descriptors

At this point, the use of appropriate, well-informed, application-specific chemical descriptors is crucial in aiding both the efficiency and efficacy of the materials discovery process. In particular, these descriptors should be both central to performance and accessible by calculation. In the case of next-generation PV absorbers, relevant material properties would include the magnitude and nature (direct or indirect) of the electronic band gap $E_g$, charge-carrier mobilities $\mu_{e/h}$, minority-carrier lifetimes $\tau_{e/h}$, absorption coefficient $\alpha$ and the photoluminescence quantum efficiency (PLQE) $\eta$ [40]. These are summarized in Table 1. Recently, chemical descriptors of carrier effective mass, static polaron binding energy and Fröhlich electron–phonon coupling were used to predict 10 candidate p-type metal oxides [41]. These properties can be measured or predicted through various experimental or theoretical techniques, each with an associated cost, researcher-effort and accuracy. Judicious choice of appropriate descriptors is thus vital to the success of the screening procedure in accurately and efficiently determining potential candidate materials for the application in question.

**Table 1.** A collection of material properties and associated measurable/calculable descriptors relevant to the study and performance-prediction of materials for application in solar energy conversion. Reproduced with permission from [42].

| *Property* | *Descriptor* | *Calculation* | *Experiment* |
|---|---|---|---|
| *Optical Absorption* | Bandgap<br>High-Frequency Dielectric Constant<br>Spectroscopically Limited Maximum Efficiency (SLME) | Electronic Structure (*e.g.* DFT, *GW*)<br>Solid-state Energy Scale (SSE)<br>Tight Binding Analysis | Absorption Spectroscopy<br>Ellipsometry |
| *Electrical Conductivity* | Effective Mass Tensor<br>Mobility (lifetimes | Electronic Structure (*e.g.* DFT, *GW*)<br>Lattice Dynamics | Hall Measurements<br>Four-Point Probe<br>Van der Pauw Measurements |
| *Bulk Radiative Recombination Rate* | Electronic Band Structure<br>Relativistic Rashba Splitting | Electronic Structure (*e.g.* DFT, *GW*) with Spin-Orbit Coupling | Time Resolved Photoluminescence |
| *Defect Tolerance / Non-Radiative Recombination* | Dielectric Constant<br>Bonding Analysis<br>Defect Transition Levels | Electronic Structure (*e.g.* DFT, *GW*)<br>Explicit Defect Calculations | Time Resolved Absorption Spectroscopy |

| Bulk Photovoltaic Effect | Shift Current | Electronic Structure (*e.g.* DFT, *GW*) | Terahertz Emission Spectroscopy<br><br>Current-Voltage Relationships |
|---|---|---|---|
| Contact Resistance | Band Offsets | Electronic Structure (*e.g.* DFT, *GW*) | Transmission Line Measurements<br><br>Four-Point Probe<br><br>Photoemission Spectroscopy |
| Surface / Interface Recombination | Surface Defect Density | Electronic Structure (*e.g.* DFT, *GW*) | Pump-Probe Reflectivity |
| Sustainability | Crystal/Elemental Abundance | Database (*e.g.* implemented in *SMACT* package)[43] | N/A |

Crucially, the level of theory, and hence computational overhead, required for sufficiently-accurate calculations can vary depending on both the property and material in question. For instance, standard DFT functionals exhibit a systematic underestimation of electronic band gaps in insulating crystals, as a consequence of the 'derivative-discontinuity' and self-interaction errors [33, 42]. Consequently, expensive techniques, *viz.* hybrid non-local DFT or many-body GW theory are necessary for the accurate determination of certain electronic properties, such as the band gap and optical absorption. At present, these costly methods are not readily applicable to high-throughput studies of large material datasets, despite recent advances in computational power. On the other hand, relatively-cheap standard DFT methods (such as LDA / GGA) are capable of accurately predicting the shape of electronic bands and thus determining related descriptors such as the charge carrier effective masses and mobilities [33].

In many cases, while the electronic band gap may be the more fundamental, performance-defining property (in relation to the specific application of *e.g.* a battery electrode material), it may be more prudent to use standard DFT methods to calculate other relevant descriptors first (*e.g.* carrier mobility), ruling out any inappropriate candidates before implementing higher-order, expensive techniques. In this manner, the most burdensome procedures, both in terms of computational demand and researcher effort, are restricted to only the most promising of candidates.

The presence of heavy elements is another case for which more sophisticated treatments are required, necessitating the description of relativistic effects. In particular, spin-orbit coupling (SOC) effects scale approximately with the square of the atomic mass, and so these contributions become increasingly significant as one moves down the periodic table [20]. Further down the line, proper accounting of non-radiative recombination for the determination of charge-carrier lifetimes and non-ideal quantum luminescence efficiencies requires explicit defect calculations. As discussed in further detail in Section 2.3, these investigations demand substantial researcher effort and computational overhead, due to the number of distinct calculations involved and the mandatory use of large simulation supercells. After calculating the chosen descriptors for the material dataset, pre-determined selection criteria are applied so that only those materials with appropriate values for each property are considered. These candidate materials which successfully emerge from the chemical-descriptor screening stage should themselves be diligently analyzed in terms of crystal structure and chemical bonding. If similar characteristics evolve independently across multiple systems, design principles may be elucidated to accelerate materials discovery efforts and guide future investigations.

### 2.1.4   *Performance Metrics and Figures of Merit*

Finally, a figure of merit or 'selection metric' is typically determined for each of the remaining candidates, such as the popular $zT$ for thermoelectrics, yielding a final prediction of their potential success in the designated application. The ultimate purpose of the high-throughput materials discovery process is to identify promising candidates from vast input chemical space, based on the limited information available with finite resources (of time and CPU power). Accordingly, intelligent choice of selection metric is imperative to the success of these preliminary investigations.

For PV, there is still ongoing work in this regard, and several potential figures of merit have been put forward, as discussed below.

**SQ Limit:** In 1961, Shockley and Queisser derived the band gap-dependent radiative efficiency limit of a photovoltaic absorber ('SQ Limit'), using the principle of detailed balance [44]. This represented a tremendously significant development in the field of PV research, enabling predictions of maximum absorber performance, solely on the basis of the fundamental electronic band gap.

However, a number of large approximations are invoked in the SQ model, such as the assumption of perfect step-function absorption above the band gap $E_g$ and neglect of non-radiative recombination (*i.e.* finite charge-carrier lifetimes) and material thickness. Consequently, several models have since been developed which go beyond the ideal SQ radiative efficiency limit [40, 45]. With each additional level of detail and complexity in these models, the predicted maximum efficiency decreases marginally from the SQ value, approaching a reasonable upper limit to the PV efficiency of a real-world device.

**SLME:** One extension was the 'spectroscopic limited maximum efficiency' (SLME) proposed by Yu and Zunger [45], which has seen considerable use in the literature since its introduction in 2012 [46, 47]. The SLME extends the SQ model to include the thickness and real, non-ideal absorbance of materials. Additionally, the SLME has the capacity to account for finite charge-carrier lifetimes, through specification of the internal quantum luminescence efficiency $Q_{lum,int}$, though this requires prior knowledge of the rate of non-radiative recombination.

**Blank *et al.*:** In 2017, Blank *et al.* went a step further by describing the solar material using bulk properties, namely the refractive index and external (rather than internal) quantum luminescence yield $Q_{\text{lum,ext}}$, thereby including the effects of light outcoupling and photon recycling [40]. The authors applied this metric to a range of emerging thin-film PV materials, including CZTS ($Cu_2ZnSnS_4$), $CuSbSe_2$ and $Sb_2Se_3$, finding that $CuSbSe_2$ exhibits the highest efficiency for ideal quantum luminescence efficiencies ($Q_{\text{lum,ext}} = 1$), but that CZTS takes the lead at more realistic values ($Q_{\text{lum,ext}} < 10^{-2}$). At present, this method entails a certain degree of manual calculation analysis, and so the development of a user-friendly open-source package to automate and streamline the process would substantially aid both the popularity of this performance metric and its applicability to high-throughput screening investigations.

**TLC:** In 2020, Kim *et al.* developed a 'trap-limited conversion' efficiency (TLC) model to explicitly incorporate equilibrium populations of native defects, their carrier-capture coefficients, and the associated non-radiative recombination rates (Figure 3) [48]. The TLC model relies on Shockley-Read-Hall (SRH) recombination statistics and the principle of detailed balance. Such an approach can provide a "gold standard" of maximum efficiency prediction for assessing candidate PV absorber materials. However, it also entails significant computational expense, and so it is only applied to the most promising materials.

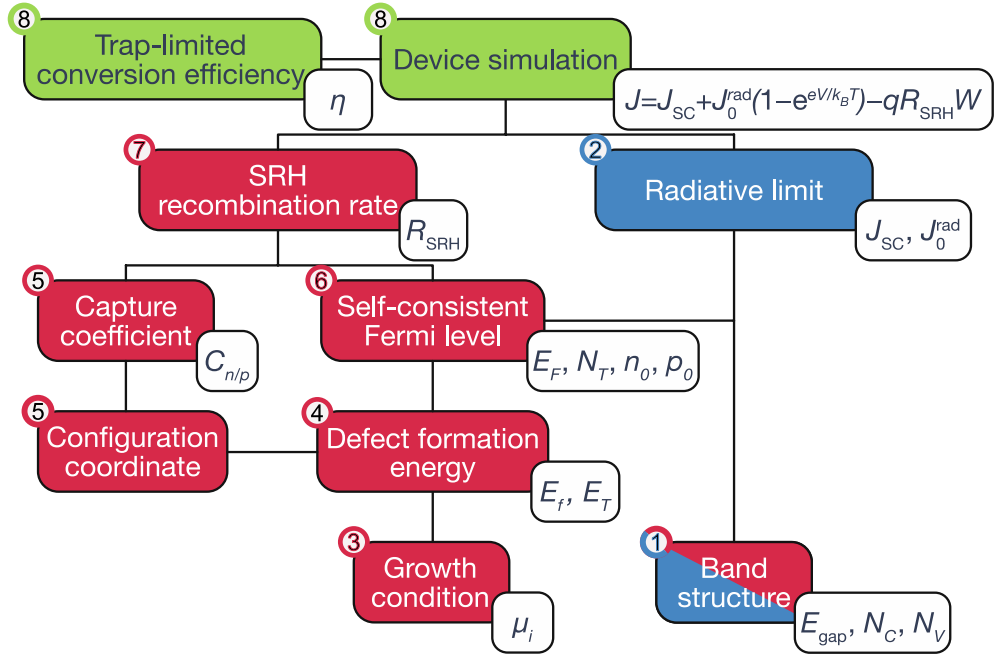

**Figure 3.** Schematic diagram of the calculation of the trap-limited conversion efficiency (TLC) metric. The dependent calculations are numbered and connected by lines, with the relevant calculated quantities appended. The red and blue boxes represent calculations for radiative and non-radiative electron–hole recombination, respectively. The combined device simulations are marked in green. Adapted from Ref. [48] by permission of the Royal Society of Chemistry.

For example, the TLC model was recently applied to the kesterite family of quaternary thin-film photovoltaic materials, including CZTS, AZTSe ($Ag_2ZnSnSe_4$) and others, revealing intrinsic defect-limited maximum efficiencies in the range of 20%. In this case, the primary culprits are native tin-based defects, in particular the tin-on-zinc antisite $Sn_{Zn}$. These defects produce electronic states deep in the band gap, which act as 'killer' recombination centers, severely reducing charge-carrier lifetimes (as further discussed in Section 2.2.1). This work offered valuable insight regarding the physical origins of poor experimentally-observed PV performance of these materials, which despite years of intense research and SQ radiative limits of ∼ 33% [49], have as-yet failed to exceed power-conversion efficiencies of 12% [50].

In light of the fact that the efficiencies of actual PV devices are in most cases limited by non-radiative recombination (the effects of which are illustrated in Figure 4) [51, 52], it is imperative that this loss mechanism is treated in first-principles investigations of PV performance. Indeed, champion materials from each respective generation of solar absorbers (c-Si, CdTe, MAPbI₃) demonstrate slow non-radiative electron-hole recombination kinetics which, in combination with large PLQE η and small Urbach tails, affords solar energy conversion efficiencies close to their radiative limits [53]. Extending the scope of materials discovery procedures to include these defect-governed properties will lead to improved predictions of successful next-generation photovoltaic materials.

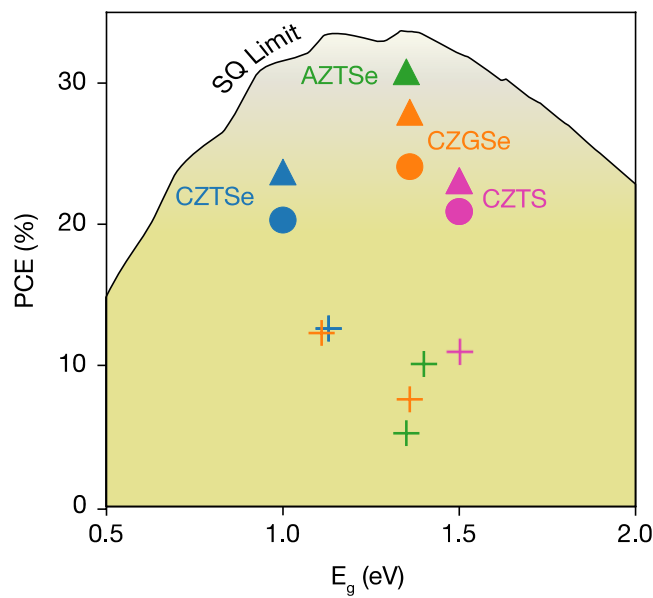

**Figure 4.** Shockley–Queisser limit and trap-limited-conversion (TLC) efficiencies for the kesterite family of thin-film PV materials. Filled symbols represent the TLC power-conversion efficiency and the black line is the SQ limit. TLCs with doping (triangles) show improved performances over TLCs without doping (circles). Plus signs indicate experimental data for kesterite solar cells. Adapted from Ref. [48] by permission of the Royal Society of Chemistry.

## 2.2 Perovskite-Inspired Design Principles and Defect Tolerance

Following the impressive optoelectronic performance of LHPs, recent efforts in the field have focused on leveraging theory to understand the physical properties underpinning their success [20]. From these chemical insights, design rules and screening criteria may be developed to guide the search for high-performance 'perovskite-inspired materials' (PIMs).

Undoubtedly, a defining feature of lead-halide perovskites (LHPs) is a remarkable tolerance to non-stoichiometry and crystalline defects, which facilitates their high PV efficiencies despite low-cost, solution-based synthesis routes. While deliberate implementation of defects plays an essential role in many relevant technological applications, such as topological insulators, quantum materials and, of course, doped semiconductors [54], the unintentional presence of defects can often trigger serious degradation in device performance [48]. In fact, as previously mentioned, provided the electronic band gap is in the appropriate range for an efficient PV material (~0.8 to 1.8 eV) [44], the most common performance-limiting factor for new absorbers is non-radiative recombination [51, 52]. By reducing the minority charge-carrier lifetime, non-radiative recombination impedes the collection of photo-generated carriers at the device terminals, thereby decreasing the open-circuit voltage.

We define defect tolerance as the tendency of a material to exhibit long minority carrier lifetimes (due to slow electron-hole recombination kinetics) despite the presence of crystallographic defects. This phenomenon has been definitively identified as a crucial factor in the success of LHPs [21, 54]. Many high-efficiency systems, such as single-crystal GaAs, are known to be extremely intolerant to the presence of point defects, necessitating a series of sophisticated and expensive procedures (in a clean room, or at high temperature) to fabricate pristine crystals for efficient PV devices. In contrast, perovskite solar cells may be synthesized via solution-based processing in normal laboratory environments at lower temperatures, which are typically up to only 100 °C. Nevertheless, LHPs manage to exhibit remarkably long minority carrier lifetimes and high PV efficiencies. Characterizing and understanding the physical mechanisms behind this extraordinary defect tolerance is thus a key research focus for the discovery, design and development of low-cost, high-efficiency next-generation solar cells.

### 2.2.1    *Shockley-Read-Hall Theory of Non-Radiative Recombination*

From Shockley-Read-Hall (SRH) theory, it is predicted that traps most likely to facilitate non-radiative recombination are those near the center of the electronic band gap [55]. Within this model, carrier-capture rates decrease exponentially with increasing energy separation between the defect transition level and the respective band-edge (*i.e.* the 'depth' of the defect level). Electron-hole recombination is a two-(quasi)particle process, and so a defect must sequentially capture both an electron and a hole for non-radiative recombination to occur, as depicted in Figure 5. The total recombination rate is limited by the slower of the two capture processes. Thus, the closer a defect level is to the mid-gap region, the more rapid the recombination kinetics expected for an intrinsic semiconductor. Beyond the depth of the trap levels, there are several other important factors that contribute to the recombination activity of a particular defect (*vide* Section 2.3), including the accessible charge states, concentration, and capture coefficient.

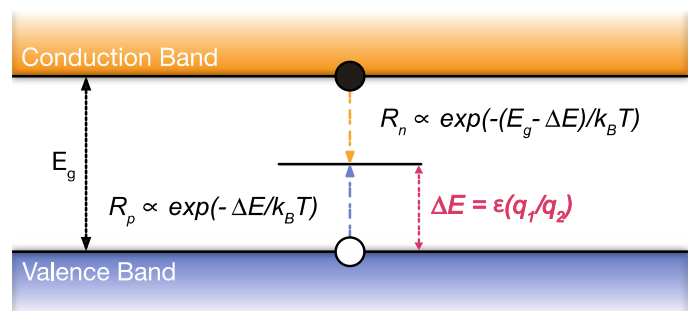

**Figure 5.** Electronic energy diagram of Shockley-Read-Hall recombination at a defect level $\Delta E = \varepsilon(q_1/q_2)$ above the valence-band maximum (VBM)(More on charge transition levels in Section 2.3.3). Electron and hole capture rates, $R_n$ and $R_p$, are exponentially dependent on the energy separation between trap level and band-edge.

*Shallow and Deep Defect States*

Point defects are typically classified as 'shallow' or 'deep' according to a variety of criteria, such as the localization of the donor/acceptor wavefunction [56], the relative magnitudes of the carrier capture and emission coefficients [57], or, most commonly, the energetic position(s) of the defect charge transition level(s) in the band gap [21, 48, 54]. Deep defects typically exhibit localized electronic wavefunctions and charge transition levels 'deep' within the band gap. Shallow defect states, on the other hand, produce delocalized wavefunctions with charge transition levels near or within the continuum electron bands, thereby remaining innocuous with respect to carrier lifetimes and PV performance.

### 2.2.2   What Makes a Material Defect Tolerant?

#### 2.2.2.1  Orbital Character

In an elemental semiconductor such as Si, there is a bonding valence band maximum (VBM) and an anti-bonding conduction band minimum (CBM), as depicted in Figure 6a. A point defect, such as a vacancy, can therefore produce non-bonding states at the center of the band gap. Initially, defect tolerance was attributed to a change in the bonding character of the band-edges [54, 58]. This concept posited that the formation of shallow traps would be favored if the VBM originated from anti-bonding interactions and the CBM from bonding interactions. Non-bonding defect states would therefore fall in the valence or conduction band continua, rather than the band gap. This type of bonding is often associated with the presence of $d_{10}$ cations (e.g. $Cu_+$), and has been, for example, observed in $Cu_3N$ [59]. In $Cu_3N$, $p$-$d$ orbital repulsion (due to the presence of filled Cu $3d$ states) produces anti-bonding character at the VBM, with similar interactions leading to a bonding-type CBM.

Analogously, a similar phenomenon has been observed for LHPs, in which orbital repulsion between the halide anion $p$ states and cation $s_2$ lone-pair acts as the source of anti-bonding interactions at the valence band-edge. However, in LHPs, the hybridization of the Pb $6p$ and anion $p$ orbitals results in the formation of an anti-bonding state at the CBM (as opposed to the bonding-character CBM of $Cu_3N$), with the bonding orbital forming within the valence band [21]. This orbital arrangement is illustrated in Figure 6b, and is conducive to shallow defect formation because the original atomic orbitals (from which non-bonding defect states arise) are close to or within the continuum bands, rather than deep within the band gap (Figure 6).

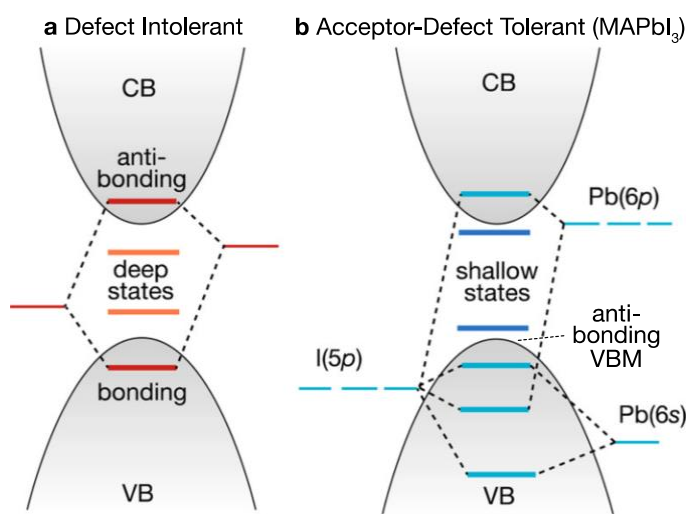

**Figure 6.** Electronic structure of typical 'defect intolerant' III–V, II–VI, or group IV semiconductors (left), prone to the formation of deep defect levels, compared to the 'defect tolerant' anti-bonding valence band of the lead halide perovskites (right). Adapted with permission from [21]. Copyright 2020 American Chemical Society.

A similar 'defect-tolerant' electronic structure can occur in the case of substantial band-edge splitting and dispersion, whether by spin-orbit or crystal-field effects. Through shifting and broadening the continuum electronic bands, the band-edges are pushed toward (if not beyond) the atomic orbital energy levels, in a manner analogous to that depicted in Figure 6. Likewise, this situation promotes the formation of shallow defect states.

### 2.2.2.2 Coordination Environment

Charge-carriers may be trapped at a deep defect state by long-range electrostatic and/or short-range (orbital overlap and hybridization) potentials, with the former dominating in ionic materials and the latter more relevant for covalent compounds [60, 61]. In covalent semiconductors, hybridization of dangling-bonds at a defect center is thus a governing factor for the depth of the resultant charge transition level [62–64], as alluded to in the original defect tolerance concept above.

In the case of an anion vacancy, which is positively-charged in the fully ionized state (e.g. $V_{I+}$ in CsPbI$_3$), the donor-type defect can trap a free electron by localizing it in the unoccupied *p* orbitals of the surrounding cations. The addition of an electron will cause these neighboring cations to slightly contract, on account of Coulomb attraction, thereby lowering the energy of the localized electron (due to increased hybridization of the dangling *p* orbitals). The strength of this electronic stabilization, relative to the strain energy from nearby stretched bonds, determines the trapping energy of the defect (*i.e.* depth of the defect transition level). If the orbital hybridization is not sufficiently strong, this localized electronic state will be unstable and the defect will instead adopt the energetically-favored shallow hydrogenic donor state, with the electron located at the defect-perturbed CBM.

The strength of this cation-cation interaction at the anion vacancy site will be dependent on both the anion coordination environment and the cation-cation distances, with low coordination numbers and large inter-cationic distances associated with weak hybridization and shallow defect levels. Thus, the presence of large anions and bond lengths with low anion coordination is predicted to aid defect tolerance, particularly in the case of chalcogenides, halides and chalcohalides, since anion vacancies are typically the primary native donor defects in these materials [63, 64].

In Ref. [64], Kim *et al.* provide an instructive example of this behavior; employing an 'inorganic frame model' to decouple the chemical and structural effects of halide substitution on the anion-vacancy defect level for inorganic lead-halide perovskites. They demonstrate that the halide vacancy ($V_X$, X = I, Br, Cl) transforms from a delocalized shallow state in [PbI$_3$]- and [PbBr$_3$]- to a localized deep state in [PbCl$_3$]- (Figure 7), as a result of lattice contraction (hence increased hybridization) from the smaller anionic radius, alongside band-shifting effects.

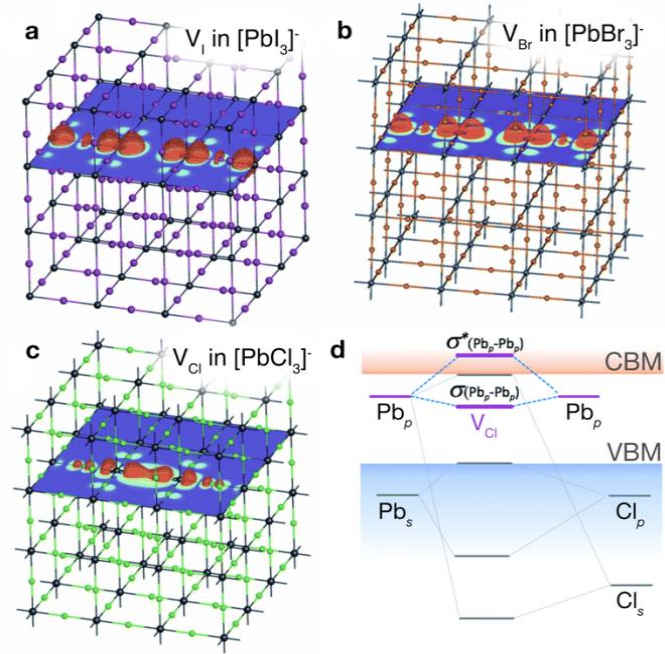

**Figure 7.** Charge density isosurfaces of V$_X$ (X = I, Br, Cl) in *(a)* [PbI$_3$]-, *(b)* [PbBr$_3$]- and *(c)* [PbCl$_3$]-. For [PbI$_3$]- and [PbBr$_3$]- in (a) and (b), the orbital configuration of the halide anion vacancy V$_X$ is primarily Pb *6p* - X *p* anti-bonding character, matching that of the CBM (i.*e.* a shallow perturbed CBM state). For [PbCl$_3$]- in (c), however, a localized Pb *6p* bonding state is formed. In (d), the orbital configuration for the formation of deep halide vacancy defects in [PbCl$_3$]- is shown. Adapted with permission from Ref. [64] with permission from the PCCP Owner Societies.

Analogous behavior could hypothetically occur for cation vacancy sites, with the strength of anion-anion dangling bond hybridization contributing to the depth of the acceptor defect level. That said, to the knowledge of the authors, this mechanism has not yet been ascribed to the formation of shallow acceptor states in a defect tolerant material. One likely reason for this is the fact that band-structure defect tolerance effects (Section 2.2.2.1) are not always independent of structural effects. For instance, in the hypothetical cubic PbI$_2$, tetrahedral Pb coordination and shorter Pb-I distances strengthen the anti-bonding orbital interaction at the VBM, yielding shallow cation vacancies, as opposed to the deep states found in ground-state layered PbI$_2$ [65]. However, this coordination environment implies greater dangling-bond hybridization, and deeper defect states. The fact that, in this case, the energy upshift of the VBM outweighs that of the dangling-bond defect states, results in overall shallow cation vacancies.

In a similar manner, defect tolerance has been partially ascribed to ionic bonding in certain perovskite materials [61, 66, 67], with reduced covalency yielding lower energy dangling bonds and hence shallower defect levels. However, as mentioned earlier, long-range electrostatics may also induce charge localization and hence deep defect formation in heteropolar/ionic systems; the strength of this interaction being mediated by the dielectric constant of the material [60]. Thus, one should be wary of suggesting decreased covalency as an indication of potential defect tolerance in a candidate material, as it is a delicate balance of several competing effects (ion coordination, bond length, hybridization, bonding character, crystal symmetry, oxidation state, Jahn-Teller distortion, etc.), which ultimately determine the energetic position of the defect charge transition level.

These points serve to highlight the intricacy of the physics governing the defect wavefunction, which must be carefully maneuvered when attempting to construct robust defect tolerance design principles.

### 2.2.2.3 Defect-Carrier Coulomb Interactions

An electron or hole in a semiconducting material may be attracted or repelled from a charged point defect due to simple Coulomb interactions, increasing or decreasing the carrier-trapping rate relative to the neutral case. This electrostatic effect is incorporated via the Sommerfeld enhancement factor $\langle s \rangle$, which for the attractive case is given by [68, 69]:

$$\langle s \rangle \ = \ 4 |\frac{Q}{q}| \sqrt{\frac{\pi E_R}{k_B T}} \propto \frac{|Q|}{\varepsilon} \sqrt{\frac{m^*}{T}} \tag{1}$$

where $|Q|$ is the absolute charge of the point defect, $q$ and $m^*$ are, respectively, the carrier charge and effective mass, $\varepsilon$ is the (low-frequency) dielectric constant, $T$ is the temperature and $E_R$ is the effective Rydberg energy:

$$E_R \;=\; \frac{m^* q^4}{2\hbar^2 \varepsilon^2} \tag{2}$$

Depending on the dielectric and conductive properties of the material, the Sommerfeld factor $\langle s \rangle$ can range from anywhere between $\sim 0.1$ to 20, therefore strongly influencing electron-hole recombination kinetics and subsequent charge-carrier lifetimes. From visual inspection of Equation (**1**), it is evident that several material properties can influence the magnitude of this carrier-capture enhancement factor, thereby contributing to the defect (in)tolerance of a material.

**Enhanced Dielectric Screening:** The dielectric constant reflects the ability of a compound to screen an electrostatic perturbation. Numerous material properties critical to the physics of a PV device, such as the rates of carrier-capture and ionized impurity scattering, the depth of shallow defect levels and exciton binding energies, are dependent on the dielectric constant [54]. For a given defect concentration, a larger dielectric constant corresponds to stronger Coulombic screening and hence diminished defect carrier-capture cross-sections ($\langle s \rangle \propto 1/\varepsilon$), thereby suppressing the rate of non-radiative recombination.

**Oxidation States:** The oxidation states of species in a material determines the relative likelihood of different charge states for a given defect [70]. For instance, in CdTe, cadmium is formally in the +2 oxidation state, and so a cadmium vacancy $V_{Cd}$ can take on the neutral and -2 charge states; $V_{Cd}x$ and $V_{Cd}$'' in the Kröger-Vink formalism [71]. For a highly-charged defect, the kinetics of carrier-capture, and hence electron-hole recombination, can be accelerated relative to the neutral case ($\langle s \rangle \propto |Q|$). Thus, low oxidation-state materials may be more likely to contain low-charge defects with reduced carrier-trapping rates.

**Effective Mass:** Spatial localization of electron and hole wavefunctions at defect sites can slow charge-carrier transport and is associated with thermal losses. A small effective mass will favor delocalized, nearly-free charge-carriers with decreased defect capture cross-sections ($\langle s \rangle \propto m^*$). Of course, a small effective mass is also beneficial for the carrier mobility and electrical conductivity of the device, as well as preventing the formation of small polarons [54].

### 2.2.2.4 Benign Complex Formation

Another proposed technique for the design of defect-tolerant materials is the passivation of 'killer' defect centers via the deliberate formation of benign defect complexes [54, 67, 72]. In this case, defect-defect interactions are leveraged to shift deep-levels closer to (or even into) the continuum bands, 'bleaching the gap' of deleterious trap states. The main adversary in this situation, however, is the entropic cost associated with defect-complex formation, which must be overcome either by an adequate enthalpy gain or the creation of metastable arrangements via intelligent synthesis strategies. Due to the inherent difficulty in engineering this state, this strategy remains largely unexplored for the design of novel defect tolerant materials.

### 2.2.2.5 Self-Inhibiting Defect Levels

If a defect center produces multiple levels in the band gap, rather than accelerate the recombination kinetics as one might initially expect, these extra levels often 'self-inhibit', reducing the overall rate of recombination [73, 74]. In this situation, it is likely that one of the defect levels is near a band-edge, far from the gap center. Due to the exponential dependence of charge-capture rate on the energetic separation of the defect level and corresponding band-edge (*i.e.* the 'depth' of the defect level), enormous imbalances in charge transition rates can arise in this case, leading to defects becoming 'stuck' in extreme charge states. Being near a continuum band, the near-edge defect level efficiently captures and re-emits the corresponding charge carrier (hole if near the VBM or electron if near the CBM). However, as depicted in Figure 8, capture of the opposite charge carrier is exceptionally slow. Consequently, the majority of defects adopt the extreme charge state, exhibiting sluggish carrier capture and recombination kinetics.

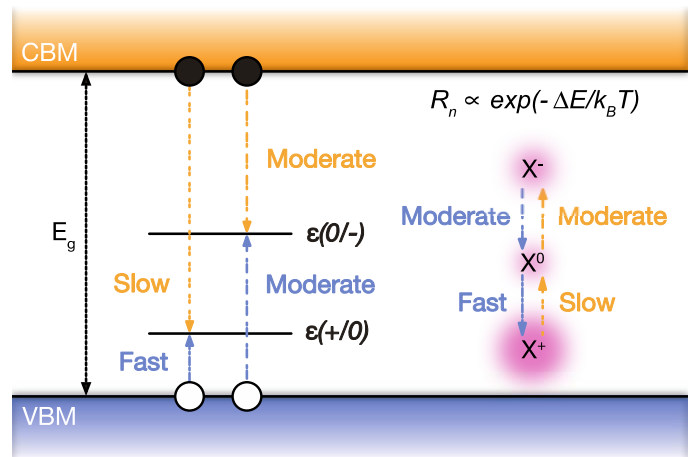

**Figure 8.** Energy level diagram of carrier capture and SRH recombination at two defect levels in the band gap, demonstrating 'self-inhibiting' behavior. (More on charge transition levels in Section 2.3.3). Under the SRH model, the exponential dependence of carrier capture rate $R_n$ on the energy separation between trap level and band-edge leads to a buildup of defects in the $X_+$ state. This defect charge state exhibits weak carrier capture kinetics, thus suppressing non-radiative recombination.

On the other hand, the highly recombination-active charge states (which produce the mid-gap transition level) have almost negligible concentration and thus minimal overall recombination activity. Another way to conceptualize this mechanism is to recall that the total recombination rate for a given defect species is dictated by the slowest capture process involved (Section 2.2.1). Hence, the incredibly slow capture rate of the extreme charge state introduces a bottleneck in the recombination cycle, greatly diminishing the overall recombination activity of the defect. Such tolerance can similarly occur in the case of strong anharmonicity, whereby transition rate imbalances are driven by a lack of symmetry in potential energy surfaces and charge capture barriers [74].

### 2.2.2.6 Related Defect Tolerant Characteristics

**Crystal Structure:** In any material, the crystal structure is a governing factor in the resultant electronic structure. Highly-symmetric crystal structures (*viz.* cubic phases such as halites, CsCl-type structures and cubic perovskites) tend to exhibit isotropic bonding and transport properties. Often, this behavior produces a direct fundamental gap (though the optical gap may still be indirect, due to strong selection rules) and relatively disperse band extrema, resulting in low effective masses [75, 76]. On the other hand, highly-anisotropic compounds such as layered materials often exhibit heavy effective masses and relatively flat electronic bands. More importantly, the crystal structure significantly influences the strength of both dangling-bond hybridization (Section 2.2.2.2) and orbital repulsion at band-edges (Section 2.2.2.1).

**Cation Electronic Configuration:** The electronic configuration of the cation is associated with a number of defect-tolerant properties in a material. In halide perovskites, an important case is the formation of the $6s^2$ lone pair configuration of $Pb_{2+}$ in $MAPbI_3$, which has a partially oxidized valence electron shell relative to the fully-ionized $6s^0 6p^0$ $Pb_{4+}$ ion. The resulting large, polarizable cation induces large Born effective charges (and thus a high dielectric constant) alongside stronger spin-orbit coupling of the unoccupied cation $p$ states in the conduction band (hence increased band dispersion, favoring shallow donor defects, and reduced electron effective mass) [76, 77]. Moreover, filled $s^2$ orbitals can partially hybridize with the valence-band-dominated anion $p$ orbitals, introducing anti-bonding character in the VBM (encouraging the formation of shallow acceptor defects) and broadening the valence bandwidth (thus reducing the hole effective mass).

This stable 'N-2' lone-pair oxidation state, and accompanying effects, is most commonly observed for the heavy post-transition metals in the sixth row of the periodic table (i.e. $Tl_+$, $Pb_{2+}$ and $Bi_{3+}$), due to the relativistic contraction (and consequent stabilization) of the valence $s$ orbital [78, 79]. Nonetheless, lighter cations (namely $In_+$, $Sn_{2+}$ and $Sb_{3+}$) may also adopt these oxidation states in PV materials [76, 79, 80]. Moreover, strongly electronegative anions ($F_-$, $Cl_-$, $O_{2-}$ etc.) are more likely to fully oxidize the cations, hence softer elements ($S_{2-}$, $Se_{2-}$, $I_-$ etc.) are often required to stabilize the cation lone-pair.

A salient point, regarding the valence band anion $p$ - cation $s$ hybridization, is that the strength of this interaction (hence degree of dispersion and anti-bonding character at the band-edge) is dependent on the alignment and overlap of the atomic orbitals. If the cation lone-pair is much deeper than the higher-energy anion $p$ orbitals, there will be minimal interaction between these states, with a consequently negligible contribution from the cation states to the band-edge. Indeed, this has been observed for certain $Bi^{3+}$ lone-pair materials which were recently investigated as potential solar absorbers, including $BiI_3$ and $Cs_2AgBiBr_6$ [21, 46, 79].

This point provides an instructive example of the specificity of certain defect tolerance mechanisms, which unfortunately prevents the decoupling of elemental contributions to this phenomenon. Rather than simply searching for materials with stable lone-pair cations, one must also consider their pairing with compatible anions to achieve the desired response.

### 2.2.3    Defect Tolerance in Lead Halide Perovskites

There has been much research on the atomistic origins of the strong defect tolerance exhibited by lead-halide perovskites (LHPs) [21, 54, 57, 64, 74–77, 81, 82]. Despite some disagreement in the literature [74, 81, 82], likely due to the multi-faceted origin of this phenomenon in LHPs, several material properties have now been definitively identified to contribute to the remarkable defect tolerance of these materials.

#### 2.2.3.1  Defect Tolerant Band Structure

Firstly, as depicted for the valence band in Figure 6, perovskites exhibit a unique band structure in which interaction between the B-site cation $ns_2$ lone-pair and the anion $p$ orbitals produces a dispersive, anti-bonding VBM. As discussed above, this situation is predicted to yield predominantly shallow acceptor-type defects, such as cation vacancies, alongside small hole effective masses (and thus a decreased hole capture rate, per (1) [21, 57, 76].

In addition, for a heavy cation, such as $Pb^{2+}$ in $MAPbI_3$, strong spin-orbit coupling broadens the bandwidth of the CBM, reducing the occurrence of deep donor defects in a similar manner to the bonding-character CBM in Figure 6. Greater conduction band dispersion also lowers the electron effective mass, thus decreasing the rate of electron trapping per (1).

### 2.2.3.2  Cationic Lone Pair

The $6s_2$ electronic configuration of $Pb_{2+}$ also aids defect tolerance in MAPbI$_3$ through a dramatic contribution to the dielectric properties of the material [62]. Born effective charges, describing the ionic polarizability, are enhanced by the presence of the $Pb_{2+}$ cationic lone-pair. The resulting increase in mixed ionic-covalent bonding character leads to a massively-increased dielectric constant ($\varepsilon_r > 25$), which is approximately three times greater than that of other thin-film solar absorbers, such as CdTe and CZTS [54, 57, 77]. By effectively screening charged defect centers and enhancing carrier mobility, the highly-polarizable lone-pair emphatically suppresses non-radiative recombination and improves photovoltaic performance.

To illustrate the influence of a small effective mass $m^*$, large permittivity $\varepsilon_r$ and low absolute oxidation states (+1, +2 and -1 respectively for the prototypical ABX$_3$ perovskite) on defect tolerance in LHPs, one can inspect the attractive Sommerfeld enhancement factor $\langle s \rangle$ for carrier capture. For a typical singly-charged defect in MAPbI$_3$ (*e.g.* V$_{I+}$), $\langle s \rangle \simeq 2$, compared to $\langle s \rangle \simeq 10$ for a triply-charged defect in GaAs (*e.g.* V$_{As3+}$), using values for $m^*$ and $\varepsilon_r$ quoted in [57].

### 2.2.3.3  Anion Coordination Environment

Beyond the effects of the valence $s_2$ lone pair, the low coordination number (2) of iodine, in conjunction with the expanded lattice constant from the methylammonium A-site cation, gives rise to weak Pb-Pb hybridization at the iodine anion vacancy center. This weak interaction is insufficient to create a deep bound state, further aiding defect tolerance in this material [62, 63]. Despite the presence of partial ionic character, long-range electrostatic effects are also unable to localize charge in this case, due to the small band gap and effective dielectric screening [60, 61].

The success of MAPbI$_3$ in combining weak cation-cation dangling-bond hybridization with a defect-tolerant electronic structure (Section 2.2.3.1) indicates that ternary or multinary compounds may provide a fruitful search space for the discovery of new defect-tolerant materials [65]. In particular, materials which exhibit 'pseudobinary' behavior, containing a spectator ion which does not contribute to the band-edge states (*cf.* MA$_+$ cation in MAPbI$_3$), may be more likely to simultaneously exhibit low ion coordination and orbital repulsion at band-edges.

### 2.2.3.4  *Inactive Deep States*

On the other hand, the few deep intrinsic defects that do still exist in LHPs, including iodine antisites I$_{Pb}$ and I$_{MA}$, lead-iodine antisite Pb$_I$ and interstitial lead Pb$_i$, have been shown to demonstrate weak recombination activity [74, 83]. The physical reason behind this inactivity has been debated. Initially, it was attributed to large predicted formation enthalpies, which would result in low concentrations with little to no adverse effect on carrier lifetimes [21, 83].

Recently however, this finding has been challenged, with 'self-inhibition' (as a result of multiple intra-gap defect levels) predicted to be the origin of slow recombination kinetics for I$_{Pb}$ and Pb$_I$ antisites [74]. The authors suggest that both anharmonic carrier-capture barriers and the presence of additional near-edge defect levels introduce bottlenecks in the non-radiative recombination cycle (*vide* Section 2.2.2.5), thereby rendering LHPs tolerant to these deep traps.

### 2.2.3.5  *Low Schottky Defect Formation Energies*

In addition, theoretical and experimental investigations have found low formation enthalpies of intrinsic Schottky-type defects (*i.e.* cation vacancy - anion vacancy complexes) [67, 72], suggesting high concentrations of benign defect complexes.

### 2.2.3.6 Self-Healing

Finally, the soft crystal framework and relatively low melting temperature of LHPs points at the possibility of appreciable ion diffusion and 'self-healing' of defects at typical processing temperatures, resulting in fewer recombination-active defect center [84, 85]. Moreover, this leads to small phonon energies in MAPbI$_3$, which also contribute to a reduction in carrier capture rates [57, 86].

### 2.2.4 Target Features for Defect Tolerance

Through thorough investigation and enhanced understanding of the physical origins of defect tolerance in LHPs, it is hoped that general design principles may be developed for 'perovskite-inspired' defect-tolerant materials.

Such target material features include: *(i)* band-edge orbital repulsion (bonding character); *(ii)* large polarizable cation(s) (aiding dielectric screening of charged defects, band dispersion and anti-bonding valence band interactions – assuming sufficient spatial overlap and energetic alignment with anionic states); *(iii)* low atomic coordination alongside large ionic radii and lattice constants (reducing dangling-bond hybridization); *(iv)* mixed ionic-covalent bonding character (strengthening dielectric screening and suppressing dangling-bond hybridization); *(v)* crystal anharmonicity and/or multiple intra-gap defect levels (promoting 'self-inhibition'); *(vi)* low formation enthalpies for innocuous defect complexes; and *(vii)* soft crystal framework (to allow 'self-healing').

There is a plethora of ongoing research focusing on the implementation of these design principles as target features for high-throughput materials screening, with the goal of discovering next-generation defect-tolerant PV materials which are capable of rivalling the performance of the LHP prototype [21, 46, 57, 64, 74–77, 81, 82]. While there are likely few materials that match the wealth of defect-tolerant properties exhibited by lead-halide perovskites, with intelligent discovery techniques driven by detailed understanding of the underlying chemistry (as depicted in Figure 9), one can hope that further successful absorber materials may be unearthed.

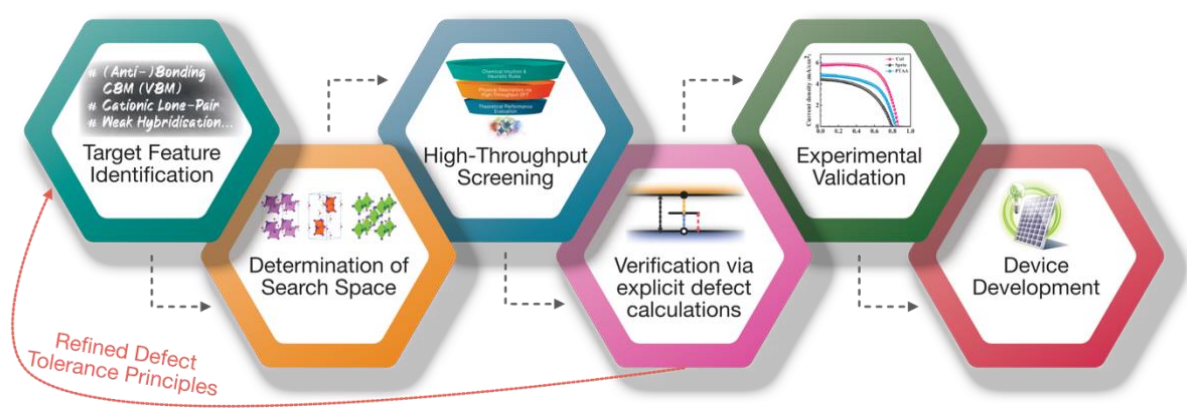

**Figure 9.** Outline of the synergistic combination of high-throughput screening, detailed theoretical investigation and experimental characterization to yield accelerated pathways to the discovery of defect-tolerant materials for low-cost high-efficiency devices. Red arrow indicates the feedback loop between target features for material screening and explicit defect investigations, affording refined design principles and screening criteria for efficient materials discovery.

In the context of materials screening, some of these material properties are relatively simple to predict *a priori*, for example the cation electronic configuration, bonding covalency and ionic radii, while others, such as those involving orbital interactions or carrier-capture kinetics, can be tremendously challenging. For instance, accurate calculation of the energetic alignment of cation lone-pair and anion *p* states (to predict anti-bonding interactions at the VBM in a perovskite-structured material) is often a computationally taxing procedure, requiring a high level of theory. Identification of 'self-inhibition', whether by crystal anharmonicity or multiple intra-gap levels, requires defect-specific, quantitative investigations [74]. Likewise, prediction of benign defect complexes is not readily amenable to high-throughput computation, due to the enormous configurational space and inherent complexity involved, on top of the substantial cost of defect calculations. This point serves to reinforce the importance of strategic organization of materials screening steps, alongside the judicious selection of chemical descriptors and success criteria, on the basis of relevance, cost, accuracy and researcher-effort.

After applying these design principles at an intermediate stage of the screening process (through the use of appropriate chemical descriptors), explicit defect calculations may then be performed on the most promising remaining candidates at the bottom of 'screening funnel' (Figure 2), in order to fully assess their defect tolerance potential and to yield an accurate performance figure of merit (*SLME, Blank et al., TLC...*). In Section 2.3, we provide a brief overview of the practical calculation of relevant defect properties from first-principles.

### 2.3 *Ab-Initio* Defect Calculations

A wide range of defect properties (including formation energies, equilibrium concentrations, charge transition energy levels, carrier capture coefficients, etc.) are now accessible to theoretical investigation, due in no small part to the aforementioned advances in theory and computational capacity. By elucidating the physical underpinnings of defect behavior in solids, chemical models of improved accuracy may be constructed in order to better predict, describe and design defect-related material properties. Such insights can enable researchers to tune material properties via strategic manipulation of defect concentrations (through carefully-controlled synthesis conditions), whether to deliberately introduce defects (*e.g.* for doping-induced conductivity or enhanced lasing action [87]) or to mitigate the effects of deleterious defects (*e.g.* for improved PV efficiency or battery lifetime [54, 88, 89]).

The purpose of this section is to give a brief flavor of the practical implementation, frontiers and challenges of this research. For a more detailed overview of the theory, state-of-the-art and practical aspects of defect calculations, the keen reader is directed elsewhere [56, 90, 91].

### 2.3.1   Supercell Approach

Density Functional Theory (DFT) has established itself as the most powerful technique for first-principles investigations in solid-state physics [92, 93], including point-defect calculations [90]. Within DFT, the most widely employed method for defect calculations is, by far, the 'supercell' approach with a plane wave basis set [56, 90, 91]. This involves placing a defect in an expanded crystal unit cell of the 'host' material (termed a 'supercell') with anywhere from ~ 50 to 200 atoms, which is then periodically repeated through space. In doing so, researchers may exploit the power and versatility of periodic DFT codes, while also guaranteeing an accurate description of the host material in most cases.

$$\Delta H_{\mathrm{X,q}}(E_F, \mu) \;=\; [E_{\mathrm{X,q}} - E_{\mathrm{H}}] \;\;-\;\; \sum_i n_i \mu_i \;\;+\;\; q E_F \;\;+\;\; E_{\mathrm{corr}}(q)$$

| Defect Formation Energy | Defect Supercell | Host Supercell | Chemical potentials from phase stability | Electron chemical potential | Finite size corrections |

**Figure 10.** Contribution of various terms in the calculation of defect formation energy.- Adapted with permission from Ref. [94]. Copyright 2017 Elsevier.

### 2.3.2   Defect Formation Energy

Both equilibrium concentrations and charge transition energy levels are determined by the formation energies of the corresponding defect states. The formation energy $\Delta H_{X,q}(E_F, \mu)$ of a defect $X$ in charge state $q$ is given by:

$$\Delta H_{X,q}(E_F, \mu) = [E_{X,q} - E_H] - \sum_i n_i \mu_i + q E_F + E_{Corr}(q) \tag{3}$$

The contribution of each of these terms to the calculated defect formation energy $\Delta H_{X,q}(E_F, \mu)$ is illustrated diagrammatically in Figure . $E_{X,q}$ is the calculated energy of the *defect-containing* finite-sized supercell, while $E_H$ is the energy of an equivalent *pristine* host supercell. The second term $(-\sum_i n_i \mu_i)$ accounts for the thermodynamic cost of exchanging $n_i$ atoms with their reservoir chemical potential(s) $\mu_i$, to form the defect $X_q$ from the ideal bulk material. Analogously, $q E_F$ represents the energetic cost of removal or addition of charge $q$ to the defect site, from the electronic reservoir of the system (*i.e.* the Fermi energy $E_F$). Finally, $E_{Corr}(q)$ is a correction for any spurious finite-size supercell effects.

As mentioned previously, the dependence of defect formation energy $\Delta H_{X,q}(E_F, \mu)$, and hence equilibrium concentration, on atomic chemical potentials $\mu_i$ allows for the manipulation of defect populations, via controlled synthesis conditions. Indeed, high-efficiency perovskite solar cells have relied on the strategic manipulation of iodine chemical potentials during crystal growth in order to suppress non- radiative recombination channels [88].

### 2.3.3    *Charge Transition Energy Levels*

Defects in semiconductors typically introduce electronic levels near or within the band gap, which determine the interaction of the defect site with charge carriers in the material. These electronic levels are the 'thermodynamic charge transition levels' $\varepsilon(q_1/q_2)$, defined as the Fermi level position for which the formation energies of the defect in charge states $q_1$ and $q_2$ are equal [90]:

$$\varepsilon(q_1/q_2) = \{\ E_F\ \mid\ \Delta H_{X,q_1}(E_F) = \Delta H_{X,q_2}(E_F)\ \} \tag{4}$$

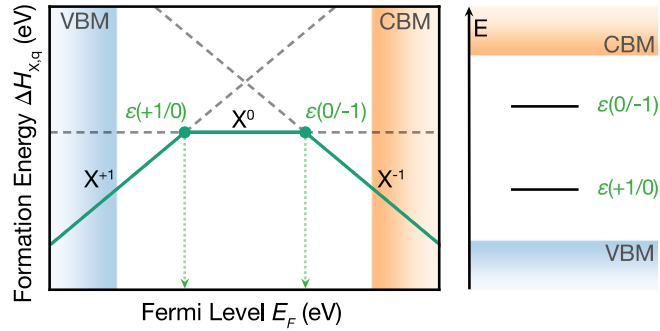

**Figure 11.** Defect charge transition levels from a formation energy (with respect to Fermi level) plot **(left)** and on an energy band diagram **(right)**. Drawn based on the defect calculation and processing tools available at Ref. [95].

As illustrated in Figure 11, charge state $q_1$ is stable for Fermi level below $\varepsilon(q_1/q_2)$, while for Fermi levels above $\varepsilon(q_1/q_2)$, charge state $q_2$ is stable ($q_1 > q_2$, by convention). Charge transition levels are crucial to the electronic behavior of defects, as their energetic position within the host band structure is a dominant factor in the recombination activity of defect centers (*vide* Section 2.2.1).

Employing the computational methods described in Sections 2.3.1 and 2.3.2, charge transition levels may be theoretically predicted via equations (**3**(4 as illustrated in Figure 11. Moreover, these levels may be experimentally detected via techniques such as Deep-level Transient Spectroscopy (DLTS) [90]. This highlights the importance of these properties in defect investigations, as they allow for the powerful combination of theory and experiment to yield enhanced understanding of defect behavior. Through comparison of theoretical and experimental results, computational methods may be validated, and experimental observations may be interpreted and understood in detail.

*2.3.4   Computational Overhead*

While standard DFT methods (*i.e.* LDA/GGA) are capable of producing reliable crystal structures (in most cases), systematic failures in the prediction of band gaps (Section 2.1.2) and charge localization render them unsuitable for investigations of defect electronic structure. These electronic properties are crucial to defect behavior, determining the position of charge transition levels in the host band structure and hence the nature of the defect trap state. This issue is tackled via the use of 'beyond-DFT' methods, the most popular choice being hybrid DFT, which inevitably entail significant inflation of the computational demand. This requirement of expensive, high-level electronic structure theory, in conjunction with large simulation supercells (Section 2.3.1), is the primary culprit behind the substantial computational cost associated with defect calculations.

### 2.3.5 Frontiers and Challenges of Defect Modelling

#### 2.3.5.1 Automation and Databases

At present, defect investigations entail exorbitant levels of manual researcher effort, due to the large number of individual calculations required. Fortunately, a number of computational packages have been released (including PyCDT [96], PyLada [94], COFFEE [97], sxdefectalign [98], CarrierCapture.jl [99], CPLAP [100], SC-Fermi [101]...), which can assist with efficient pre- and post-processing of these calculations. Through continued development of comprehensive, flexible, user-friendly computational tools, researchers hope to accelerate the efficiency of defect studies and expand their applicability to high-throughput materials discovery.

Furthermore, powered by a rapid growth in computational power and availability of automation frameworks such as AiiDA [102] and FireWorks [103], databases of material simulations are being extended to include *ab-initio* defect calculations. These invaluable resources will provide the opportunity to gain powerful chemical insights, especially when combined with machine learning models and data science techniques.

*2.3.5.2  Carrier Capture*

A crucial step in the process of non-radiative electron-hole recombination is the capture of delocalized charge carriers by defect traps, via electron-phonon coupling. Recent years have seen the development of theory and computational tools to quantitatively model this process [99, 104, 105], though further improvements in efficiency, versatility, and reliability are necessary to extend this approach to general application.

*2.3.5.3  Additional Hurdles*

Further challenges faced by defect studies include transferability issues (due to inconsistent realizations of electronic structure theory and finite-size corrections), accurate treatment of finite-temperature effects (demanding substantial computational and researcher effort [90]) and modelling of diffusion/transport properties (limited by the length- and time-scales of molecular dynamics simulations).

## 3.  Classes of Perovskite-Inspired Materials Explored

Halide perovskites have the general formula $ABX_3$, where A is a monovalent organic or inorganic cation, B is an octahedrally coordinated divalent cation, and X is a monovalent halide such as Cl, Br, or I. Chemical substitution of $Pb_{2+}$ for alternative, less-toxic B-site cations has been widely explored. However, the perovskite crystal structure can only be maintained if the cation has a +2 oxidation state and if the ionic radius can fit within the crystal structure. The latter is described by the Goldschmidt tolerance factor, $t = \frac{(r_A + r_X)}{\sqrt{2}(r_B + r_X)}$, where $r_A$, $r_B$, and $r_X$ are the ionic radii of the A, B, and X site ions, respectively. Empirically, stable perovskite structures have *t* ranging from 0.8-1.1 [106], which can only be achieved by a limited number of compositions [107]. Substitution of Pb(II) for other metal Group 14 cations with sufficiently large ionic radii (*e.g.*, $Sn_{2+}$ and $Ge_{2+}$) can result in *t* values within the required range, forming perovskite structures (*i.e.*, $ASnX_3$ and $AGeX_3$), as shown in Figure 12. An alternative that allows metal cations without a stable +2 oxidation state to be used is to combine a monovalent and trivalent cation (*e.g.*, $Ag_+$ and $Bi_{3+}$), which occupy alternating octahedra, to form double perovskites. But beyond identifying materials with perovskite crystal structures, efforts have also focused on the inverse design of defect-tolerant materials, as described in Section 2.

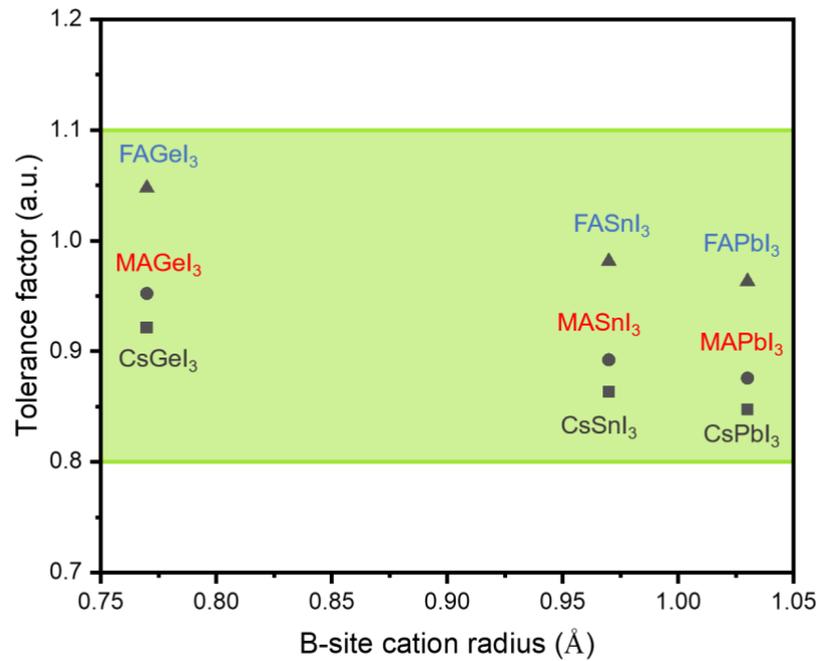

**Figure 12**. The geometric tolerance factors of three classes of perovskites with different B-site cations based on their ionic radii [107–109].

This section discusses the synthesis routes and optoelectronic properties of several classes of PIMs considered, including Sn- and Ge-based perovskites, double perovskites, as well as other potentially defect-tolerant compounds ($A_3B_2X_9$, $ABZ_2$, binary halides, and V-VI-VII materials). Crystal structures and some important features of the PIMs discussed here will be illustrated in Figure 13 below and Table 2 at the end of this section, respectively.

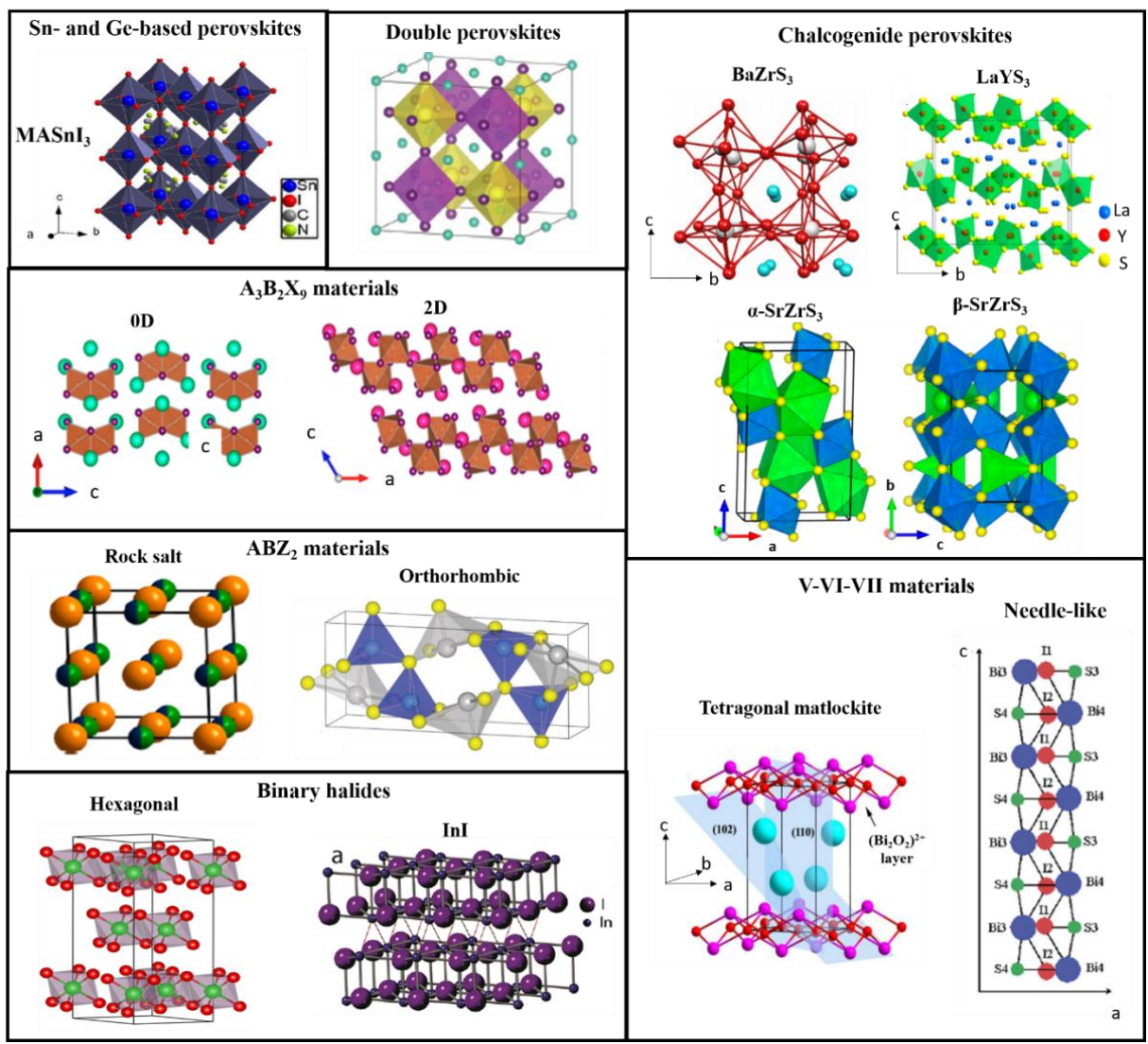

**Figure 13**. Crystal structures of various PIMs discussed in this Review. The axes for the crystal structures of each material follows the axes in the upper left corner if not otherwise marked. Reprinted with permission from Ref. [22] (Copyright 2014 Royal Society of Chemistry), [23] (Copyright 2019 Wiley), [24] (Copyright 2018 American Chemical Society), [25] (Copyright 2018 American Chemical Society), [26] (Copyright 2019 American Chemical Society), [27] (Copyright 2016 Royal Society of Chemisitry), [28] (Copyright 2007 Elsevier), [110] (Copyright 2018 American Chemical Society), [111] (Copyright 2020 MDPI), [112] (Copyright 2014 Royal Society of Chemistry), [113] (Copyright 2009 American Physical Society).

## 3.1 Sn-based perovskites

As with Pb-based perovskites, Sn-based perovskites can have $CH_3NH_{3+}$ (MA), $CH(NH_2)_{2+}$ (FA), $Cs_+$ and other alkali cations in the A-site. Though most Sn-based perovskites have a perovskite structure with a high symmetry α cubic phase (Figure 13), some $CsSnX_3$ perovskites may transform into the γ orthorhombic phase with distorted octahedra at room temperature [114], which is mainly due to their smaller tolerance factors.

Although many fabrication methods have been developed for LHPs, these methods are usually not directly compatible with their Sn-based counterparts. Sn-based perovskites crystallize much faster than Pb-based perovskites due to the lower solubility of $SnI_2$ in typical solvents such as $N,N$ – dimethylformamide (DMF) [115]. Therefore, Sn-based perovskite thin films are often fabricated by modified solution methods with the addition of reagents to the precursors.

The one-step solution method was first developed by Hao *et al.* [116]. It has been found that replacing DMF with dimethyl sulfoxide (DMSO) as the solvent can reduce the crystallization rate of Sn-based perovskites by the formation of the intermediate product $SnI_2 \cdot 3DMSO$ (Figure 14a), leading to pinhole-free films [115]. Film morphology has been further enhanced using solvent engineering methods (Figure 14b). In this method, perovskite precursors are dissolved in mixed solvents, such as γ-butyrolactone (GBL)/DMSO [117] and GBL/1-methyl-2pyrolidinone (NMP) [118]. During spin-coating, a non-polar "antisolvent" is dropped. Common antisolvents include toluene and chlorobenzene, which the ionic perovskite precursors are insoluble in. Dropping the antisolvent causes the intermediate phase to nucleate and form. After annealing, a uniform thin film can be obtained. Faster reaction between precursors and following nucleation can significantly improves the uniformity of thin films. The morphology of fabricated thin films is also strongly related to the antisolvent used and how it is dropped onto the spinning substrate.

Sn-based perovskites can also be fabricated by vapor deposition methods [119, 120], which are advantageous for depositing multiple layers of over large areas. For instance, $MASnI_3$ showing complete coverage over the substrate has been grown through co-evaporation of MAI and SnI$_2$ [121]. Vapor deposition can also be combined with solution deposition to make films with higher quality. For example, low-temperature vapor-assisted solution processing (LT-VASP) [122] has been developed to deposit ultrasmooth MASnI$_3$ thin films (Figure 14c). Two steps were involved in this method. Firstly, SnI$_2$ thin films were spin-coated onto mesoporous TiO$_2$ substrates. Secondly, the substrates kept at 60 °C were placed over MAI powder heated at 150 °C and MASnI$_3$ films would then form as MAI was evaporated onto SnI$_2$ films.

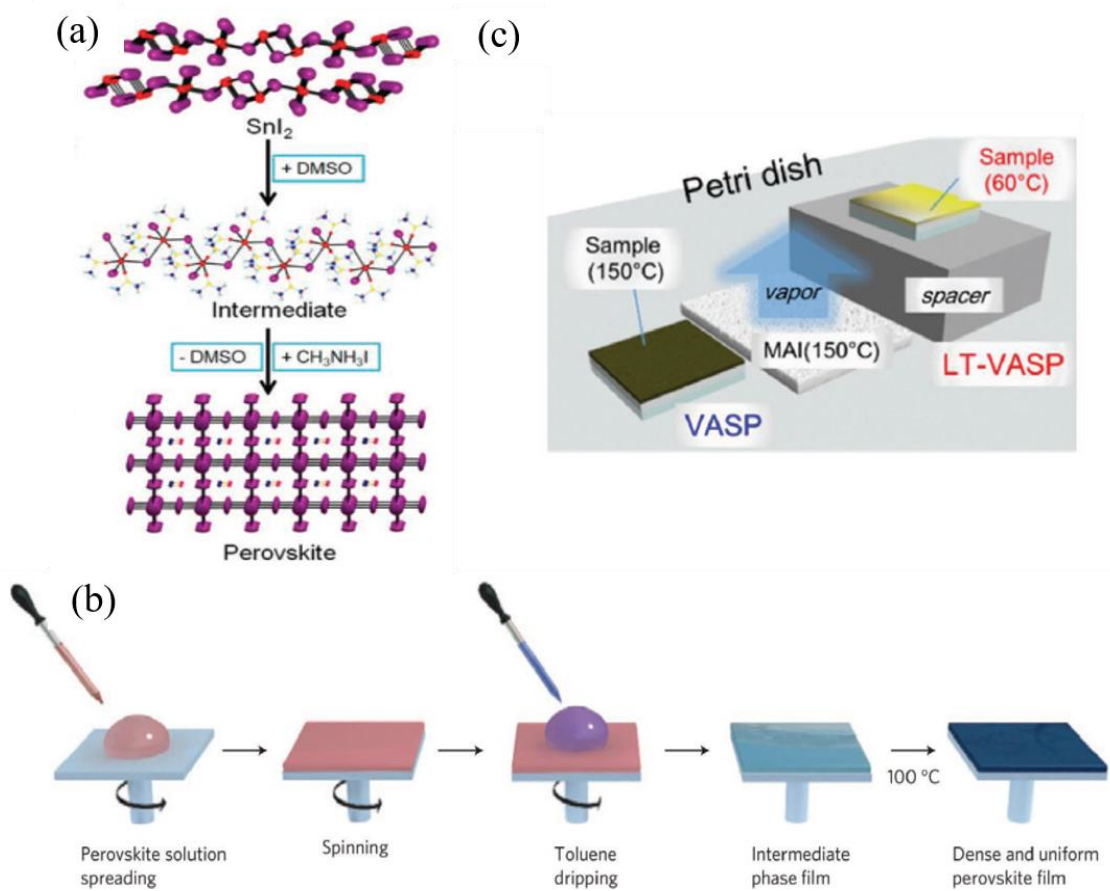

**Figure 14**. The scheme of (a) one-step solution method, (b) solvent engineering method, and (c) LT-VASP method. Reprinted with permission from Ref. [29]. Copyright 2018 Wiley.

In addition, some tin-halide additives such as SnF$_2$ and SnCl$_2$ have been introduced into the above methods to promote the film quality in many aspects. For example, SnF$_2$ added in CsSnI$_3$ can reduce the Sn vacancies and enhance the stability of perovskites without changing the lattice parameters [123, 124]. On the other hand, SnCl$_2$ may form an ultrathin hole transport layer at the ITO/CsSnI$_3$ interface and improve the carrier extraction [125].

Similar to Pb-based perovskites, Sn-based perovskites also have direct band gaps ranging from 1.3-2.1 eV with high absorption coefficients of around 10$_4$ cm$_{-1}$ in visible region [126]. Band gap tunability is crucial for the design of tandem solar cells, where semiconductors with different band gaps are stacked sequentially in order to absorb different regions of the solar spectrum. It has been shown that different X-site halides can greatly affect the band gap of Sn-based perovskites [116, 127]. For instance, the band gap can be tuned from 1.3-3.61 eV from by changing the X anion in MASnX$_3$ from I$_-$ to Cl$_-$. Also, Sn/Pb hybrid perovskites were found to exhibit significant band gap bowing, where the alloy band gaps are smaller than the band gap of both end compounds. Recently, MA(Pb$_{1-x}$Sn$_x$)I$_3$ has demonstrated a band gap below 1.2 eV at $x$ = 0.5 [128]. Goyal *et al.* showed through computations that this extraordinary feature is primary due to the energy mismatch between the *s* and *p* orbitals between Sn and Pb and their non-linear mixing (Figure 15) [129].

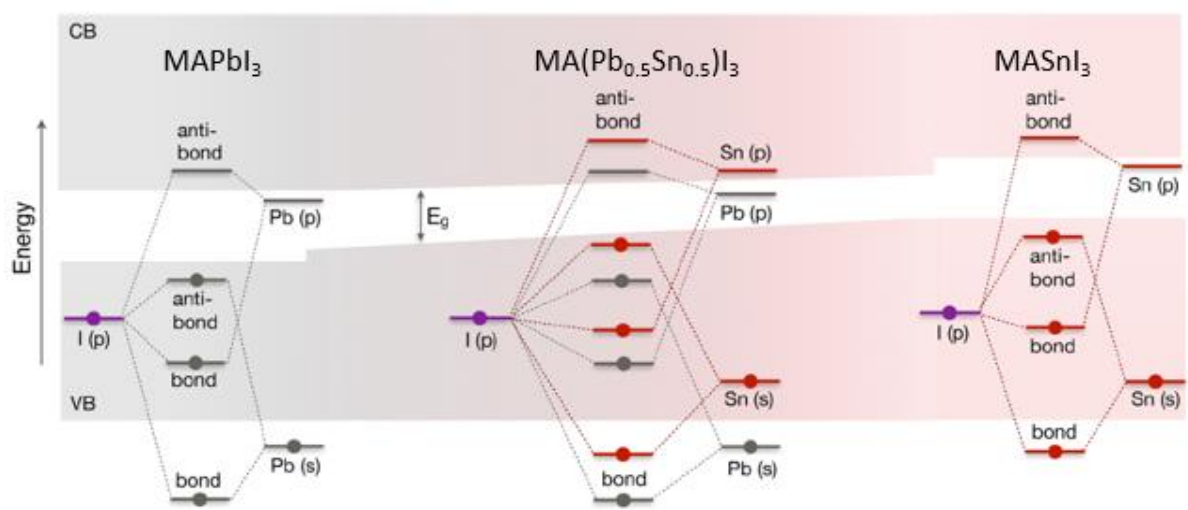

**Figure 15.** Schematic illustration of band gap bowing in MA(Pb$_{1-x}$Sn$_x$)I$_3$ using a molecular orbital diagram to approximate the band structure of the crystal. Reprinted with permission from Ref. [129]. Copyright 2018 American Chemical Society.

Sn-based perovskites tend to become *p*-type due to the self-doping. Although the oxidation of Sn$_{2+}$ to Sn$_{4+}$ should release electrons ($n$-doping), the lower formation energies of Sn vacancies [130] and some processes such as the reaction of Sn$_{2+}$ with O$_2$ may create holes and lead to $p$-doping [114]. However, the mechanism behind $p$-type doping is not fully understood yet. $p$-type self-doping leads to high mobilities and metal-like conductivity of Sn-based perovskites. Relatively small effective masses (0.01-0.11$m_0$, $m_0$: rest mass of an electron) caused by the dispersive VBM further improve their carrier transport. A hole mobility of 585 and 322 cm$_2$ V$_1$ s$_{-1}$ have been reported in CsSnI$_3$ and MASnI$_3$ polycrystalline ingots, respectively [130, 131], based on Hall-effect and resistivity measurement. Interestingly, Stoumpos $et$ $al.$ claimed that the $p$-type character of Sn-based perovskites is related to the formation of Sn$_{4+}$ centers, which could be suppressed by the introduction of a powerful reducing agent ($e.g.$, H$_3$PO$_2$). Eventually, they showed that $n$-type polycrystalline CsSnI$_3$ and MASnI$_3$ could be obtained and also exhibit high electron mobilities of 2320 and 536 cm$_2$ V$_{-1}$ s$_{-1}$ (from Hall-effect measurement), respectively [131].

The carrier lifetime has been reported to be 59 ps and 6.6 ns in FASnI$_3$ thin films with 20 mol% SnF$_2$ additives and CsSnI$_3$ single crystals [132, 133], respectively. These relatively short carrier lifetimes compared to their Pb-based counterparts were mainly related to the electron recombination with large background hole densities ($10_{14}$-$10_{19}$ cm$_{-3}$) [130, 134]. However, the incorporation of ethylenediammonium (en) or SnF$_2$ in some Sn-based perovskites has been found to reduce Sn vacancies and extend carrier lifetimes [135, 136]. For instance, MASnI$_3$ thin films with 20 mol% SnF$_2$ have shown a lifetime longer than 10 ns, which is almost 5-10 times of that of pristine MASnI$_3$ [136]. Moreover, the addiction of ZnI$_2$ could even improve the carrier lifetimes of FAPb$_{0.5}$Sn$_{0.5}$I$_3$ perovskites up to ~ 1 μs [137]. The presence of these sub-gap defects also results in the low PLQEs of Sn-based perovskites. As a result, the PLQEs of mixed Pb-Sn perovskites under solar illumination were typically less than 1% [138]. It is worth mentioning that even with such short carrier lifetimes, Sn-based perovskites can still exhibit long diffusion length over 500 nm [136] mainly due to their high mobilities.

Poor stability is one of the main issues of Sn-based perovskites because $Sn^{2+}$ are prone to oxidation into more stable $Sn^{4+}$ after exposure to ambient air [29]. A significant reduction in PL lifetime was observed in $FAPb_{0.5}Sn_{0.5}I_3$ films after one-minute air exposure [137], indicating Sn-based perovskites should be encapsulated immediately after fabrication in order to maintain their performance. In addition, thermogravimetric analysis (TGA) showed that $MASnI_3$ prepared in air would start to decompose at lower temperature (~150 °C) compared to that prepared in a sealed tube or solution (> 300 °C) [131], which is related to the presence of $Sn^{4+}$ and liberation of volatile $SnI_4$. Nevertheless, both decomposition temperatures are significantly higher than standard testing temperature for solar cells (85 °C), indicating that $MASnI_3$ is thermally stable.

### 3.2 Ge-based perovskites

There have been few experimental reports of Ge-based perovskites focusing on $AGeX_3$ (A = $Cs^+$, $MA^+$, or $FA^+$), in which X = $I^-$ or a mixture of $I^-$ and $Br^-$. The Ge-based perovskites discussed here exhibit trigonal structure (Figure 13), where the distorted $GeX_6$ octahedra are arranged in a corner-sharing pattern with A cations align along the c-axis.

Solution-based methods are used to synthesize most Ge-based perovskites. For example, $MAGeI_{3-x}Br_x$ ($x = 0$-0.6) solution could be prepared by dissolving MAI, MABr, and $GeI_2$ in DMF [139]. On the other hand, various Ge-based perovskite precipitates can be obtained by adding different iodide salts (CsI, MAI, or FAI) into the solution containing HI, $H_3PO_2$ and $GeO_2$, where $H_3PO_2$ and $GeO_2$ acts as the reduction agent and Ge source, respectively [140, 141]. Ge-based perovskite precipitates can be dissolved into DMF to make solutions, and thin films can be fabricated via one-step methods [141] or by using solution engineering methods [139] to improve the film qualities. Unsatisfactory solubility of the precursors in polar solvents is a limiting factor of synthesizing Ge-based perovskites, which may lead to poorer film qualities [141].

$CsGeI_3$, $MAGeI_3$, and $FAGeI_3$ have direct band gaps varying from 1.6-2.35 eV with increasing A cation size [141]. These Ge-based perovskites also have strong absorption in the wavelength range between 400 and 600 nm, along with sharp edges. From thermogravimetric (TGA) measurements, Ge-based perovskites were found to exhibit high thermal stability since they tended to decompose at temperature > 250 °C [141]. However, similar to Sn-based perovskites, Ge-based perovskites also face stability issues in air, which is attributed to the oxidation of $Ge^{2+}$ to the more stable $Ge^{4+}$. The tendency of $Ge^{2+}$ to oxidize is supported by X-ray photoemission spectroscopy (XPS) measurements, which show the presence of $Ge^{4+}$ in the Ge 2$p$ orbital scan [141]. Under ambient conditions, it was shown the absorption of $MAGeI_3$ thin films would drop by over 50% within one day. Kopacic *et al.* showed that the addition of Br into $MAGeI_3$ can improve its ambient stability [139], though the films could only maintain one third of its initial absorption after one day.

*3.3 Double perovskites*

Although In, Sb and Bi are promising low-toxicity substitutes for Pb (see Section 2), they tend to form a +3 oxidation state. Direct substitution of these metal cations into the B-site in perovskites would result in a non-perovskite crystal structure forming. This issue can be solved by using a double perovskite structure with the general formula: $A_2B'B''X_6$, where A and X are monovalent cations and halides, respectively. But two $Pb^{2+}$ cations are replaced by a pair of monovalent B' and trivalent B" cations. This allows charge neutrality to be maintained in a perovskite crystal structure, which is then comprised of alternating $B'X_6^{5-}$ and $B''X_6^{3-}$ octahedra (Figure 13). Examples of B' cations that have been commonly used experimentally include $Ag_+$ and $Na_+$, which are paired with $In_{3+}$, $Sb_{3+}$ or $Bi_{3+}$ B" cations. Although double perovskites have a 3D crystal structure, in some cases, the $[B'X_6]^{5-}$ octahedra do not contribute to the density of states at the band-edges and result in 0D electronic dimensionality [75], which may lead to large indirect band gaps and hinder carrier transport. Due to this reduced electronic dimensionality, effective masses for double perovskites usually display anisotropic behavior, though most calculated values are still small (between 0.1-0.5 $m_0$) [23].

Double perovskites can be synthesized through solution processing methods. For instance, polycrystalline $Cs_2AgBiX_6$ can be obtained by dissolving AgX, $BiX_3$, and CsX in hydrohalic acid HX/hypophosphorous acid mixed solvents [142]. Double perovskite thin films can then be fabricated via spin-coating or thermal evaporation of as-synthesized powders [143]. On the other hand, the synthesis of hybrid halide double perovskites is more difficult compared to that of inorganic double perovskites owing to the low decomposition and vaporization energies of the organic precursors. Therefore, most hybrid double perovskites were synthesized either by solid-state reaction or hydrothermal methods [144, 145], though these two synthesis routes can apply to other inorganic double perovskites as well [142, 146]. Apart from the bulk crystals, double perovskites can be also synthesized as nanocrystals through hot-injection methods. For example, $Cs_2AgBiBr_6$ nanocrystals would form when Cs-oleate is injected into the organic solvent (oleic acid or oleylamine) containing $BiBr_3$ and $AgNO_3$ at 200 °C [147]. The impurities can be reduced by introducing additives (*e.g.*, HBr) as well as controlling the injection temperature. Particularly, $Cs_2AgBiI_6$ nanocrystals could be obtained via anion-exchange process by injecting TMSI to convert $Cs_2AgBiBr_6$ nanocrystals into the target material [148], even though $Cs_2AgBiI_6$ is expected to be thermodynamically unstable compared to $Cs_2AgBiBr_6$. Double perovskite solutions can be prepared by dissolving the synthesized products in solvents, such as dimethyl sulfoxide (DMSO), and films are fabricated by spin-coating these solutions following annealing processes. Annealing at high temperature or under low pressure [149] (low-pressure assisted annealing) can give crystalline films without phase impurities. Also, the utilization of antisolvents, such as IPA, during spin-coating has been shown to improve film quality and reduce the root-mean square roughness [150]. The main challenge for preparing double perovskite samples is the high temperature requirement involved in dissolving some insoluble precursors [151] (*e.g.*, AgCl), solid-state reaction [152], or annealing process [153]. This challenge may limit the application of double perovskite devices on substrates intolerant to high temperature (*e.g.*, ITO glass) or their compatibility with other materials.

Most double perovskites have low absorption coefficients ($10_2 - 10_4$ cm$_{-1}$ in visible region [154]) at the band-edge due to an indirect band gap (*e.g.*, $Cs_2AgBiBr_6$ or $MA_2AgBiBr_6$) or a parity forbidden band gap (*e.g.*, $Cs_2AgInCl_6$ [155]), which limits their SLMEs. For example, $Cs_2AgBiBr_6$, which is the most common double perovskite explored experimentally, has an SLME of 7.9% [46]. Band gaps of double perovskites range from around 2 to 3.4 eV [148], which are too large for PV applications. Band gap engineering is therefore important. Doping or elemental substitution can tune the band gaps. Tran *et al.* demonstrated that the band gaps of $Cs_2Ag(Sb_xIn_{1-x})Cl_6$ can decrease with increased Sb content and experience a transition from direct to indirect character [156], which stems from the dominant *p*-orbitals contributed by $Sb_{3+}$ for the hybridization at band-edges, as illustrated in Figure 16. In addition, Li *et al.* showed that band gap bowing occurs in $Cs_2Ag(Sb_xBi_{1-x})Br_6$ [79]. The alloy double perovskites exhibited a band gap value of 2.08 eV at $x = 0.9$, which is smaller than $Cs_2AgSbBr_6$ (2.18 eV) and $Cs_2AgBiBr_6$ (2.25 eV). The mechanism for band gap bowing is similar to that of $MA(Pb_{1-x}Sn_x)I_3$ (refer to Section 3.1 and Figure 15). That is, the lower band gap exhibited by the alloys of $Cs_2Ag(Sb_xBi_{1-x})Br_6$ is due to the energy mismatch between $s$ and $p$ orbitals from Bi and Sb cations, and their non-linear mixing.

The investigations of lifetimes and carrier mobilities for double perovskites were mainly focused on $Cs_2AgBiBr_6$. Though trap densities on the order of $10_{16}$ cm$_{-3}$ are present in $Cs_2AgBiBr_6$ bulk crystals [157], they can be reduced by thermal annealing in $N_2$ [158]. Annealed $Cs_2AgBiBr_6$ thin films have demonstrated long lifetimes of 1.4 μs [154], which is approaching that of multi-crystalline Si used in commercial solar cells [159]. However, the carrier transport of double perovskites is limited by larger effective masses than in lead-halide perovskites, as well as strong electron-phonon coupling (Refer to Section 3.9). Consequently, $Cs_2AgBiBr_6$ single crystals were found to simply exhibit moderate mobilities of 11.81 cm$_2$ V$_{-1}$ s$_{-1}$ [158].

$Cs_2AgBiBr_6$ can maintain a stable phase in ambient air (with a relative humidity of ~60%) and in the dark for 3 months [158], though significant surface discoloration can be observed after 15 days of light exposure, which possibly stems from the photosensitivity of silver [160]. TGA results also showed that $Cs_2AgBiBr_6$ has higher thermal stability of up to 430 ℃, while $(MA)_2AgBiBr_6$ would decompose at around 277 ℃ [145] due to its organic A-site cations. Particularly, double perovskites such as $Cs_2InBiX_6$, $Cs_2CuInX_6$ or $Cs_2AgInX_6$ that include $Cu_+$, or $In_+$ as B'-site cations have been confirmed theoretically to be relatively unstable owing to the small ionic radii, higher $d$ orbital levels, and redox tendency of these B' cations [161, 162].

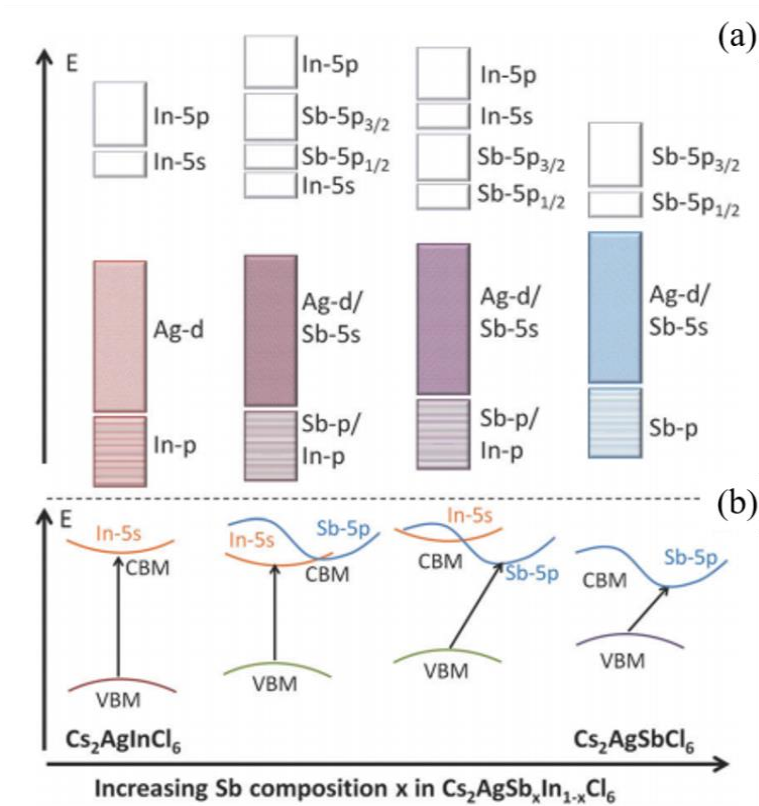

**Figure 16**. (a) Model band diagrams for Cs₂AgSbₓIn₁₋ₓCl₆ and (b) illustration of how the band gap changes due to the contributions from In 5*s* and Sb 5*p* orbitals in the conduction band. Reprinted with permission from Ref. [156]. Copyright 2017 Royal Society of Chemistry.

## 3.4 Chalcogenide perovskites

Chalcogenide perovskites also have the ABX₃ chemical formula, but have the X-site anions replaced by S₂₋. Owing to their stable crystal structure and smaller band gaps (1.5-2 eV), BaZrS₃, SrZrS₃, LaYS₃ are materials investigated most among chalcogenide perovskites. Due to the small electronegativity between B-site cations and sulfur, these materials tend to have strong covalent B-X bonding elements, which is beneficial to narrowed band gaps and carrier transport. The stronger Coulomb attraction between A₂₊ and [BS₆]₈₋ can also enhance the ambient stability of chalcogenide perovskites. For example, it has also been shown that both BaZrS₃ and SrZrS₃ samples stored in air showed no significant degradation in structure and physical properties after one year. It is worth mentioning that the -2 charge of the X-site anion now allows B-site cations that are more stable in +4 oxidation state to be used (*e.g.*, Sn₄₊ and $Ge_{4+}$). As shown in Figure 13, $BaZrS_3$ exists in distorted perovskite phase with corner sharing $BX_6$ octahedra. $SrZrS_3$ may form similar distorted perovskite phase (β phase) or needle-like phase with edge-sharing $BX_6$ octahedra (α phase) at room temperature depending on the synthesis routes. The crystal structure of $LaYS_3$ is different from 3D perovskite since the oxidation states of A- and B-sites are both 3+. Therefore, $LaYS_3$ tend to form a 2D structure with $[Y_3S_9]_{9-}$ layers extended in the [bc] plane and separated by $La_{3+}$ ions.

Phase pure $BaZrS_3$ can be obtained by sulfurization of $BaZrO_3$ with $CS_2$ at 1050 ℃ for 4 hours [163]. Alternatively, heating the stochiometric mixtures of ground and cold-pressed BaS and $ZrS_2$ at 800-1000 ℃ for 15 hours under high vacuum can also make $BaZrS_3$ product, though further purification will be required [164]. Niu *et al.* also made $BaZrS_3$ powders by heating BaS, Zr, and sulphur pieces with stochiometric quantities in a quartz tube at 600 ℃ for 60 hours [165]. $I_2$ was introduced in this method to act as a catalyst to enhances the reactivity of transition metals. Experimentally, $BaZrS_3$ has direct band gaps of 1.75-1.85 eV [26, 164] with large absorption coefficients was on the order of $10_5$ cm-1 from UV to visible range. Ti-alloying has been found to effectively reduce the band gap of $BaZrS_3$ to 1.51 eV [166], resulting a theoretical conversion efficiency up to 32%. $BaZrS_3$ was found to exhibit n-type characteristic with high carrier density between $10_{19}$-$10_{20}$ cm-3, which may arise from excess electrons released when S vacancies form [167]. The Hall mobilities of $BaZrS_3$ thin films were reported to range from 2.1 to 13.7 cm2 V-1 s-1 [167], which have been comparable to those of $MAPbI_3$ [168]. The higher mobilities also correspond to larger grain size and better crystallinity of $BaZrS_3$ thin films sulfurized at higher temperature. In addition, a PL at the wavelength accounting for the band gap and a lifetime of around 400 ns were observed in $BaZrS_3$ thin films [165, 167], indicating relatively lower defect density within this material.

SrZrS$_3$ can be synthesized via mixing SrS, Zr and sulfur at stoichiometric quantities along with iodine as catalyst in a quartz tube [165]. α-SrZrS$_3$ and β-SrZrS$_3$ would be obtained after heating the mixture at 850 and 1100 °C for 60 hours, respectively. A similar synthesis process was conducted by heating Sr, Zr powders and sulfur at 850 and 1000 °C for 3 weeks to prepare α-SrZrS$_3$ and β-SrZrS$_3$, respectively [169]. α-SrZrS$_3$ and β-SrZrS$_3$ individually exhibits a direct band gap of around 1.53 and 2.05-2.13 eV, and both phases have large absorption coefficients on the order of $10^4 - 10^5$ cm$_{-1}$ in the UV to visible range [26, 165]. Especially, polycrystalline β-SrZrS$_3$ showed a high PL intensity comparable to that of single crystalline CdSe and InP and its highest EQE is even higher than that of CIGS solar cells [165]. A high $V_{OC}$ was thus expected in SrZrS$_3$-based solar cells based on these features.

LaYS$_3$ thin films can be fabricated by the two-step process [26]. At the first step, La and Y atoms were deposited on a substrate via co-sputtering in order to form LaY films. The second step is to sulfurize the deposited LaY films at 950-1000 °C with a H$_2$S flow at atmospheric pressure. Crovetto *et al.* showed that the introduction of O$_2$ at the first step could result in LaY films with improved morphology and fewer impurities. LaYS$_3$ thin films were found to display a band gap of 2 eV with absorption coefficients on the order of $10^4 - 10^5$ cm$_{-1}$ in the UV to visible range [26]. LaYS$_3$ films also showed a PL intensity higher than that of high-efficiency CZTS thin films, implying their lower defect densities [26]. However, based on the time-resolved microwave conductivity (TRMC) measurement, LaYS$_3$ films demonstrated a low mobility of only 0.009 cm$_2$ V$_{-1}$ s$_{-1}$ (lower limit) and a relatively long carrier lifetime of 30 ns (upper limit) [170]. These results could be attributed to the presence of highly localized defect states, which has been observed in spatially resolved PL images [170]. Additionally, LaYS$_3$ was found to be an *n*-type material with a low doping density (~$10_{14}$ cm$_{-3}$) and similar band position with MAPbI$_3$ [170].

Complicated processing and the high processing temperatures are two of the main challenges for synthesizing chalcogenide perovskites and integrating them with other transport layers, where contact materials may diffuse into active layers and form unwanted materials after heating [170]. Consequently, although many potential chalcogenide perovskites have been predicted theoretically, most of them are not yet realized experimentally.

### 3.5 $A_3B_2X_9$ materials

Owing to the low toxicity and the similar electron configuration with $Pb_{2+}$, $Bi_{3+}$ and $Sb_{3+}$ might serve as alternatives for the B-site cations in perovskites. To satisfy charge balance, however, an $A_3B_2X_9$ arrangement would be needed. These $A_3B_2X_9$ materials can be considered as defect-ordered perovskites, where only 2/3 of the B-site cations of the perovskites are occupied, thus forming an $A_3B_2\square X_9$ structure. These $A_3B_2X_9$ materials have been found to form two polymorphs. One is the 0D dimer-phase structure consisting of isolated bioctahedral clusters, while the other is 2D layered structure, which forms from general $ABX_3$ arrangement with the removal of every third B layer along the crystal axis (Figure 13). While some materials have a fixed structure, other materials (*e.g.*, $Cs_3Sb_2I_9$) have been found to capable of switching between 0D and 2D structures depending on the processing and composition [171, 172]. Several $A_3B_2X_9$ materials have been investigated as potential absorber layers in future PV devices. These include (i) 0D: $MA_3Bi_2X_9$ (X = Cl, Br, I), $FA_3Bi_2I_9$, $Cs_3Bi_2I_9$, $K_3Bi_2I_9$, $Rb_3Bi_2I_9$, and $(NH_4)_3Bi_2I_9$, (ii) 0D/2D: $Cs_3Sb_2I_9$, (iii) 2D: $MA_3Sb_2I_9$, $Rb_3Sb_2I_9$, and $K_3Sb_2I_9$.

Most $A_3B_2I_9$ materials can be synthesized via solution processing methods, which resemble those used for preparing LHPs. For instance, $MA_3Bi_2I_9$ solutions can be prepared by mixing MAI and $BiI_3$ in DMF/DMSO mixed solvents [173, 174]. $A_3B_2X_9$ films can be fabricated through spin-coating followed by annealing or solvent engineering methods [24, 175] to promote film crystallization. Some modified methods can be applied to further improve film quality. For instance, the PL lifetime of $MA_3Bi_2I_9$ films can be extended by using vapor-assisted conversion, where high-quality $MA_3Bi_2I_9$ films would form by exposing $BiI_3$ films to MAI vapor [176]. Various $A_3B_2X_9$ powders can also be prepared by solid-state synthesis, where $AX$ and $BX_3$ powders can react to form $A_3B_2X_9$ product under ball-milling [177]. To grow $A_3B_2X_9$ single crystals, the Bridgman method is a common synthesis route. In this method, precursors are sealed in an ampoule and placed into in a two-zone furnace, where raw materials would melt in the hot-zone and crystallize in the cold-zone [178]. Other than these general methods, some 0D phase materials can also be synthesized by different methods. $(NH_4)_3Bi_2I_9$ single crystals would precipitate when heating $(NH_4)I$ in a mix of $BiI_3/HI$ for 3 hours. Also, high-quality $Rb_3Bi_2I_9$ and $Cs_3Bi_2I_9$ powders could be obtained by solvothermal reactions [179]. Particularly, for materials that can transform between the 0D and 2D phase, 2D polymorphs are usually more desirable for optoelectronic applications since their stacked planes can exhibit higher mobilities. Therefore, although 0D $Cs_3Sb_2I_9$ can be easily synthesized by simple solution processing with antisolvent engineering, more researchers have worked on achieving 2D $Cs_3Sb_2I_9$ via various routes. Singh *et al.* employed a vapor-assisted solution processing method to deposit 2D $Cs_3Sb_2I_9$ films [180], which were annealed inside an enclosed bottle filled with $SbI_3$ vapor to prevent the decomposition of material, as illustrated in Figure 17. A similar synthesis was taken in 2D $Rb_3Sb_2I_9$ films [181] as well. Umar *et al.* also found that HCl could be used as a coordinated additive to promote the formation of 2D phase $Cs_3Sb_2I_9$ [182]. Recently, another approach found to favor the formation of the 2D phase of $Cs_3Sb_2I_9$ was to alloy with Cl [171]. In addition, by alloying with smaller halides, such as Cl or Br, 0D $Cs_3Bi_2I_9$ was found to be transformed into 2D polymorphs [183, 184] , where a complete anion ordering has been shown in some work [183]. 2D $Cs_3Bi_2I_6Br_3$ has been prepared by heating the DMF/DMSO mixture solvent containing stoichiometric mixtures of CsI, BiI$_3$, and BiBr$_3$ [185]. On the other hand, 2D Cs$_3$Bi$_2$I$_6$Cl$_3$ ingots can be synthesized via a stoichiometric melt reaction of CsCl with BiI$_3$ at 750 °C [184].

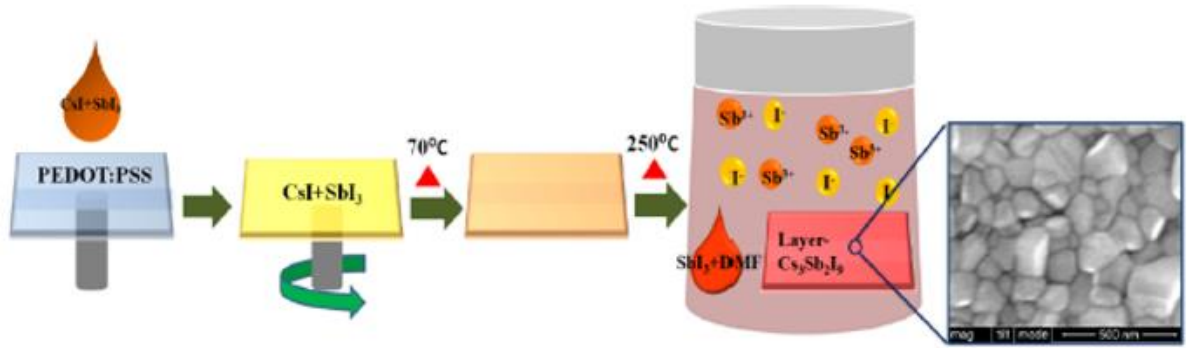

**Figure 17.** The scheme of vapor-assisted solution-processed method. Reprinted with permission from [180]. Copyright 2017 American Chemical Society.

Although A$_3$B$_2$X$_9$ materials generally have wide band gaps larger than 1.9 eV, large absorption coefficients of up to $10^5$ cm$_{-1}$ in the UV-visible region (wavelength < 600 nm) have been presented in some of them (*e.g.*, Cs$_3$Sb$_2$I$_9$ and MA$_3$Bi$_2$I$_9$ [172, 173]). In addition, materials in distinct phases can also demonstrate different band gap characteristic. For instance, it has been shown that 2D Cs$_3$Sb$_2$I$_9$ exhibits a direct band gap of 2.05 eV, while its 0D polymorph has an indirect band gap of 2.4 eV [172]. In addition, 2D Cs$_3$Bi$_2$I$_6$Br$_3$ also shows a reduced band gap of 2.03 eV compared with that of 0D Cs$_3$Bi$_2$I$_9$ (2.2 eV), which is related to a transition in crystal structures [185]. A gradual decrease in band gap from over 2.6 eV to 2.06 eV has been observed in 0D Cs$_3$Bi$_2$Br$_9$ when alloyed with iodide (Cs$_3$Bi$_2$Br$_{9-x}$I$_x$). This in turn transformed the 0D material into 2D Cs$_3$Bi$_2$Br$_3$I$_6$ [183].

Apart from MA$_3$Bi$_2$I$_9$ and (NH$_4$)$_3$Bi$_2$I$_9$ single crystals, which have demonstrated a high mobility of 70 cm$_2$ V$_{-1}$ s$_{-1}$ (hole mobility based on the Hall-effect measurement) and 213 cm$_2$ V$_{-1}$ s$_{-1}$ (sum mobility based on the SCLC method), respectively [186, 187], most A$_3$B$_2$X$_9$ materials suffer from poor carrier transport due to the strong electron-phonon coupling (refer to Section 3.9) and their low dimensional crystal structures, where carriers can be trapped within the isolated regions of materials. Not surprisingly, 0D $A_3B_2X_9$ materials usually have lower carrier mobilities than the 2D ones. For example, based on the space charge limited current (SCLC) method, 0D $Cs_3Bi_2I_9$ single crystals showed a hole mobility on the order of $10^{-2}$ $cm^2$ $V^{-1}$ $s^{-1}$ [188], while 2D $MA_3Sb_2I_9$ single crystals were found to have an electron and hole mobility over 10 $cm^2$ $V^{-1}$ $s^{-1}$ (or over 40 $cm^2$ $V^{-1}$ $s^{-1}$ with Sn doping) [189, 190]. Similar differences can be observed between 0D and 2D $Cs_3Sb_2I_9$. It has been verified by the SCLC method that 2D $Cs_3Sb_2I_9$ thin films have a hole mobility of 6.81 $cm^2$ $V^{-1}$ $s^{-1}$, which is almost double that in 0D polymorphs [182]. In addition, carrier transport in low dimensional materials can experience significant anisotropy along different axes. For instance, estimated mobility-lifetime ($\mu\tau$) product could vary by over two times under the photoconductivity measurement along b- and c-axis of $MA_3Bi_2I_9$ single crystals [186]. Additionally, the low dimensionality also leads to large exciton binding energies since excitons could be trapped as well. The exciton binding energies in most $A_3B_2X_9$ materials have been estimated to vary from 70 to 270 meV [24, 173], which are significantly larger than those reported from Pb-based perovskites (25-50 meV) [191, 192] and could make carrier extraction from these materials relatively challenging.

Several $A_3B_2X_9$ materials have shown promising carrier lifetimes. Evaluating from its $\mu\tau$ product, single crystalline $MA_3Bi_2I_9$ and $(NH_4)_3Bi_2I_9$ may have carrier lifetimes up to 40 and >51.6 $\mu$s, respectively [186, 187]. Also, a PL lifetime of 271 ns was reported in $MA_3Sb_2I_9$ single crystals [189]. The different polymorphs of $Cs_3Sb_2I_9$ were also found to have different carrier lifetimes due to differences in their defect densities. Based on the SCLC model, 2D $Cs_3Sb_2I_9$ was found to have a trap density of $1.1 \times 10^{15}$ $cm^{-3}$, which is almost five times smaller than that of 0D $Cs_3Sb_2I_9$ [182]. Furthermore, 2D $Cs_3Sb_2I_9$ also shows a lower Urbach energy (134 meV) in comparison with 0D $Cs_3Sb_2I_9$ (162 meV), which verifies its less structurally disordered feature [180] (although these values are still significantly higher than that of Pb-based perovskites (~15 meV [193])). However, only a PL lifetime of 6 ns has been reported in 2D $Cs_3Sb_2I_9$ single crystals [180], which is shorter than that of 0D $Cs_3Sb_2I_9$ thin films (54.29 ns) [24]. This result may be related to the presence of deep-level states in 2D $Cs_3Sb_2I_9$ [172].

$A_3B_2I_9$ materials are found to be stable in air. The absorption spectra of $Cs_3Bi_2I_9$ and $MA_3Bi_2I_9$ show negligible changes after one-month storage in dry air in the dark [173]. TGA measurements also shows that $MA_3Bi_2I_9$ and $(NH_4)_3Bi_2I_9$ hardly decomposes until 250 °C and 240 °C [174, 176], respectively, indicating they are thermally stable as well. However, though the PXRD patterns for $Cs_3Bi_2I_9$ in ambient atmosphere could maintain unchanged for a few months, that of $K_3Bi_2I_9$ could not keep well after a few hours, which implies that the larger cation sizes may strengthen the stability of crystal structures [179].

### 3.6 ABZ₂ materials

$ABZ_2$ materials here refer to ternary chalcogenides, where A = Na, Ag, Cu and other cations with stable +1 charge states, and B-site cations and Z-site anions are generally Sb/Bi and S/Se. These materials are expected to be promising in PV application due to their usage of earth-abundant and non-toxic elements. In this section, we will mainly focus on the discussion of $NaBiS_2$, $NaSbS_2$, $AgBiS_2$, $CuSbS_2$, and $CuSbSe_2$, which have been experimentally investigated. Among these materials, $NaBiS_2$, $NaSbS_2$, and $AgBiS_2$ have a rock salt structure with mixed cations octahedrally coordinated by anions (Figure 13). Interestingly, different cation ratios can lead to various superstructures known as "coloring patterns", which can change their crystal structures and band gaps [25]. Instead of a rock salt structure, $CuSbS_2$ and $CuSbSe_2$ exhibits layered orthorhombic structure composed of $SbZ_5$ units (Figure 13).

NaBiS$_2$ can be obtained by melt synthesis [194], in which powders of S, Bi, and Na$_2$S are mixed in a sealed tube, followed by a series of heating steps at high temperature for very long reaction time (~9 days). The hydrothermal method [195] can reduce the reaction time by heating Bi(NO$_3$)$_3$·5H$_2$O, L-cysteine and NaOH in an autoclave at high temperature for around 3 days. Solution-based routes have also been used. For instance, Rosales *et al.* used S, NaH, and triphenylbismuth as precursors, and NaBiS$_2$ nanocrystals (NCs) with different sizes can be synthesized with different solvents and reaction parameters [25]. Particularly, NaBiS$_2$ quantum dots (QDs) can be synthesized by the successive ion layer adsorption and reaction (SILAR) technique [196], where QD sizes can be controlled by different concentrations of the precursor solution. NaBiS$_2$ has been found to have an indirect band gap of about 1.2-1.45 eV, in agreement with computations [25]. Absorption measurements of NaBiS$_2$ QDs under different conditions showed that their band gaps increase with the decrease of QD sizes, implying the quantum confinement effect of this material [196], which occurs when QD sizes become comparable with the excitonic Bohr radius of the material. Also, large extinction coefficients over $10_4$ cm$_{-1}$ M$_{-1}$ in visible region [25] shows the great potential of NaBiS$_2$ NCs on PV application. On the other hand, Zhong *et al.* theoretically proved that NaBiS$_2$ is a ferroelectric material with a large polarization (~33 μC cm$_{-2}$) [197], which may improve the carrier separation by strong induced field.

NaSbS$_2$ thin films can be synthesized through spray pyrolysis [198], where the precursor solution (Na$_2$S and Sb$_2$S$_3$ in water) is sprayed onto a substrate heated at 230 °C to reacted into products. Also, NaSbS$_2$ QDs can be deposited on the mesoporous TiO$_2$ electrodes via Successive Ionic Layer Adsorption and Reaction (SILAR) followed by post-annealing procedure [199]. NaSbS$_2$ has an indirect band gap of around 1.5-1.8 eV along with large absorption coefficients of about $10_4$-$10_5$ cm-1 in the visible range [198, 199]. Rahayu *et al.* showed that the overall band gap of QDs-coated $TiO_2$ will decrease significantly with the increase of SILAR cycles, which is also a result of quantum confinement effect [200]. In addition, Xia *et al.* used UPS and Hall effect measurements to demonstrate that $NaSbS_2$ is a weakly n-type material with an electron mobility between 14 to 22 cm2 V-1 s-1 [198].

$AgBiS_2$ has been synthesized via several methods in the literature. Earlier reports were usually based on SILAR methods [201] and spray pyrolysis [202]. In addition, $AgBiS_2$ has been grown as nanocrystals (NCs) by hot injection method [203] involving the rapid injection of a sulfur source into the oleic acid solution containing bismuth and silver salts at high temperature. However, hot injection needs to be performed under vacuum, or within inert gas, which is inconvenient and costly. Some researchers thus worked on preparing $AgBiS_2$ NCs at room temperature and under ambient atmosphere. Pejova *et al.* showed that $AgBiS_2$ NCs could be easily synthesized through dissolving $AgNO_3$, $Bi(NO_3)_3$, and $Na_2S_2O_3$ in nitric acid, and NC sizes could be reduced under high-intensity ultrasonic irradiation [204]. Recently, Akgul *et al.* used air-stable precursors such as AgI and $BiI_3$ dissolved in amines, and $AgBiS_2$ NCs would form after quickly injecting another sulphur-amine solution [205]. $AgBiS_2$ has been found to have indirect band gaps of about 1.15-2.7 eV, and exhibits high absorption coefficients of around $10_4$-$10_5$ cm-1 from visible to near-infrared range [203]. Similar to $NaBiS_2$, tunable band gaps have been observed in $AgBiS_2$ QDs with different sizes [201]. This tunability is also ascribed to the quantum confinement effect when QD sizes approached the excitonic Bohr radius of $AgBiS_2$ (4.6 nm [206]). $AgBiS_2$ also displays disperse band-edges and hence small effective carrier masses [206], which should favor carrier transport. However, only a small carrier mobility less than 0.1 cm2 V-1 s-1 and a short diffusion length of 150 nm were reported for $AgBiS_2$ NC thin films by the TRMC method [207]. These poor performances may be resulted from the presence of deep-level defects, which has been predicted to arise from the minor off-stoichiometry of AgBiS$_2$ NCs [208]. Nevertheless, an extremely long carrier lifetime up to 1.67 ms was still seen in AgBiS$_2$ QD thin film based on photoconductivity measurement [209], implying that the defect physics of this material is not yet fully understood.

CuSbS$_2$ thin films can be fabricated by many techniques including thermal evaporation [210], sputtering [211], chemical bath deposition (CBD) [212], electrodeposition [213], ALD [214], solvothermal synthesis [215] and so on. On the other hand, solvothermal [216] and hot injection methods [217] can be also applied on the synthesis of CuSbS$_2$ nanoparticles. This material has been shown to have large absorption coefficients over 10$^4$ cm$^{-1}$ in the visible region as well as direct band gaps ranging from around 1.4-1.9 eV, depending on various deposition processes [218–220]. DFT calculation also predicted that CuSbS$_2$ tends to exhibit $p$-type characteristic owing to the presence of dominant Cu vacancies [221] acting as electron acceptors. It also exhibits small carrier effective masses of around 0.3 $m_0$ [221], leading to a high hole mobility of 64.6 cm$_2$ V$^{-1}$ s$^{-1}$ [222]. Yang $et$ $al.$'s work also demonstrated that CuSbS$_2$ may be defect tolerant owing to higher formation enthalpies for deep-level defects and the shallow transition level of Cu vacancies [221]. However, CuSbS$_2$ thin films were found to have a carrier lifetime of only 0.7 ns even after the improvement on film quality via thermal treatment [223, 224].

CuSbSe$_2$ thin films can be deposited through evaporation [225], co-sputtering of Cu$_2$Se and Sb$_2$Se$_3$ [226], sulfurization of Cu-Sb alloy [220], and electrodeposition [227]. Particularly, Yang $et$ $al.$ also fabricated CuSbSe$_2$ by the hydrazine method, which involved the preparation of precursors such as Cu$_2$S, Sb$_2$Se$_3$ in hydrazine with specific stoichiometry [228]. In addition, it has been shown that CuSbSe$_2$ single crystals could also be synthesized by the hot-injection technique [229]. One main challenge within these methods is that several competitive phases may occur concomitantly, making pure phase $CuSbSe_2$ can be obtained only under strictly controlled environment [228, 230]. $CuSbSe_2$ has direct band gaps of 1-1.2 eV [228], which is close to the optimal band gap for single-junction solar cells. Moreover, Tiwari *et al.* showed that $CuSbSe_2$ also exhibits large absorption coefficients up to $6.6 \times 10^6$ cm-1 in the visible region [231], indicating its potential as a solar absorber. Similar to $CuSbS_2$, a *p*-type characteristic caused by Cu vacancies is also observed in $CuSbSe_2$ with a high hole concentration over $10_{17}$ cm-3 being reported [226]. However, the larger effective hole mass (0.9 $m_0$ [232]) of $CuSbSe_2$ may lead to a lower Hall mobility of 20.17 cm2 V-1 s-1 [233] compared to $CuSbS_2$. Though Xue *et al.* predicted that $CuSbSe_2$ would have benign defect properties [233], Welch *et al.* indicated that Se vacancies could serve as deep-level defect states that may influence the carrier lifetime, which has been confirmed by a short carrier lifetime of 190 ps measured via transient THz spectroscopy technique [230].

In terms of stability, it has been claimed that $AgBiS_2$ and $NaBiS_2$ are stable in air for at least several weeks [25, 203]. No obvious degradation issues have been reported in the literature for other $ABZ_2$ materials, but Peccerillo *et al.* believed that Cu-related instability resulted from ions migration under continuous DC operation may occur in Cu-based $ABZ_2$ materials [234].

*3.7 Binary halides*

Binary halides are compounds made from heavy metals and halides, which contain cations with valence $s_2$ electrons, making them promising for potentially exhibiting defect tolerance (refer to Section 2). InI, $BiI_3$, and $SbI_3$ have been investigated most as PV materials among all binary halides, and we will focus our discussion on them. InI has an orthorhombic crystal structure with layers stacked along the *b*-axis [235] (Figure 13), and every two atom sheets are staggered in this structure, making In atoms bonded more closely with adjacent In atoms. On the other hand, BiI$_3$ and SbI$_3$ exhibit layered structures [30, 236] with each layer composed of edge-sharing metal-halide octahedra bonded together by van der Waals interactions.

InI films can be deposited by thermal evaporation [110], or fabricated from the melt [21], where the In metal-I pellets mixture was placed into a quartz ampoule and heated at 400 ℃ for 12 hours to ensure the complete reaction into InI. Alternatively, Shah *et al.* used the Bridgman method to grow InI single crystals by placing purified InI into a vertical furnace kept at 450 ℃ but with a sharp temperature drop at the bottom, where melted InI can crystallize as molten InI moved down [237]. InI has a direct band gap of around 2 eV, which has been verified in calculation and experiment [110]. It has been reported that InI can exhibit reasonably long PL lifetime over 6 ns [21] and large mobility-lifetime products corresponding to a diffusion length of around 16 μm [110]. Defect calculations also suggest that the main vacancies and anti-site defects do not have transition levels resonant within the band gap, or have high formation energy transition levels that involve a large gradient change (which would have a low capture cross-section) [238]. However, In is not stable in the +1 oxidation state and tends to oxidize to the +3 oxidation state. InI films have been found to decompose under ambient conditions within 3 hours [110].

BiI$_3$ thin films have been synthesized by physical vapor transport (PVT) [239], physical vapor deposition (PVD) [240], and solution processing [241]. On the other hand, BiI$_3$ single crystals can be obtained by the vertical Bridgman method [242] similar to that was described in InI section. It is worth noting that BiI$_3$ tends toward dissociation at 250-300 ℃ [243, 244], which limits its synthesis to some extent. BiI$_3$ has indirect band gaps of around 1.7-1.8 eV with large absorption coefficients (>10$_5$ cm$_{-1}$ above 2 eV) [239, 242]. BiI$_3$ single crystals have demonstrated the electron mobility as high as 260 cm$_2$ V$_{-1}$ s$_{-1}$ (or 1000 cm$_2$ V$_{-1}$ s$_{-1}$ with Sb doping) [245, 246], though the hole mobility was estimated to be much lower due to relatively large effective hole mass (2.01 $m_0$). Furthermore, a long electron diffusion length of 4.9 μm estimated from the large μτ product as well as a PL lifetime of 1.3-1.5 ns was obtained from $BiI_3$ single crystals [239], indicating the potential of $BiI_3$ as a great PV material. $BiI_3$ films are also stable in air. For instance, Hamdeh *et al.* have demonstrated that they could stand against oxidation for several months or withstand several hours of annealing without degradation [241]. In addition, Du *et al.*'s work showed that the large Born effective charge of $BiI_3$ could result in its large dielectric constant of around 54 [247], which can Coulombically screen charged defects.

$SbI_3$ single crystalline plates could be grown from vapor phase [248] directly or by Bridgman method [249]. Particularly, $SbI_3$ microcrystals could be obtained by the incorporation with Y-zeolite faujasites via sublimation [249]. In addition, Mohan *et al.* also fabricated $SbI_3$ films by iodizing evaporated Sb films [250]. Kepinska *et al.* showed that $SbI_3$ has a temperature dependent indirect band gap [248], which is around 2.1 eV at room temperature. Virko *et al.* proved that this band gap would be blue-shifted when the deposited $SbI_3$ clusters became smaller, which implies the quantum confinement effect of this material [249]. $BiI_3$ also demonstrates rather large absorption coefficients in visible region on the order of $10_5$-$10_6$ cm$_{-1}$ [248]. However, $SbI_3$ single crystals were found to have an Urbach energy over 200 meV at room temperature [248], indicating that several defect states may exist within the band gap.

*3.8 V-VI-VII materials*

V-VI-VII materials are materials made from elements of V, VI, and VII group, which may be Bi/Sb, O/S/Se, and halides, respectively. Similar to binary halides, V-VI-VII materials also have antibonding orbital character at the band-edges along with strong spin-orbit coupling, which are therefore promising for potentially exhibiting defect tolerance. At present, BiOI, BiSI, and SbSI are the compounds studied the most. Both BiOI and BiSI have a layered structure (matlockite tetragonal system [251, 252]), where I-Bi-O/S-Bi-I layers are stacked along the c-axis and linked by weak van der Waals interactions (Figure 13). On the other hand, SbSI has a needle-like crystal structure with long chains growing along the (001)-axis [253]. Strong anisotropy can be seen in V-VI-VII materials due to their low dimensionality and may be reflected in many features. For example, the effective hole mass can differ by over 3 times as along different axes of BiOI [252], which also leads to the anisotropy in its carrier transport.

BiOI polycrystalline thin films can be fabricated by chemical vapor deposition (CVD) [254, 255] or the SILAR technique [256]. Owing to the 2D structure, BiOI thin films exhibit a textured morphology. For example, CVD BiOI grown on solution-processed $NiO_x$ have a {012} preferred orientation. It was recently shown that this could be controlled through the vaporization temperature of the $BiI_3$ precursor and the growth temperature, which influence whether growth is nucleation or kinetically-dominated [257]. The preferred orientation of BiOI films can then be changed from a/b-axis to c-axis, which results in a denser morphology.

BiOI has an indirect band gap of 1.8-1.9 eV with absorption coefficients larger than $10^4 cm^{-1}$ from UV to visible range [254, 258]. Computed defect diagrams showed that the defects with the lowest formation energy have shallow transition levels [254], which was consistent with experiments. It has been showed that the electronic structure, PL intensity and charge-carrier lifetime are robust against percent-level defects [259]. Furthermore, a high dielectric constant of 45 was found through computations [254], and this can help to screen charged defects. The PL lifetime measured was 2.7 ns [254]. Although this is sufficient for further exploration in solar absorbers [21], it is shorter than the lifetime of lead-halide perovskites. Recently, photoinduced current transient spectroscopy measurements showed that BiOI films have deep traps located at 0.3 and 0.6 eV from the band-edge [259], but their origin have not yet been identified. BiOI has improved air stability over lead-halide perovskites, with no change in phase over the entire 197 day testing period [254].

BiSI polycrystalline thin films can be synthesized by various methods such as single-precursor solution processing [260], spray pyrolysis [261], asynchronous ultrasonic spray pyrolysis (APUSP) [262], hydrothermal [263] and solvothermal methods [264]. APUSP here used two ultrasonic humidifiers to deliver two nebulized solution onto a hot substrate subsequently to improve the film morphology. On the other hand, BiSI single crystals can be synthesized by a gel process [265] or the sublimation of BiSI polycrystals [266]. One key challenge of depositing BiSI films is the competition from other Bi-S-I phases [267], which is verified from the isothermal section of this ternary system. BiSI has a nearly direct band gap of around 1.6 eV with absorption coefficients of around $10^4$ cm$^{-1}$ in UV and visible range [261]. Though the effective electron mass is small for BiSI [260], Sasaki *et al.* have shown that this material may exhibit an electron drift mobility of only 1 cm$^2$ V$^{-1}$ s$^{-1}$ along the c-axis [266]. Moreover, Hahn *et al.* also estimated a mean hole diffusion length of only $5 \times 10^{-6}$ cm [261], which arises from its larger hole effective mass (0.95 $m_0$ [266]). In addition, a short PL lifetime of about 1.03 ns was also obtained from BiSI thin films [260], implying the presence of deep-level sulfur vacancies. BiSI is also a semiconductor exhibiting ferroelectricity, piezoelectric, photoconductivity and hence is an appealing material in various fields [268].

SbSI films can be prepared by hydrothermal methods [269] or simple solution processing methods [270]. In addition, SbSI single crystals can be synthesized via the vertical Bridgeman technique [271] or CVT [272]. Due to the anisotropy of SbSI, it is still difficult to fabricate large-area and uniform single crystalline samples. SbSI has indirect band gaps of around 2 eV [270, 273], which may be blue-shifted when the temperature decreases [271]. The ferroelectricity of SbSI also leads to the large static dielectric constant of the order of $10_4$ at 292 K [274], which is beneficial for screening charged defects. In addition to the ferroelectricity, SbSI also has high photoconductivity and piezoelectricity [275] and is as appealing as BiSI.

## 3.9 Strong electron-phonon interaction and self-trapped excitons

Although Sn-based perovskites have demonstrated high mobilities exceeding 500 cm$_2$ V$_{-1}$ s$_{-1}$ in single crystals [168], the mobilities of other PIMs are more modest. This is partly due to less dispersive bands being present in many cases, particularly compounds with layered, 1D or 0D structures. But another important factor is electron-phonon coupling. Charge carriers could couple with phonons (the quanta of lattice vibrations) by polarizing the intrinsic ions inside lattices, inducing a localized structural distortion. Meanwhile, the movement of charge carriers would be constrained by the distortions, and the term polarons could be used to describe the charge carrier coupled to the distorted crystal vibrations. It has been verified that strong electron-phonon coupling occurs in $A_3B_2X_9$ materials [178], double perovskites [276, 277], and binary halides [278]. Owing to these materials being polar, Fröhlich interactions are expected to occur, in which the charge-carriers couple to longitudinal optical (LO) phonons [279]. Optical phonons involve out-of-phase lattice movements, which results in an electric field due to electric polarization that interacts with the charge-carrier [280]. The strength of coupling is described by a dimensionless coupling constant, α, which is given by Eq. 5 [281]:

$$\alpha = \frac{q^2}{\hbar c} \sqrt{\frac{m_b c^2}{2\hbar \omega_{LO}}} \left( \frac{1}{\varepsilon_\infty} - \frac{1}{\varepsilon_0} \right) \tag{5}$$

In Eq. 5, $q$ is the charge of an electron, $\hbar$ the reduced Planck constant, $c$ the speed of light, $m_b$ the effective mass of the charge-carrier in the absence of lattice interactions, $\omega_{LO}$ the rotational frequency of the longitudinal optical phonon mode, and $\varepsilon_\infty$ and $\varepsilon_0$ the optical and static dielectric constants, respectively [281]. A coupling constant larger than 2 is considered to be strong, and larger coupling constants result in the effective mass of the polaron being larger, thereby significantly limiting the mobilities. In $Cs_2AgBiBr_6$, the Fröhlich coupling constant has been calculated to be 2.54 for electrons and 2 for holes [276], showing electron-phonon coupling to be strong. This can also be quantified through the Huang-Rhys factor, $S$, which indicates the average number of vibrations emitted after an optical transition between excited and ground states [276]. The Huang-Rhys factor for $Cs_2AgBiBr_6$ has been measured to be high, at 11.7 – 15.4 [276, 282]. As a result, the mobilities of $Cs_2AgBiBr_6$ single crystals has been measured to only reach 11.81 $cm_2$ $V_{-1}$ $s_{-1}$, well below lead- and tin-based perovskites [154, 168]. Beyond reducing the mobility, Fröhlich interactions also result in a temperature-dependent widening in the photoluminescence peak [279].

Excitons can also interact with phonons. This coupling is similar to electron-phonon coupling discussed above, but an important difference is that excitons are quasi-particles with no charge. Such interactions can result in the exciton localizing in the lattice in the absence of defects due to distortions in the lattice [283] (*e.g.*, Jahn-Teller distortions [277]). This is referred to as intrinsic self-trapping (Figure 18a). These self-trapped excitons (STEs) can be considered to be defect states, but unlike lattice defects, STEs are only created after excitation. Moreover, if the exciton occurs in the vicinity of a permanent defect, it will form an extrinsic STE that is trapped at different energies compared to those of intrinsic STEs. Schematic images of three trapping conditions are displayed in Figure 18a. On top of the restriction to carrier movement, multiple excited states relative to the ground state will be also generated as coupling occurs, which is illustrated in Figure 18b. These excited states will exhibit a large equilibrium position offset relative to the free exciton (FC) state, and this offset is proportional to the Huang-Rhys parameter. As a result, a broadband white light emission with large Stokes shift is usually observed in the materials [277, 284].

Beyond $Cs_2AgBiBr_6$, strong coupling has also been identified in $A_3B_2X_9$ materials. For example, $Cs_3Sb_2I_9$ was found to have a Huang-Rhys factor as high as 42.7. It is proposed that this strong electron-phonon coupling resulted in the formation of STEs and in turn gave the broad PL peaks observed [178]. Luo *et al.* fabricated an alloyed $Cs_2(Ag_{0.6}Na_{0.4})InCl_6$ double perovskite [277] and observed that its PL showed a linear dependence on the excitation power with the PLQEs almost unchanged. This result verified the existence of STEs since permanent defects will encounter the saturation effect, where none of carrier transitions occurs after all defect states are filled at high power excitations. They also demonstrated that though Na ions could break the inversion symmetry of $Cs_2AgInCl_6$ and improve the PLQE to some extent, too many Na ions might force the excited states to cross ground state, leading to non-radiative recombination through phonon emission. Other important features regarding white light emission have been investigated in low-dimensional perovskites. For instance, Cortecchia *et al.* indicated that the white light emission intensity will increase at lower temperature due to the thermal energy is not comparable to the de-trapping energy ($E_{a,detrap}$ in Figure 18b), but it will decrease when the temperature is down to below ~80 K since the thermal energy is even less than self-trapping energy ($E_{a,trap}$ in Figure 18b). These materials featuring white light emission with large Stokes shift will not suffer PL self-absorption and thermal quenching, which are suitable for the white light LED applications.

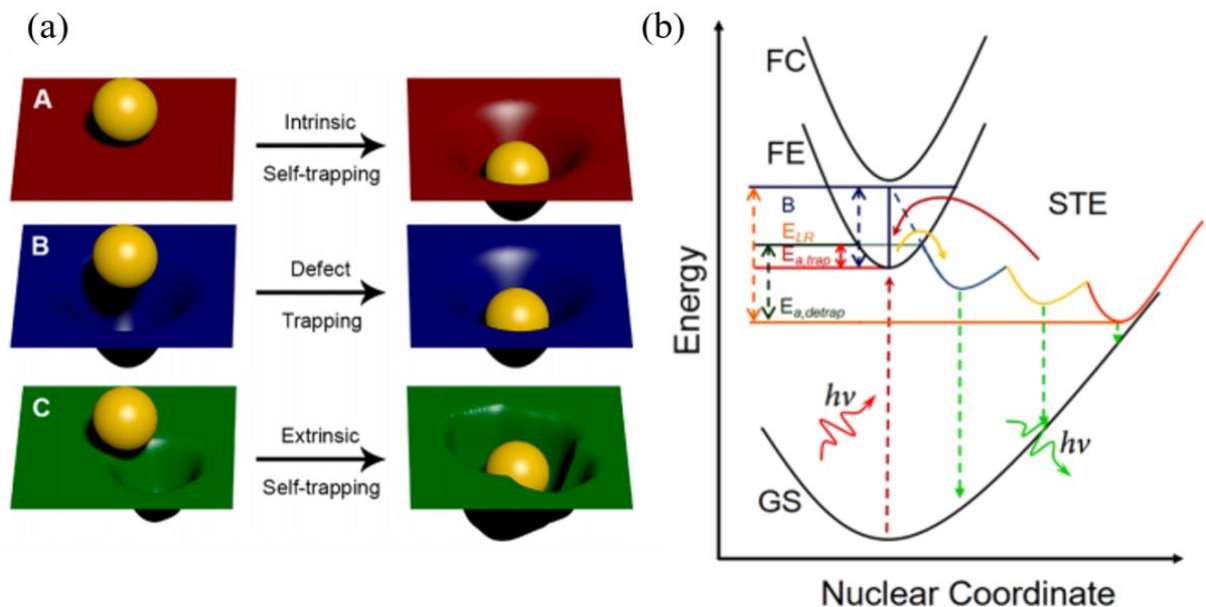

**Figure 18.** (a) Schematic illustrations of (A) intrinsic self-trapping, which is created as an exciton induces a lattice distortion, (B) defect trapping, which is created as an exciton is trapped within a permanent defect and (C) extrinsic self-trapping, which is created as an exciton induces a lattice distortion near the edge of a permanent defect. Reprinted with permission from [283]. Copyright 2018 American Chemical Society. (b) Band diagrams for STE states, where FC, FE, and GS refers to the state of free carrier, free exciton, and ground state, respectively. Reprinted with permission from Ref. [284]. Copyright 2019 American Chemical Society.

**Table 2.** Important properties of PIMs discussed in this review. RT = room temperature; $m_e^*/m_h^*$ = the smallest effective electron/hole masses reported in the corresponding references; $\mu/\mu_e/\mu_h$ = the highest sum/electron/hole mobilities reported in the corresponding references; SC = single crystals; PC = polycrystalline ingots; TF = thin films; QD = quantum dots; NC = nanocrystals

| Material | | Space group @ RT | $m_{e^*}/m_{h^*}$ ($m_0$) | Band gaps (eV) | Mobility (cm$_2$ V$_{-1}$ s$_{-1}$) | Carrier lifetime (ns) | Highest PCE (%) | Reference |
|---|---|---|---|---|---|---|---|---|
| Sn-based perovskites | MASnI$_3$ | Pm3m | 0.1/0.11 | 1.23-1.4 | $\mu_h \sim 322$ (PC) $\mu_e \sim 2320$ (PC) | 10.2 (TF) | 6.63 | [22, 131, 136, 285–288] |
| | FASnI$_3$ | Amm2 | 0.02/0.05 | 1.35-1.45 | $\mu_h = 67$ (TF) $\mu_e = 103$ (PC) | 0.59 (TF) | 12.4 | [131, 132, 288–291] |
| | CsSnI$_3$ | Pnma | 0.041/0.069 | 1.3 | $\mu_h = 585$ (SC) $\mu_e = 356$ (PC) | 6.6 (SC) | 3.56 | [125, 130, 131, 133, 277, 292] |
| Ge-based perovskites | MAGeI$_3$ | R3m | 0.27/0.29 | 1.9-2 | - | - | 0.68 | [139, 293] |
| | CsGeI$_3$ | R3m | 0.22/0.23 | 1.6 | - | - | 0.11 | [141, 293] |
| | FAGeI$_3$ | R3m | 0.66/0.76 | 2.35 | - | - | - | [293, 294] |
| Double perovskites | Cs$_2$AgBiCl$_6$ | Fm$\overline{3}$m | 0.53/0.15 | 2.5-2.77 | - | - | - | [142, 150, 295] |
| | Cs$_2$AgBiBr$_6$ | Fm$\overline{3}$m | 0.37/0.14 | 1.95-2.5 | $\mu \sim 1$ (TF) $\mu \sim 11.81$ (SC) | 1400 (TF) | 2.84 | [142, 150, 154, 157, 158, 296, 297] |
| | Cs$_2$NaBiI$_6$ | Fm$\overline{3}$m | 0.28/0.58 | 1.66 | - | - | 0.42 | [146, 298] |
| | Cs$_2$AgInCl$_6$ | Fm$\overline{3}$m | 0.29/0.28 | 3.3-3.53 | - | - | - | [150, 299] |
| | (MA)$_2$AgBiI$_6$ | Fm$\overline{3}$m | 0.47/0.7 | 1.02-1.96 | - | - | - | [150, 300] |
| | (MA)$_2$AgBiBr$_6$ | R3m | 0.15/0.24 | 1.25-2.02 | - | - | - | [145, 301] |
| A$_3$B$_2$X$_9$ materials | MA$_3$Bi$_2$I$_9$ | C 2/c | 0.54/0.95 | 1.94-2.1 | $\mu_h = 70$ (SC) | ~40000 (SC) | 3.17 | [173, 176, 177, 186, 302, 303] |
| | FA$_3$Bi$_2$I$_9$ | P6$_3$mc | - | 1.85-1.94 | $\mu = 0.224$ (TF) | - | - | [177, 304] |
| | Cs$_3$Bi$_2$I$_9$ | C 2/c | 1/1.1 | 1.9-2.2 | $\mu_h = 1.7 \times 10^{-2}$ (SC-TF) $\mu_h = 4.4 \times 10^{-7}$ (PC-TF) | - | 3.2 | [173, 175, 176, 179, 188, 303] |
| | (NH$_4$)$_3$Bi$_2$I$_9$ | P 2$_1$/n | - | 2.04-2.05 | $\mu = 213$ (SC) | ~51600 (SC) | - | [174, 187] |
| | Cs$_3$Sb$_2$I$_9$ | P6$_3$/mmc (0D) P$\bar{3}$m1 (2D) | 1.4/1.55 (0D) 0.34/0.42 (2D) | 2.3-2.5 (0D) 2.02-2.05 (2D) | $\mu_h = 6.81$ (2D-TF) $\mu_h = 3.54$ (0D-TF) | 54.29 (0D-TF) 6 (2D-TF) | 1.49 | [24, 180, 182, 305] |
| | MA$_3$Sb$_2$I$_9$ | P6$_3$/mmc | - | 1.92 | $\mu_e \sim \mu_h = 16.68$ (SC) | 271 (SC) | - | [189, 190] |
| | Rb$_3$Sb$_2$I$_9$ | P 2$_1$/n | 1.11/1.55 | 2-2.24 | $\mu = 0.26$ (TF) | 8.97 (TF) | 1.37 | [24, 181, 306-309] |
| | K$_3$Sb$_2$I$_9$ | P$\bar{3}$m1 | 3.15/1.83 | 2.03 | - | 29.98 (TF) | - | [24] |
| ABZ$_2$ materials | NaBiS$_2$ | Fm$\bar{3}$m | 0.36/0.25 | 1.2-1.45 | - | - | 0.05 | [25, 196, 310] |
| | NaSbS$_2$ | C 2/c | 0.34/0.35 | 1.5-1.8 | $\mu_e = 14 - 22$ (TF) | - | 3.18 | [198, 200, 311] |
| | AgBiS$_2$ | Fm$\bar{3}$m | 0.35/0.722 | 1.15-2.7 | $\mu_e = 0.07$ (NC) $\mu_h = 0.032$ (NC) | 1670 (QD) | 6.4 | [201, 207, 209, 312-314] |
| | CuSbS$_2$ | Pnma | ~0.3/0.3 | 1.4-1.9 | $\mu_h$ = 64.6 (TF) | 0.7 (TF) | 3.22 | [218–220, 222, 223, 315] |
| | CuSbSe$_2$ | Pmna | 0.71/0.9 | 1-1.2 | $\mu_h$ = 20.17 (TF) | 0.19 (TF) | 4.7 | [226, 228, 230, 232, 233, 316, 317] |
| Binary halides | InI | Cmcm | 0.215/0.118 | 2 | - | 6 (SC) | 0.39 | [21, 110, 235, 318] |
| | BiI$_3$ | R$\overline{3}$ | 0.68/2.01 | 1.67-1.8 | $\mu_e$ = 260 (SC) | 1.3-1.5 (SC) | 1.21 | [236, 239, 245, 318, 319] |
| | SbI$_3$ | R$\overline{3}$ | 0.62/1.54 | 2.1 | - | - | - | [248, 318, 320] |
| V-VI-VII materials | BiOI | P4/nmm | 0.19/0.2 | 1.8-1.93 | - | 2.7 (SC) | 1.8 | [251, 252, 254] |
| | BiSI | Pnam | 0.53/0.95 | 1.57 | $\mu_e$~1 (SC) | 1.03 (TF) | 1.32 | [260, 266, 321] |
| | SbSI | (approx.) P2$_1$P2$_1$P2$_1$ | 0.65/0.34 | 1.8-2.72 | - | - | 3.62 | [322–324] |
| Chalcogenide perovskites | BaZrS$_3$ | Pnma | 0.43/0.75 | 1.75-1.85 | $\mu_e$ = 2.1 − 13.7 (TF) | 400 (TF) | - | [26, 167, 325] |
| | SrZrS$_3$ | Pnma | 3.11/0.64 | 1.52-1.53 (α-phase) 2.05-2.13 (β-phase) | - | - | - | [26, 165, 169, 325] |
| | LaYS$_3$ | P2$_1$/m | 0.49/0.67 | 2 | $\mu$~0.009 (TF) | ~30 (TF) | - | [26, 170, 325] |

## 4. Engineering of Photovoltaic Devices – Strategies and Challenges

Two common device structures are used: *n-i-p* and *p-i-n*. In the *n-i-p* structure, the active layer is grown over an electron transport layer (ETL) covering the transparent conducting oxide. A hole transport layer (HTL) and a high-work function top electrode is deposited over the active layer (*vice versa* for *p-i-n*). These device structures can be divided into planar and mesoporous structures. The planar structure only needs a compact ETL, but the mesoporous structure uses a mesoporous scaffold (typically comprised of $TiO_2$ nanoparticles) deposited on a thin compact ETL ($TiO_2$ thin film). The active layer will then incorporate into the mesoporous scaffold, which improves carrier transfer. These device structures are illustrated in Figure 19.

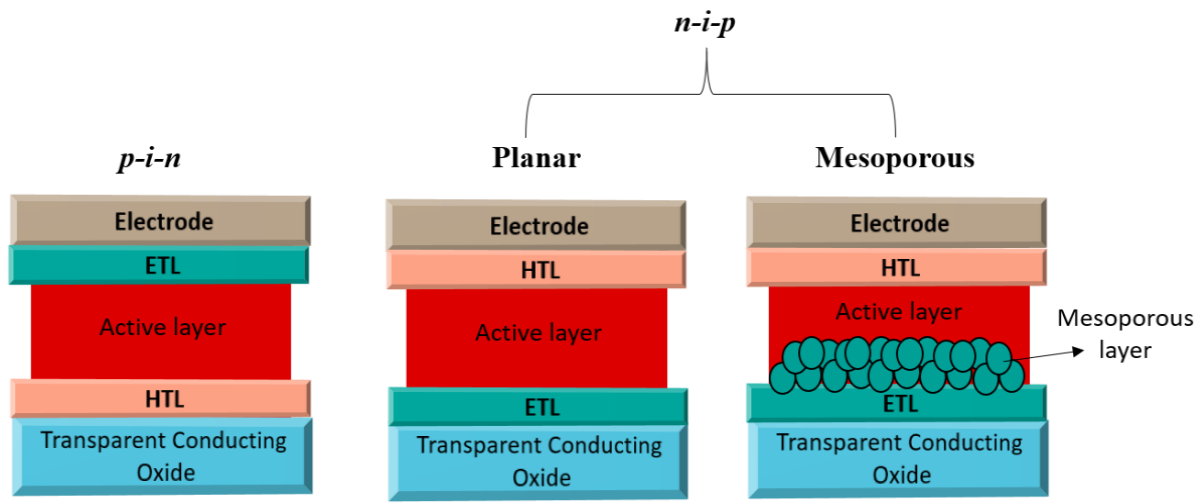

**Figure 19.** Illustration of the most common device structures used for perovskite-inspired materials.

The stability of PV devices is a critical consideration for commercial applications. However, publications usually lack consistency in terms of testing procedures and parameters reported. To resolve this issue, protocols have been developed for stability testing. For commercial silicon solar cells, the protocol IEC (International Electrochemical Commission) 61215-1-3:2016 is commonly used (refer to the Introduction) [326]. At the lab level, protocols developed by the International Summit on Organic Photovoltaic Stability (ISOS) are commonly used [327]. These were initially developed for organic solar cells, but were more recently adopted for perovskite solar cells as well. The ISOS protocols are categorized into 5 stress conditions (dark storage testing, outdoor testing, light soaking testing, thermal cycling testing, light-humidity-thermal cycling testing). Additional testing procedures specific to PSCs, such as light-dark cycling and testing under continuous bias in the dark, have been added to these protocols. It is suggested that the stability of PIM-based solar cells can be evaluated according to similar protocols as well to resolve the large discrepancies in testing procedures that are present in the literature.

In this section, we will firstly introduce the structures used in PV devices of each class of PIMs along with the corresponding device performances and improvements. Next, the device stability will be discussed, followed by a brief review on current challenges or limiting factors.

## 4.1 Sn-based perovskites

PV devices with Sn-based perovskites have been fabricated in the *n-i-p* (planar and mesoporous) and *p-i-n* structures. For the planar structure, although the diffusion length of Sn-based perovskites can be very long, it is still challenging to make devices based on this structure owing to the requirement of high film quality (compact film morphology and low pinhole density). Trimethylamine (TMA) was found to be an additive that can improve the film quality of $FASnI_3$, and a PCE (power conversion efficiency) of 4.34% with a $V_{OC}$ (open-circuit voltage) of 0.31 V, and a $J_{SC}$ (short-circuit current density) of 21.65 mA cm$^{-2}$ in planar $FASnI_3$ solar cells was achieved [328].

On the other hand, the most common architectures used in Sn-based perovskite PV devices is the mesoporous structure. The so-called hollow solar cells based on mesoporous structure were found to demonstrate improved performance over planar devices [286, 329]. In these devices, ethylenediammonium (en) was incorporated to replace parts of A-site cations in Sn-based perovskites, which would create several $SnI_2$ vacancies and change the electronic structures of perovskites. As a result, hollow solar cells could demonstrate better film morphology and lower defect densities (estimated from the trap-filled limited voltage from SCLC measurement) compared to neat solar cells, leading to an improved PCE and $V_{OC}$. Consequently, hollow {en}FASnI$_3$, {en}MASnI$_3$, and {en}CsSnI$_3$ solar cells have shown the highest PCE of 7.23%, 6.63%, and 3.79%, respectively among mesoporous PV devices [286, 329].

Several *p-i-n* Sn-based perovskite solar cells have demonstrated high PCEs. Shao *et al.* fabricated *p-i-n* FASnI$_3$ solar cells with poly(3,4-ethylenedioxythiophene):polystyrene sulfonate (PEDOT:PSS) as the HTL and $C_{60}$ + 2,9-dimethyl-4,7-diphenyl-1,10-phenanthroline (BCP) as the ETL, achieving a PCE of 9% with a $V_{OC}$ of 0.53 V, and a $J_{SC}$ of 24.1 mA cm$_{-2}$ [330]. Jiang *et al.* also proposed *p-i-n* PEA$_x$FA$_{1-x}$SnI$_{3-x}$ (PEA = $C_6H_5CH_2CH_2NH_{3+}$) solar cells with indene-$C_{60}$ bisadduct (ICBA) and PEDOT:PSS as the ETL and HTL, respectively [290]. They showed that the use of ICBA, which exhibits a lowest unoccupied molecular orbital (LUMO) closer to the vacuum level compared to PCBM, could effectively increase the $V_{OC}$. Introducing NH$_4$SCN as an additive to the perovskite can help to reduce defect densities as well. As a result, their champion device showed a record PCE of 12.4% and $V_{OC}$ of 0.94 V along with a $J_{SC}$ of 17.4 mA cm$_{-2}$. It should be noted that simpler devices without hole transport layers have also been reported. For instance, HTL-free *p-i-n* CsSnI$_3$ solar cells (with SnCl$_2$ added to the perovskite to improve film quality) demonstrated a PCE of 3.56%, along with a high fill factor of ~70% [125].

The efficiency for Sn-based perovskite solar cells is limited by nonradiative recombination

(with carrier lifetimes < 1 ns for most thin film samples) and unmatched band alignment with common transport layers [29]. Other than poor film morphology during processing, the high recombination rate of Sn-based perovskites is mainly due to the creation of Sn vacancies from self-doping (refer to section 3.1). Recently, it has been shown that the incorporation of additives such as enI$_2$ and PEAI [135, 330] or using high-impurity precursors might help to decrease the density of these background recombination centers. On the other hand, Sn-based perovskites have larger band mismatch to common ETLs and HTLs compared to MAPbI$_3$, as shown in Figure 20. This mismatch may impede the carrier extraction and cause the recombination loss at the interface of the transport layers and the perovskite layer, which will reduce $V_{OC}$ significantly. Searching for transport layers with suitable band-edges is therefore important as fabricating Sn-based perovskite solar cells.

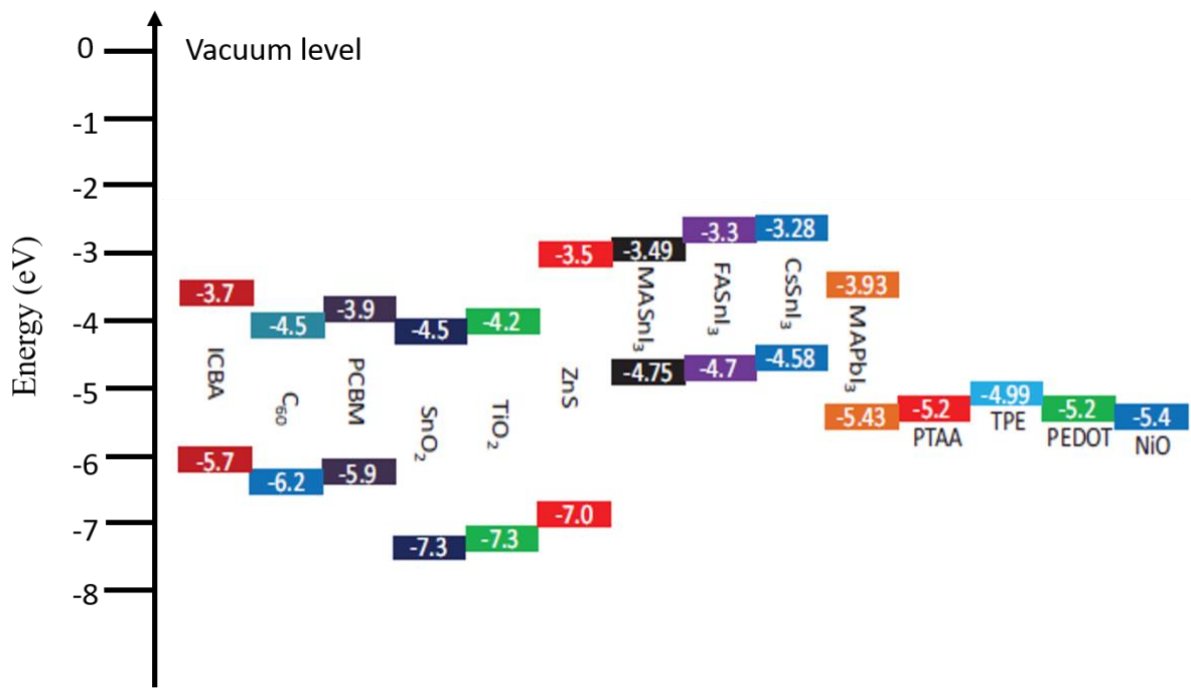

**Figure 20**. Energy diagram of Sn- and Pb-based perovskites along with some common ETLs and HTLs. Unit is in eV. The energy levels are estimated from Tauc plots and photoemission spectroscopy in air. Reprinted with permission from Ref. [29]. Copyright 2018 Wiley.

Sn-based perovskite solar cells are unstable in air owing to the oxidation of Sn$_{2+}$, as discussed in Section 3.1. Typically, neat Sn-based perovskite solar cells may completely lose their initial

PCEs in air within a few hours [135, 331, 332]. In addition to processing under an oxygen-free or hydrazine-based atmosphere, many works have been attempted a wide range of strategies to improve the stability of Sn-based perovskites. For instance, the hollow structures incorporated with en [135] or using TMA as additives have been shown to enhance the air stability of $FASnI_3$ [328]. Recently, solar cells composed of 2D Sn-based perovskites were found to exhibit better ambient stability than their 3D counterparts. For example, it has been shown that the PCE of 2D layered $(BA)_2(MA)_3Sn_4I_{13}$ (BA: butylammonium) solar cells degraded much slower than that of 3D $MASnI_3$ in air [332]. However, no Sn-based perovskite solar cells can retain their ambient performances over one day unless proper encapsulation is employed. By encapsulating devices with glass and epoxy or glue under inner gas atmosphere, the encapsulated cells could extend their device lifetimes up to a few months [333].

## 4.2 Ge-based perovskite

Ge-based perovskite PV devices have been fabricated in the mesoporous and *p-i-n* structures. Krishnamoorphy *et al.* fabricated various mesoporous Ge-based perovskite solar cells by using 2,2′,7,7′-tetrakis(N,N-di-p methoxyphenylamine)-9,9′-spirobifluorene (spiro-OMeTAD) as the HTL [141]. However, the PCEs of mesoporous $CsGeI_3$ and $MAGeI_3$ solar cells are limited by their low $V_{OC}$ in the mV range, which is far below the band gaps.

Champion Ge-based perovskite solar cells were based on the *p-i-n* structure with PEDOT:PSS as HTL and $PC_{70}BM$ as ETL. Kopacic *et al.* showed that the efficiency and stability of *p-i-n* structured $MAGeI_3$ solar cells could be improved with the substitution of a few percent of bromine [139]. Consequently, the optimal device with $MAGeI_{2.7}Br_{0.3}$ demonstrated a record PCE of 0.68% along with an improved $V_{OC}$ of 460 mV.

There are currently few reports of Ge-based perovskite PV devices mainly due to their limited stability, as discussed in section 3.2. Even MAGeI$_{2.7}$Br$_{0.3}$ solar cells with improved stability still lose two-thirds of their initial performance in air within a few hours after preparation [139]. It has also been claimed that the formation of 2D perovskites with the addition of bulky cations such as PEA$_+$ [331] could be a promising solution to address this issue.

Ge-based perovskite PV devices also suffer from low $V_{OC}$s, which limits PCEs, and which are arise in part due to the high density of defect states during the oxidation of Ge$_{2+}$. Improved synthesis methods would be needed to address this issue, but the poor solubility of some precursors in most polar organic solvents is also a challenge that needs to be overcome. Krishnamoorthy *et al.* proposed that preparation methods without the addition of hypophosphorous acid as well as the strict control of synthesis environment and precursors may help improve the quality of fabricated Ge-based perovskite films [141].

*4.3 Double perovskite*

Most double perovskite solar cells investigated have used Cs$_2$AgBiBr$_6$, while most groups reported only ~1% PCE in PV devices based on this material. However, the first report of this material in photovoltaics by Greul *et al.* led to higher PCEs of 2.43% and external quantum efficiencies (EQEs) reaching 60% [153]. This was achieved through careful optimization of the film quality. Recently, Yang *et al.* introduced an N179 dye (di-tetrabutylammonium cis-bis(isothiocyanato) bis(2,2'-bipyridyl-4,4' dicarboxylato) ruthenium (II)) interlayer between the active layer and spiro-OMeTAD HTL, which could improve hole transfer and enhance light absorption. As a result, the optimized mesoporous Cs$_2$AgBiBr$_6$ solar cell exhibited a record PCE of 2.84% with a $V_{OC}$ of 1.06 V and a $J_{SC}$ of 5.13 mA cm$_{-2}$ [296]. On the other hand, Gao *et al.* showed that *p-i-n* Cs$_2$AgBiBr$_6$ solar cells could also deliver a PCE of 2.23% with a $V_{OC}$ of 1.01 V and a $J_{SC}$ of 3.19 mA cm$^{-2}$ after using IPA as the anti-solvent during film synthesis [150], which resulted in smooth and uniform films with micron-sized grains. In addition, $Cs_2AgBiBr_6$ solar cells have demonstrated improved stability in ambient air over Sn-, Ge- and Pb-based perovskites. The stability of $Cs_2AgBiBr_6$ solar cells have been tested by storing cells in the dark under ambient condition (temperature: 20-30 °C, relative humidity: 40-60% [149]), which is actually consistent with the dark storage study in ISOS protocols (ISOS-D-1). Eventually, the optimal $Cs_2AgBiBr_6$ solar cells showed no PCE degradation for at least 30 days.

As mentioned in Section 3.3, the performance of double perovskite devices is still significantly limited by their intrinsic optoelectronic properties (large band gaps and poor carrier mobilities) and synthesis challenges (low solubility precursors).

### 4.4 $A_3B_2X_9$ materials

$MA_3Bi_2I_9$ and $Cs_3Bi_2I_9$ are the two $A_3B_2X_9$ materials that are most commonly investigated in reports of defect-ordered perovskite PVs. $MA_3Bi_2I_9$ devices are generally based on the mesoporous structures with spiro-OMeTAD as the HTL. Most PV devices, however, have demonstrated very low PCEs, which are believed to be related to the poor film morphology of $MA_3Bi_2I_9$, in addition to the high exciton binding energy and large effective masses. Many investigators have worked on improving the film morphology. Zhang *et al.* fabricated $MA_3Bi_2I_9$ films via a two-step approach with the $BiI_3$ films evaporated under high vacuum and transformed into $MA_3Bi_2I_9$ films under a low vacuum MAI atmosphere [334]. Their $MA_3Bi_2I_9$ solar cells consist of the compact, pinhole-free, and large-grained films, demonstrating a PCE of 1.64% with a $V_{oc}$ of 0.81 V and a $J_{SC}$ of 2.95 mA cm$^{-2}$. Jain *et al.* further optimized the $MA_3Bi_2I_9$ film quality via the vapor assisted solution process by exposing solution-processed $BiI_3$ films to MAI vapors, and replaced spiro-OMeTAD with poly(3-hexylthiophene-2,5-diyl) (P3HT) as the HTL [302]. A PCE of 3.17% with a $V_{OC}$ of 1.01 V and a $J_{SC}$ of 4.02 mA cm$^{-2}$ was achieved in their champion solar cells.

Similarly, most $Cs_3Bi_2I_9$ PV devices were fabricated based on the mesoporous structure as well. The PCEs of these devices were also low due to their relatively low photocurrents. Ghosh *et al.* showed that the photocurrents of $Cs_3Bi_2I_9$ solar cells could be enhanced when the precursor solution was prepared with excess $BiI_3$, which may passivate the deep-level defects of $Cs_3Bi_2I_9$ [335]. They investigated the influence of different structures (planar and *p-i-n*), along with the mesoporous solar cells with different HTLs, while none of the PCEs of these devices exceeded 1%. A PCE of 0.21% with a $V_{OC}$ of 0.49 V and a $J_{SC}$ of 0.67 mA cm$^{-2}$ was achieved in their best device. A breakthrough was made by Bai *et al.* [175]. They reported a dissolution-recrystallization method to fabricate high-quality $Cs_3Bi_2I_9$ films. In this method, the mixed DMF/$CH_3OH$ solvent, which can thoroughly dissolve $Cs_3Bi_2I_9$ and evaporate, will be dropped onto the annealed as-deposited $Cs_3Bi_2I_9$ films, and the ultrathin $Cs_3Bi_2I_9$ nanoplates will recrystallize after the second annealing. Eventually, the devices with CuI as the HTL achieved a record PCE of 3.2% with a significantly improved $J_{SC}$ of 5.78 mA cm$^{-2}$, which is ascribed to the smaller grains and voids within fabricated films.

Sb-based $A_3B_2X_9$ material solar cells also drew the attention of many researchers due to their higher dimensionality (2D) compared to Bi-based counterparts (0D), which could lead to improved mobilities. In particular, $Cs_3Sb_2I_9$ can be 0D or 2D depending on the synthesis method (refer to section 3.5). $Cs_3Sb_2I_9$ solar cells have been fabricated based on *n-i-p* or *p-i-n* structures. Umar *et al.* deposited high-quality 2D $Cs_3Sb_2I_9$ films by using IPA as an antisolvent and introducing HCl additive into the precursor solutions, which can decrease the reaction time and suppress the formation of Sb-I-Sb clusters [182]. HTL-free planar $Cs_3Sb_2I_9$ solar cells with only $TiO_2$ as the ETL were realized with a PCE of 1.21%, $V_{OC}$ of 0.61 V and a $J_{SC}$ of 3.55 mA cm$^{-2}$. By contrast, 0D $Cs_3Sb_2I_9$ devices with a similar structure only obtained a PCE of 0.43%. Sign *et al.* fabricated *p-i-n* 2D $Cs_3Sb_2I_9$ solar cells with PEDOT:PSS as the HTL and [6,6]-Phenyl-C71-butyric acid methyl ester ($PC_{70}BM$) as the ETL [180], achieving a PCE of 1.49% with a $V_{OC}$ of 0.72 V and a $J_{SC}$ of 5.31 mA cm$^{-2}$. On the other hand, *p-i-n* 0D $Cs_3Sb_2I_9$ solar cells exhibited a larger $V_{OC}$ (0.77 V) while much smaller $J_{SC}$ (2.96 mA cm$^{-2}$) and PCE (0.89%), which is mainly ascribed to the larger band gaps of 0D $Cs_3Sb_2I_9$. Alternatively, Peng *et al.* demonstrated mesoporous 0D $Cs_3Sb_2I_9$ solar cells with poly(N,N'-bis-4-butylphenyl-N,N'-isphenyl)benzidine (poly-TPD) as the HTL [171] could only achieve a low PCE of 0.24%. However, the incorporation of Cl was found to convert 0D $Cs_3Sb_2I_9$ into 2D $Cs_3Sb_2Cl_xI_{9-x}$, and mesoporous 2D $Cs_3Sb_2Cl_xI_{9-x}$ solar cells (with LZ-HTL-1-1 as the HTL) showed a significantly improved PCE of 2.15% with a $V_{OC}$ of 0.6 V and a $J_{SC}$ of 6.46 mA cm$^{-2}$. Additionally, some promising results have been obtained with $Rb_3Sb_2I_9$ solar cells. Weber *et al.* deposited single crystallite $Rb_3Sb_2I_9$ thin films by using the antisolvent vapor diffusion crystallization method [309], where single crystals could grow as the antisolvent was evaporated and condensed onto the coated precursor solutions. A PCE of 1.37% and peak EQE of 26% was achieved in their champion mesoporous $Rb_3Sb_2I_9$ with spiro-OMeTAD as the HTL. They also found that the incorporation of Br led to a preferred orientation of $Rb_3Sb_2Br_{9-x}I_x$ films parallel to the substrate, which would impede carrier transport. Recently, Li *et al.* employed high temperature annealing under $SbI_3$ atmosphere to improve the morphology of $Rb_3Sb_2I_9$ films and increase the grain size [181]. Based on the planar structure with poly(N,N'-bis-4-butylphenyl-N,N'-bisphenyl)benzidine) (poly-TPD) as the HTL, optimized $Rb_3Sb_2I_9$ solar cells from Li *et al.*'s study demonstrated a comparable PCE (1.35%) but a much higher peak EQE (65.4%), which was attributed to the improved mobilities and reduced defects in the devices.

As discussed in Section 3.5, most PV devices based on $A_3B_2X_9$ materials are stable in air. For instance, the stability of $Rb_3Sb_2I_9$ solar cells have been tested by storing cells in the dark under inert gas atmosphere, which is consistent with the suggested protocol ISOS-D-1I for PSCs [327]. As a result, the PCE of $Rb_3Sb_2I_9$ solar cells could retain 84% of their initial values after 150 days [309]. However, planar $Cs_3Bi_2I_9$ solar cells with CuI and spiro-OMeTAD as the HTL were found to maintain only 57% and 28% of their initial PCEs, respectively after storing in a constant temperature-humid chamber (temperature: 25 °C, relative humidity: 45%) for 38 days [175], which implies that the device stability may be influenced by the interactions between transport layers and active layers.

The factors limiting the performance of $A_3B_2X_9$ material-based devices are the wide band gaps, high exciton binding energies, poor film morphology, and the presence of deep defects. As discussed in section 3.5, wide band gaps and high exciton binding energies can lead to weak absorption and poor carrier extraction, respectively. Though some works have attempted to tune the band gaps by mixing different halides with $A_3B_2X_9$ materials at different proportions [173], there are not yet sufficiently promising results reported. In addition, high exciton binding energies are partly due to the lower dimensionality of $A_3B_2X_9$ materials, and synthesis routes which can convert 0D structure into 2D counterpart for some materials (*e.g.*, $Cs_3Sb_2I_9$) may help to address this issue. Finally, many researchers have managed to resolve the latter two issues via various processing techniques. Shin *et al.* also proposed a general method to fabricate compact thin films of most Bi-based $A_3B_2X_9$ precursors [336], which involves the use of solvent complexes to increase the solubility of Bi-based precursors and a subsequently rapid nucleation process via an antisolvent dripping. It should be noted that film orientation should be taken into account as well when developing synthesis routes since strong anisotropy can be seen in $A_3B_2X_9$ materials.

*4.5 ABZ2 materials*

ABZ$_2$-based PV devices can be fabricated as *n-i-p* or *p-i-n* structures. Nanocrystals of ABZ$_2$ could also be mounted onto a mesoporous scaffold, giving a sensitized structure. At present, devices based on CuSbS$_2$, CuSbSe$_2$, and AgBiS$_2$ are investigated most. Although some early progress has been made in NaBiS$_2$ and NaSbS$_2$ quantum dot sensitized solar cells [196, 200], investigations into both materials are still rare.

CuSbS$_2$ PV devices were largely fabricated based on the *p-i-n* and mesoporous structures. *p-i-n* CuSbS$_2$ solar cells typically used Mo as the HTL and ZnO as the ETL with a CdS buffer layer between CuSbS$_2$ and ZnO to form a type-II staggered heterojunction and promote carrier separation [221]. Septina *et al.* prepared purified CuSbS$_2$ films with the precursors preheated before sulfurization, fabricating *p-i-n* CuSbS$_2$ solar cells achieving a PEC of 3.1% with a $V_{OC}$ of 0.49 V and a $J_{SC}$ of 14.73 mA cm$_{-2}$ [218]. Banu *et al.* demonstrated *p-i-n* CuSbS$_2$ solar cells prepared from hybrid inks [219]. The cells they made showed a record PCE of 3.22% with a $V_{OC}$ of 0.47 V and a $J_{SC}$ of 15.64 mA cm$_{-2}$. Mesoporous CuSbS$_2$ solar cells were found to exhibit comparable PV performance compared to that of *p-i-n* solar cells. Chio *et al.* fabricated mesoporous CuSbS$_2$ solar cells with poly(2,6-(4,4-bis-(2-ethylhexyl)-4H-cyclopenta[2,1-b;3,4-b']dithiophene)-alt-4,7(2,1,3-benzothiadiazole)) (PCPDTBT) as the HTL, achieving a PCE of 3.1% with a $V_{OC}$ of 0.34 V and a $J_{SC}$ of 21.5 mA cm$_{-2}$ [337].

On the other hand, almost all the CuSbSe$_2$ solar cells were built in the *p-i-n* structure by using Mo or Mo/MoO$_x$ and ZnO (with CdS buffer layers) as the HTL and ETL, respectively. Although there are fewer reports of CuSbSe$_2$ PV devices than CuSbS$_2$ devices, CuSbSe$_2$ PV devices have achieved higher performance. Yang *et al.* optimized the morphology and crystalline orientation of hydrazine solution processed $CuSbSe_2$ film by tuning the annealing temperature [228]. However, $CuSb(Se,S)_2$ usually occur due to the reaction with competitive phases. As a result, the optimal $CuSb(Se_{0.96}S_{0.04})_2$ solar cells only showed a PCE of 2.7%. Welch *et al.* controlled the evaporation flux ratio of $Sb_2Se_3$ and $Cu_2Se$ to form pure phase $CuSbSe_2$ films, but found a trade-off between $J_{SC}$ and $V_{OC}$ exists in $CuSbSe_2$ solar cells [230]. It has been shown that as the $Sb_2Se_3/Cu_2Se$ ratio increased, the $J_{SC}$ would be enhanced due to the extended depletion width, while the $V_{OC}$ would be suppressed owing to the reduced hole concentration (and hence the reduced quasi Fermi level splitting). Consequently, the highest PCE of 4.7% along with a $V_{OC}$ of 0.336 V and a $J_{SC}$ of 26 mA cm$_{-2}$ were achieved in their devices.

In addition to competitive impurity phases and the $J_{SC}$-$V_{OC}$ trade-off for $CuSbSe_2$ solar cells, lower $J_{SC}$ arising is a common challenge for $CuSbS_2$ and $CuSbSe_2$ solar cells. This could be attributed to the strong absorption losses of CdS buffer layer [234], which leads to a significant drop in the short wavelength range below 520 nm. Moreover, CdS layers are known to have some detrimental effects, such the introduction of deep hole defects [338]. Nevertheless, none of PV devices with CdS being replaced by other materials have shown promising performance to date.

Most $AgBiS_2$ solar cells were fabricated as the planar structure with ZnO as the ETL and PTB7 (or P3HT)/$MoO_3$ as the HTL. Bernechea *et al.* displayed $AgBiS_2$ nanocrystal solar cells achieving a PCE of 6.3% with a $V_{OC}$ of 0.45 V and a $J_{SC}$ of 22 mA cm$_{-2}$ [203]. The cells in this work also demonstrated high stability in air with the PCE almost unchanged under ambient condition for up to four months. Recently, Burgues-Ceballos *et al.* have shown that the $AgBiS_2$ nanocrystal sizes can be enlarged by using the double-step hot-injection technique [314], making thin films with higher carrier mobilities and lower defect densities. As a result, their champion AgBiS$_2$ nanocrystal solar cell exhibited a PCE of 6.4%, a $V_{OC}$ of 0.46 V along with an improved $J_{SC}$ of 22.68 mA cm$^{-2}$, which can be ascribed to better carrier extraction in this device.

Incomplete current collection is one of the main challenges for AgBiS$_2$ solar cells, which could be confirmed by the nonlinear dependence of $J_{SC}$ of on light intensity [203]. Improvements in the ligands used or the surface passivation of the nanocrystals would be needed to suppress the trap recombination during carrier transport.

*4.6 Binary halides*

Apart from BiI$_3$, binary halides have not yet been investigated much into PV devices. Most BiI$_3$-based PV devices were fabricated based on the planar structure. Hamdeh *et al.* used TiO$_2$ and V$_2$O$_5$ as the ETL and HTL [241], respectively, and found that the devices processed in air showed better performance than those in a glovebox, which is attributed to the facilitated hole transport due to the formation of BiOI layers. Moreover, solvent vapor annealing could also improve film quality and boost the PCE of PV devices. A PCE of 1.02% with a $V_{OC}$ of 364 mV and a $J_{SC}$ of 7 mA cm$^{-2}$ were displayed in their work. Tiwari *et al.* showed that using HTLs with higher ionization potential such as poly(9,9-di-noctylfluorenyl-2,7-diyl) (F8) can exhibit better match to the VBM of BiI$_3$, raising the $V_{OC}$ of PV devices [319]. As a result, their champion planar BiI$_3$ solar cell demonstrated a slightly improved PCE of 1.21% along with a record $V_{OC}$ of 0.607 V. Additionally, hole extraction also plays an important role in determining PCEs of BiI$_3$-based PV devices, which could be seen from the substantial influence of the HTL on device performance. Searching for HTLs with matched energy levels is thus crucial for BiI$_3$-based PV devices. Furthermore, the softness of the material makes it challenging to handle without scratching the film and causing pinholes to occur [339]. Finally, BiI$_3$ devices have exhibited improved stability over Sn-, Ge- and Pb-based perovskites. It has been verified that mesoporous BiI$_3$ solar cells stored in ambient air with a relative humidity of 50%, which is also consistent with the test requirement of the protocol ISOS-D-1, can retain above 90% of their initial PCEs for over 30 days [340]. In addition, these cells could also maintain ~70% of their initial PCEs and showed mineral change in their J-V curves after the heat stress (temperature: 100 °C in ambient air, similar to the environment requirement of the protocol ISOS-D-2 with a higher temperature) for 2 hours and light-soaking test (by a solar simulator, consistent with the environment requirement of the protocol ISOS-L-1) for 20 minutes, respectively [340].

Recently, planar InI solar cells with CdS and P3HT as the ETL and HTL, respectively, were reported by Mitzi's group [110]. SnI$_2$ was used as an interface modification layer between the CdS and InI layer to control the preferred orientation of InI films. Eventually, a PCE of 0.39% with a $V_{OC}$ of 450 mV and a $J_{SC}$ of 2.14 mA cm$_{-2}$ were demonstrated in the champion cell. Although Brandt *et al.* have showed that InI predominantly forms shallow defects [21], Dunlap-Stohl *et al.* indicated that the weak PL emission from InI-based PV devices suggests the presence of non-radiative recombination centers [110]. In addition, significant hysteresis was also observed in InI-based PV devices [110]. Improvements in the performance of InI devices will require further understanding of the composition and role of defects in this material, as well as control over the oxidation of In$_+$ to In$_{3+}$.

*4.7 V-VI-VII materials*

V-VI-VII materials have been applied as photocatalysts (refer to Section 5.2), but have gained increasing attention for solar absorber applications more recently. Owing to their original application in photocatalysts, early BiOI solar cells were fabricated as liquid-state sensitized architectures with iodide redox couple as electrolytes. However, most devices showed very low efficiencies. Zhang *et al.* deposited BiOI flakes onto mesoporous $TiO_2$ as the working electrodes, but their devices only reached 0.38% efficiently [341]. Sfaelou *et al.* replaced mesoporous $TiO_2$ with a compact layer of $TiO_2$ and fabricated BiOI solar cells with a PCE over 1% and a $V_{OC}$ of 0.61 V [256]. Recently, Hoye *et al.* demonstrated solid-state BiOI solar cells with a *p-i-n* structured comprising of $NiO_x$ and ZnO as the HTL and ETL, respectively [254]. A record PCE of 1.8% with a $V_{OC}$ of 0.75 V and a $J_{SC}$ of 7 mA cm$^{-2}$ was achieved in the champion device. They showed that the band bending caused by the energy level mismatch (Figure 21) between BiOI and the HTL could introduce unwanted energy barriers to hole collection. From this, they claimed that future progress on performance could be made by finding HTLs with deeper work functions [254]. Furthermore, BiOI has a layered structure with anisotropic carrier transport. The CVD BiOI grown on solution-processed $NiO_x$ by Hoye *et al.* had an {012} preferred orientation, which allowed the top and bottom electrodes to be connected by the high-mobility planes. This enabled effective charge extraction, with a peak EQE up to 80% at 450 nm wavelength [254]. At the same time, the films had an open structure, which limited the shunt resistance. It was later shown the preferred orientation of the BiOI platelets grown by CVD could be controlled through the temperature of the $BiI_3$ precursor and the temperature of the substrate [257], as discussed in Section 3.8. Growing c-axis oriented BiOI platelets resulted in compact film morphology, which improved the $V_{OC}$ from 0.7 V (a/b-axis oriented) to 0.9 V (*c*-axis oriented).

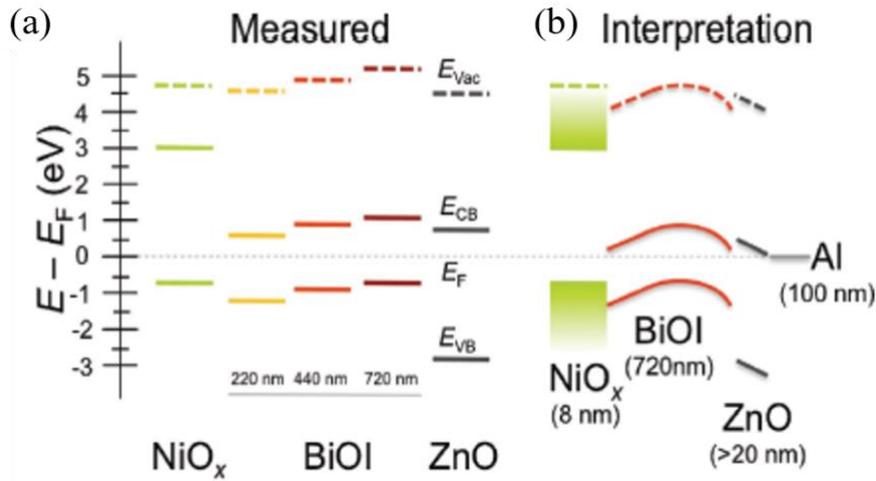

**Figure 21**. (a) The measured energy alignment of device stack and (b) schematic energy band diagrams for inverted BiOI solar cells. Reprinted with permission from Ref. [238]. Copyright 2017 Wiley.

Owing to the *n*-type characteristic of BiSI, Hahn *et al.* combined BiSI films with *p*-type CuSCN to make single junction solar cells [342]. They tested the PV performance of BiSI/CuSCN solar cells immersed in the electrolyte containing iodide compounds and achieved a 0.25% PCE. However, the poor charge separation and transport within BiSI films led to an abnormal I-V curve in their devices, where dark currents increased slowly under reverse bias. In addition, solid-state BiSI/CuSCN solar cells were also fabricated, but a small PCE of only 0.012% was reached, which was mainly ascribed to scattering losses from CuSCN layer as well as non-radiative recombination within the BiSI layer. It is worth mentioning that though they attempted to introduce Se-dope BiSI films to reduce the band gaps, this resulted in a reduction in the performance, possibly due to increased defect densities. Recently, Tiwari *et al.* fabricated planar BiSI solar cells with $SnO_2$ and F8 serving as the ETL and HTL, respectively [260]. A PCE of 1.32% with a $V_{OC}$ of 0.445 V and a $J_{SC}$ of 8.44 mA cm$^{-2}$ was obtained from their optimized device. It was claimed that one key challenge is the presence of other Bi-S-I phases in the films [267]. Moreover, BiSI devices also suffer from short carrier lifetime due to the presence of deep defects [343] and poor carrier transport, which arises from the strong anisotropy of BiSI films [260].

Nie *et al.* used CBD to fabricate mesoporous SbSI solar cells which demonstrated a PCE up to 3.05% along with a $V_{OC}$ of 0.58 V and a $J_{SC}$ of 9.11 mA cm-2 [270]. Their great success was attributed to the use of poly[2,6-(4,4-bis-(2-ethylhexyl)-4H-cyclopenta [2,1-b;3,4-b']dithiophene)-alt-4,7(2,1,3-benzothiadiazole)] (PCPBTBT) as the HTL, which can not only transport holes effectively but also harvest light not absorbed by SbSI. Mesoporous SbSI solar cells in this work retained over 93% of their initial PCEs after 15 days of storage under ambient condition in the dark. Similar stability could be maintained even under highly humid condition (humidity: 60%). Choi *et al.* demonstrated planar SbSI solar cells fabricated by a simple solution processing method [323], which may help resolve the issues of CBD-synthesized SbSI films such as the formation of impurity phases and an uncontrollable Sb/S ratio. However, the best planar SbSI solar cell only showed a PCE of 0.93% with a $V_{OC}$ of 0.548 V and a $J_{SC}$ of 5.45 mA cm-2. Recently, Nie *et al.* proposed a vapor processing method, where more uniform SbSI films could form when CBD-processed $Sb_2S_3$ were annealed under vaporized $SbI_3$ atmosphere [322]. As a result, a record PCE of 3.62% along with a $V_{OC}$ of 0.6 V and a $J_{SC}$ of 9.62 mA cm-2 was achieved in their mesoporous SbSI solar cells (with PCPDTBT as the HTL). Compared to the CBD-processed solar cells, the vapor processed solar cells exhibited better crystallinity (stronger XRD peaks), lower series resistance, and better stability (vapor-processed cells could maintain 91% of their initial PCEs while CBS-processed cells would lose over 70% of theirs after 3 hours under illumination) which is due to the compactness of the fabricated films. In this study, they also used halide treatment on $Sb_2S_3$ films to passivate surface defects and form thin SbSI layers above (Figure 22). It has been claimed that these SbSI thin films can exert an external driving force to promote the carrier transport. The optimized mesoporous SbSI-interlayered $Sb_2S_3$ solar cell demonstrated a high PCE of 6.08% with a $V_{OC}$ of 0.62 V and a $J_{SC}$ of 14.92 mA cm$^{-2}$.

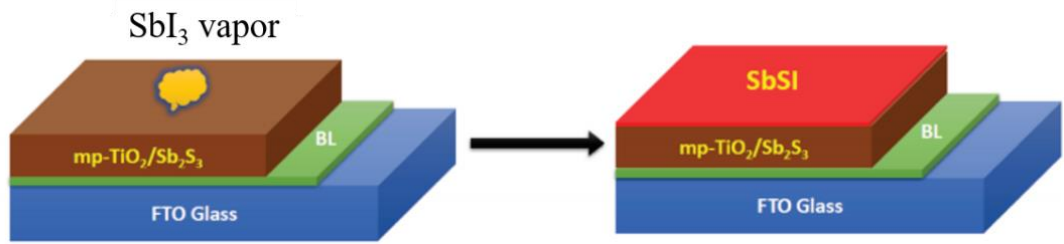

**Figure 22.** Configuration of SbSI-interlayered $Sb_2S_3$ solar cell (the HTL is not shown). Reprinted with permission from [322]. Copyright 2019 Wiley.

*4.8 Overview of perovskite-inspired materials in photovoltaics*

Figure 23 illustrates the highest PCEs and corresponding $V_{OC}$ reported in various PIM-based PV devices discussed in this Review. For comparison, the highest values achieved in Pb-based perovskite (($(FAPbI_3)_{0.95}(MAPbBr_3)_{0.05}$)) solar cells is also plotted [8]. These are compared to the SQ limit for different band gaps, along with the radiative limit in the $V_{OC}$. It can be seen from Figure 23a that except for Sn-based perovskites, all the other PIMs exhibited efficiencies below 25% of SQ limit. Furthermore, the $V_{OC}$ of the PIM devices are well below their radiative limits (Figure 23b). This is due to a combination of non-radiative recombination (due to traps or possibly self-trapped excitons in some cases), high Urbach energies, or shunting due to a sub-optimal morphology. There is therefore great room for improvements in PIM-based PV devices through the identification of new classes of materials that could tolerate defects and engineering the quality of the thin films grown.

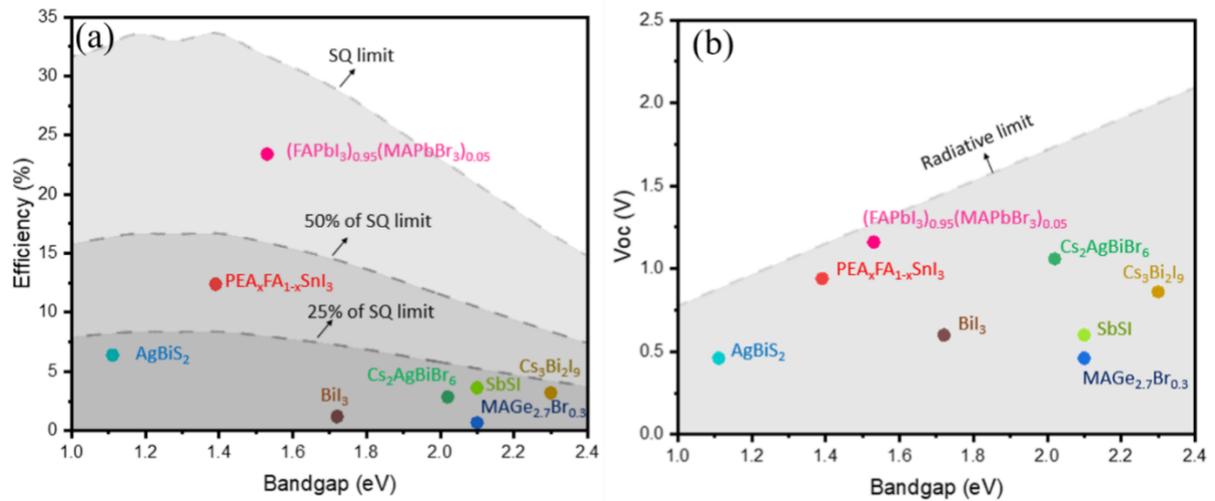

**Figure 23**. (a) The highest efficiencies and (b) the corresponding $V_{OC}$ from various PIM-based PV devices reported in the literature to date. The highest values reported in Pb-based perovskite (($FAPbI_3)_{0.95}$ ($MAPbBr_3)_{0.05}$) counterpart are also depicted.

## 5. Broader Applications of Perovskite-Inspired Materials

There is a strong overlap between the required properties of solar absorbers with the ideal properties of materials in wide range of other types of electronics. For example, strong absorption of visible light, long transport lengths and low dark currents are also needed in radiation detectors, photocatalysts and memristors. This section explores the opportunities in applying lead-free PIMs beyond PVs and the new insights into the materials that could be gained in doing so.

### 5.1 Light-emitting applications

Luminescence is one of the critical properties that influences the performance of materials in photovoltaics through the open-circuit voltage [53]. It follows that efficient photovoltaics with a direct allowed band gap are promising candidates for light-emission applications. This has been particularly well demonstrated in III-V materials [344], and more recently with lead-halide perovskites [345, 346]. There are two common applications of emitters: 1) phosphors, in which the active layer is optically excited and spontaneously emits; and 2) light-emitting diodes (LEDs), in which electrons and holes are injected into the active layer, where they radiatively recombine to give electroluminescence (EL). For both applications, one of the most important properties of the emitter is the photoluminescence quantum efficiency (PLQE), which quantifies the percentage of recombination events that are radiative (refer to Ref. [346] for more details of the other properties influencing the efficiency of LEDs). LHPs have now achieved PLQEs approaching 100% across a variety of emission wavelengths [347–349]. Green-, red- and near-infrared-emitting perovskite LEDs have now also achieved external quantum efficiencies (EQEs) >20% [345, 349, 350]. Critically, pure-halide perovskites have demonstrated low Urbach energies <30 meV, and the FWHM of the PL and EL peaks are <30 nm [53, 351]. This is narrower than achievable in typical organic and inorganic emitters, and the high color saturation, coupled with the ability to tune the emission over the entire color gamut, makes both perovskite phosphors and LEDs highly suited for ultra-high definition displays [352, 353]. But there is a strong motivation to find lead-free alternatives for display and lighting applications, owing to regulations limiting the lead content in electrical products (*e.g.*, to a maximum of 0.1 wt.% by the European Restriction of Hazardous Substances Directive [354]).

Lead-free perovskite-inspired materials, however, have not been close in matching the performance and color-purity of LHPs [346, 355–358]. One of the main reasons is the low PLQEs of most lead-free materials, many of which have indirect band gaps (refer to Section 3). But an important exception to the low luminescence in perovskite-inspired materials are $A_3B_2X_9$ nanocrystals. These include $Cs_3Sb_2Br_9$ nanocrystals (46%– 51.2% PLQE at 410 nm wavelength [359, 360]) and $Cs_3Bi_2Br_9$ nanocrystals (19.4% PLQE at 410 nm wavelength [361]). The latter has been used as blue-phosphors, optically-excited with a UV-emitting GaN LED. These phosphors were combined with a broad-emitting $Y_3Al_5O_2$ (YAG) phosphor to achieve white-light emission, with Commission Internationale de l'éclairage (CIE) coordinates of (0.29, 0.30) (close to the (0.33,0.33) coordinates for pure white) and a cool color temperature of 8477 K [361]. An important advantage of $Cs_3Bi_2Br_9$ is that it forms a passivating shell of BiOBr, which improves its stability against moisture and acid. As a result, $Cs_3Bi_2Br_9$ nanocrystals could be mixed with tetraethyl orthosilicate and react with HBr and water to form a nanocrystal-silica composite. This composite encapsulated the nanocrystals, resulting in improved stability under UV lighting (retaining 72% of the PL after 16 continuous UV illumination), and under heat stress (retaining 75% of the PL after 16 h continuous heating at 60 °C [361]). Similar results were achieved with the vacancy-ordered perovskite $Cs_2SnCl_6$. This perovskite is advantageous over regular tin-based perovskites because the tin cation is already in the more stable +4 oxidation state. By doping with bismuth, a PLQE of 78.9% could be achieved, with a blue emission wavelength of 455 nm. $Cs_2SnCl_6$:Bi was combined with yellow phosphors ($Ba_2Sr_2SiO_4$:$Eu_{2+}$ and $GaAlSiN_3$:$Eu_{2+}$) and excited with a UV GaN LED, giving white emission with CIE coordinates of (0.36, 0.37) and a color temperature of 4486 K [362].

Lead-free double perovskites have also been demonstrated as white-light-emitting phosphors. The materials investigated have been $Cs_2A(I)B(III)Cl_6$, in which A(I) is $Ag_+$ or $Na_+$ (or a mixture of both), and A(III) is $In_{3+}$ or $Bi_{3+}$ or a mixture of both. White-light emission from these materials stems from broad luminescence below the optical band gap. $Cs_2AgBiCl_6$ has a wide indirect band gap [142], whereas $Cs_2NaInCl_6$ has a direct, parity-forbidden band gap [155], across which absorption is forbidden but PL could still occur. Alloying In into $Cs_2AgBiCl_6$ resulted in a broad orange luminescence peak below the optical band gap of $Cs_2AgBiCl_6$ appearing. This was attributed to the formation of a lower parity-forbidden band gap after alloying with In, from which luminescence occurred with a PLQE up to 36.6% with 90% In addition [363]. Sub-band gap emission in $Cs_2NaInCl_6$ was attributed to self-trapped excitons. Undoped $Cs_2NaInCl_6$ has no luminescence, which was believed to be due to the self-trapped excitons being dark. Alloying with $Ag_+$ improved the PLQE to 31.1% [364]. Alloying $Cs_2(Ag_{0.6}Na_{0.4})InCl_6$ with 0.04% Bi further increased the PLQE to 86±5%, with broad emission centered at 570 nm wavelength [277]. On the other hand, $Cs_2NaInCl_6$ alloyed with $Bi_{3+}$ remained dark, with only blue PL due to free excitons. But broad yellow emission was achieved through alloying with $Mn_{2+}$, with a PLQE of 44.6% being achieved, and this was attributed to the dark self-trapped exciton transferred to an excited state of $Mn_{2+}$ [365]. $Cs_2(Ag_{0.6}Na_{0.4})InCl_6:Bi_{3+}$ was used as white-light phosphors, owing to its high white-light PLQE. With excitation from a GaN LED, emission with CIE coordinates of (0.396,0.448) was achieved with a color temperature of 4054 K. Critically, the double perovskite was air stable, and even more so when encapsulated with silica. As a result, the emission was stable after 1000 h of illumination in air [277].

Currently, there is only a handful of reports of EL from PIMs, and most are on Sn-based perovskites. Lai *et al.* reported near-infrared LEDs from $MASnI_3$, which emitted at 945 nm wavelength, but with EQEs only reaching 0.72% [366]. An important limitation was the low PLQE of 5.3% [366]. Rand *et al.* subsequently improved the EQE to 5.0% (at 917 nm EL wavelength) by alloying Pb with Sn and introducing a long-chain organoammonium iodide ligand to the precursor solution. The inclusion of the ligand resulted in smaller grains with passivated surfaces [367]. Ruddlesden-Popper tin-based perovskites have also been investigated, but an early investigation into 2D phenethylammonium tin iodide resulted in inefficient red-emitting devices (0.15 cd $m_{-2}$ luminance and 0.03 cd $A_{-1}$ current efficiency [368]). Subsequently, Rogach *et al.* investigated tin-based Ruddlesden-Popper perovskites with bromide as the halide, and oleylammonium as the A-site cation. They were able to directly inject into the self-trapped exciton, resulting in broad orange sub-band gap EL (centered at 620 nm), with 0.1% EQE and 350 cd m$_{-2}$ luminance [369]. Notably, a PLQE of 88% was achieved in colloidal suspensions, and 68% in thin films [369]. Beyond these tin-based perovskites, there has recently been a demonstration of Sb-based LEDs. Shan *et al.* achieved violet EL from Cs$_3$Sb$_2$Br$_9$ (408 nm wavelength) with an EQE of 0.2%. The PLQE of the Cs$_3$Sb$_2$Br$_9$ quantum dots was 51.2%, and the quantum dots were stable against heat, UV illumination and moisture. In particular, the LEDs retained 90% of their initial EL intensity after 6 h of operation at 7 V (~70 mA cm$_{-2}$ current density [359]). This is more stable than their Sn- and Pb-based perovskite counterparts. But the FWHMs of all tin- and antimony-based emitters discussed here are significantly wider than their lead-based counterparts, typically 50 – 100 nm [359, 366–368]. This makes them less competitive for display applications, compared to quantum dot, organic and inorganic LEDs already commercially available [352].

## 5.2 Photocatalysis

Leaves absorb sunlight to convert $CO_2$ and water to sugar and oxygen through photosynthesis. Similarly, semiconductors can absorb sunlight to enable the production of fuels, precursors for value-added products, or the degradation of organic contaminants through photocatalysis [370, 371]. The production of $H_2$ from water was first achieved by Fujishima *et al.* in 1969, in which $TiO_2$ was used to absorb UV light. Holes generated in $TiO_2$ oxidized water to $O_2$ gas, while the electrodes transferred to the Pt counter electrode reduced $H_+$ to $H_2$ [372, 373]. This spurred several decades of research into semiconductor-based photocatalysis, and was motivated both by the promise of the clean solar-to-chemical energy conversion, as well as the limitations of $TiO_2$ in only being able to absorb a small part of the solar spectrum due to its wide band gap [370, 371, 374]. The materials and devices structures for photocatalysis, as well as the operating principles, are reviewed in Ref. [370, 371, 374, 375]. The requirements for the semiconductor are: 1) good light absorption, 2) high chemical stability in the presence of the electrolyte, 3) suitable band-edge positions for enabling the redox reactions required, 4) efficient charge transport, and 5) to be low-cost. Requirements 1, 4 and 5 overlap with the requirements for photovoltaic materials. Lead-halide perovskites have therefore been considered for photocatalysis, owing to their high absorption coefficients, strong overlap in their absorption with the solar spectrum, and long diffusion lengths, coupled with the ability to achieve high-quality films when processed at low temperature. But lead-halide perovskites are limited by their low chemical stability [371]. This necessitates tailoring of the solution to minimize degradation to the perovskite (*e.g.*, using a saturated halo acid solution for water splitting), or encapsulating the perovskite to isolate them from the polar solvent [371, 376–379]. But ensuring the long-term stability of the photocathode or photoanode, as well as avoiding potential lead contamination of the solution motivate the investigation of lead-free and more stable alternatives.

In photocatalytic applications, the wider band gap of most perovskite-inspired materials is advantageous. Murphy *et al.* found that the optimal band gap for solar-driven water splitting is 2.03 eV, which would enable a theoretical maximum solar-to-hydrogen conversion efficiency of 16.8% [380]. This is because the quasi-Fermi levels of the semiconductor determine the oxidation/reduction potential of the photoelectrochemical cell. Therefore, in the case of water splitting, the band gap needs to at least be equal to the difference in redox potentials for $H_+/H_2$ (hydrogen reduction) and $O_2/H_2O$ (water oxidation). This has a value of 1.23 eV, but the semiconductor band gap needs to be larger in order to account for thermodynamic losses (0.3-0.4 eV) and overpotentials for fast reactions (0.4-0.6 eV). A band gap of >1.9 eV therefore should be used [374], indicating that many of the double perovskites, $A_3B_2X_9$ compounds and V-VI-VII compounds are suitable, particularly given their improved environmental stability compared to lead-halide perovskites (refer to Section 3). The electronic band positions of these perovskite-inspired materials compared to the redox potentials for water splitting, as well as for $CO_2$ reduction [371] are shown in Figure 24. It is noted that the redox potentials and band positions are given on a relative hydrogen electrode (RHE) scale. $H_+/H_2$ under standard conditions is typically taken as a reference point (standard hydrogen electrode), which occurs at pH 0 and has a value of 4.44±0.02 eV relative to vacuum level according to IUPAC [374]. But according to the Nernst equation, the redox potential for $H_+/H_2$, as well as $O_2/H_2O$, $CO_2/CO$ and $CO_2/CH_4$ all vary with -0.059pH [V] [371]. The potential across the Helmholtz double layer on the semiconductor surface also changes with -0.059 pH [V]. The RHE scale is therefore used, in which all potentials are given relative to $H_+/H_2$ at a given pH [374].

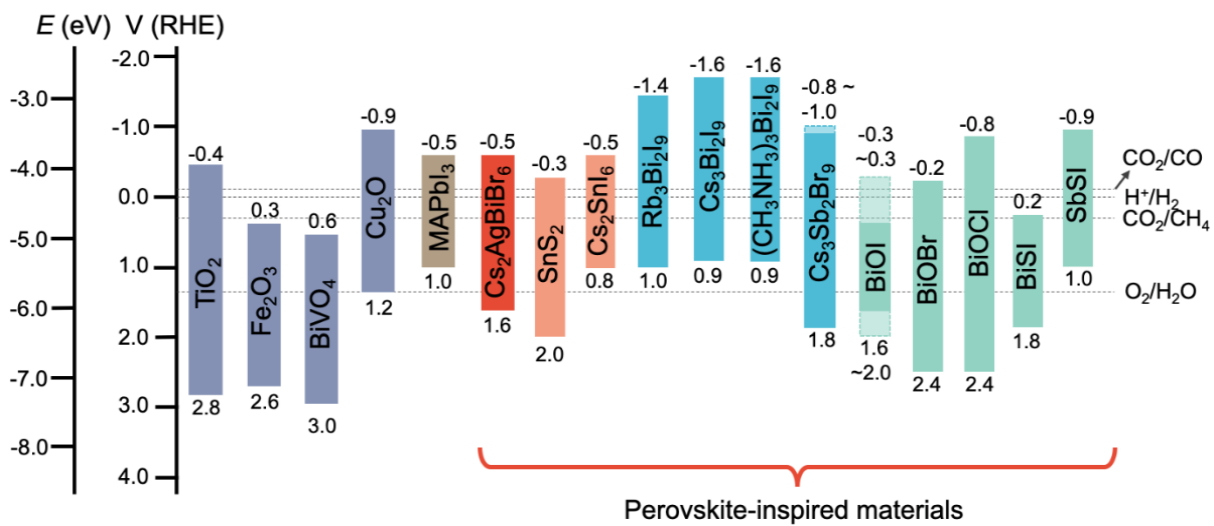

**Figure 24**. Band-edge positions of perovskite-inspired materials, which are compared to methylammonium lead iodide (MAPbI$_3$) and traditional photocatalytic materials. The band positions are shown on a relative hydrogen electrode (RHE) potential scale (energy scale shown as a comparison). The RHE redox potentials of common half-reactions are given [371, 374]. The band positions are obtained from Ref. [151, 154, 384, 385, 254, 255, 359, 371, 374, 381–383].

Many recent works on lead-free perovskite-inspired materials for photocatalytic applications have focused on double perovskites, vacancy-ordered perovskites and $A_3B_2X_9$ materials. These materials have been demonstrated to be more phase-stable than methylammonium lead iodide in ambient air [147, 151, 176, 383]. Groups have also demonstrated $Cs_2AgBiBr_6$ to remain the same phase after illumination with 70 mW cm$_{-2}$ white light for 500 h [147], while $Rb_3Bi_2I_9$, $Cs_3Bi_2I_9$ and $(CH_3NH_3)_3Bi_2I_9$ was found to maintain the same phase after 12 h of illumination with a UV lamp (80.4 µW cm$_{-2}$ [384]). Despite these promising stability results, efforts have focused on the application of these perovskite-inspired materials as photocatalysts under benign environments (*e.g.*, halo acid solutions or gaseous precursors), rather than in photoelectrolytic devices in more aggressive aqueous environments. Nevertheless, these works have found the photocatalysts to be stable under operation. Many works have focused on $CO_2$ reduction to form CO or $CH_4$ (or a mixture of both), and most materials have suitable electron affinities to induce the reduction of $CO_2$ (Figure 24). Zhou *et al.* synthesized $Cs_2AgBiBr_6$ double perovskite nanocrystals through hot-injection [147]. However, the yield of CO and $CH_4$ were 5.5 µmol g$_{-1}$ and 0.65 µmol g$_{-1}$ respectively. This was improved by washing the nanocrystals in ethanol to remove the oleic acid and oleylamine ligands prior to the application of the nanocrystals as photocatalysts, which resulted in the yields improving to 14.1 µmol g$_{-1}$ and 9.6 µmol g$_{-1}$ for CO and $CH_4$ respectively. However, the external quantum efficiency was only 0.028% at 398 nm wavelength. Nevertheless, the washed double perovskite nanocrystals were found to not change in phase or surface composition or agglomerate after 6 h operation, implying that the nanocrystals were stable in the ethyl acetate solution the $CO_2$ was dissolved in [147]. Wang *et al.* obtained comparable product yields using $SnS_2$ nanosheets decorated with $Cs_2SnI_6$. This was achieved by reacting the $SnS_2$ nanosheets with CsI dissolved in ethanol. The improvement in photocatalytic activity by forming $Cs_2SnI_6$ was attributed to the smaller band gap of the vacancy-ordered perovskite (1.3 eV [383]), which led to an increased population of photogenerated carriers [381]. Higher product yields were obtained by Bhosale *et al.* using $A_3B_2I_9$ photocatalysts, especially $Cs_3Bi_2I_9$, which was found to have high photocatalytic activity and gave yields of 78 µmol $g_{-1}$ and 15 µmol $g_{-1}$ for CO and $CH_4$ respectively after 10 h [384]. The $Cs_3Bi_2I_9$ was synthesized as nanocrystals with oleic acid and oleylamine ligands and dispersed in trichloromethane after purification. The trichloromethane had a mixture of $CO_2$ gas and $H_2O$ vapor introduced to it, and the nanocrystals were illuminated with a UV lamp [384]. Yet higher product yield was achieved using $Cs_3Sb_2Br_9$ nanocrystals, which reached 516 [385] µmol $g_{-1}$ for CO after 5 h, with illumination under 1 sun radiation. By contrast, $CsPbBr_3$ nanocrystals produced only 53.8 µmol $g_{-1}$ CO after 5 h under the same conditions. The $Cs_3Sb_2Br_9$ nanocrystals were synthesized by hot-injection, and were exposed to $CO_2$ gas dissolved in octadecene. The same nanocrystals were used three times, and it was found that with each subsequent usage, there was a decrease in the product yield. But after the third usage, the yield was approximately 30% of that of the first usage. Despite this reduction in activity, the yield remained higher than the yield of fresh $CsPbBr_3$ nanocrystals [385].

$Cs_2AgBiBr_6$ and $(CH_3NH_3)_3Bi_2I_9$ have also been investigated for the photocatalytic reduction of $H_+$ to $H_2$ [386, 387]. $H_2$ is an important clean fuel that combusts to only produce $H_2O$, is used for hydrogen fuel cells, and is a convenient option for storing excess solar energy to compensate for the intermittent nature of terrestrial solar radiation [370, 388]. Wang *et al.* synthesized $Cs_2AgBiBr_6$ powder, and a mixture of HBr and $H_3PO_2$ was used as the $H_+$ source. The $Cs_2AgBiBr_6$ photocatalyst was illuminated with white light (300 W), and $H_2$ was generated with a rate of 0.6 µmol $g_{-1}$ $h_{-1}$, averaged over 10 h. The photocatalytic activity was improved by 80 times to 48.9 µmol $g_{-1}$ $h_{-1}$ by adding reduced graphene oxide (rGO) to the double perovskite. Moreover, it was found that the photocatalytic activity of the composite remained stable after 12 cycles (120 h total). It is believed that the improved catalytic activity was due to 1) photogenerated electrons from the double perovskite being injected to the rGO, which contains active sites for $H_+$ reduction, and 2) a reduction in the recombination rate because of the separation of the photogenerated electrons and holes at the perovskite/rGO junction [386]. A higher $H_2$ production rate of 169.2 µmol g-1 h-1 was reported by Guo *et al.*, who used $(CH_3NH_3)_3Bi_2I_9$ powder coated with Pt. The electrolyte was HI solution, and it was found that $(CH_3NH_3)_3Bi_2I_9$ remained phase-stable in this solution. While the energy of the photogenerated holes was insufficient to oxidize water, they were believed to oxidize I- to I3-. $H_3PO_2$ was added to the solution to reduce I3- to prevent I3- from accumulating and blocking light absorption in the $(CH_3NH_3)_3Bi_2I_9$/Pt photocatalyst. The stability was tested through 7 repeat cycles (each 10 h long), with only a small decrease in the catalytic activity [387]. Finally, $Cs_2AgBiBr_6$ powders have also been used as a photocatalyst for the degradation of ionic dyes in ethanol solution by forming oxygen radicals from $O_2$. The catalytic activity was again enhanced by coating with Pt. The double perovskite/Pt composites were observed to maintain their photocatalytic activity and phase after five cycles, which each had a duration of 1 h [389].

V-VI-VII materials have been investigated as both photocatalysts and photoelectrodes before the work on lead-halide perovskite photocatalysts began. These V-VI-VII materials are appealing due to their strong absorption of visible light. Two examples are BiOI and BiSI, which have band gaps of 1.6–1.9 eV [254, 390] (Figure 25), and have been investigated as photocatalysts for degrading organic contaminants in water, such as methyl orange and crystal violet [390, 391]. BiOI itself has been found to have low photocatalytic activity due to its short charge-carrier lifetimes [391]. The photocatalytic activity has been improved through heterostructures, such as $BiOI/MoS_2$, $BiOI/WO_3$, BiOI/AgI, BiOI/Ag, BiOI/Ag/AgI, $BiOI/Bi_2WO_6$, $BiOI/Ag_3PO_4$, $BiOI/Bi_2MoO_6$, $BiOI/ZnSn(OH)_6$, BiOI/BiOBr [391]. Similarly, the photocatalytic activity of BiSI was improved by forming a composite with $MoS_2$ [390].

MoS$_2$ forms a type II heterojunction with both BiOI and BiSI, and this is thought to improve the separation of photogenerated electrons and holes, which can then take part in the photodegradation of the organic contaminants [391]. High photocatalytic activity was achieved from SbSI without needing to form a heterostructure. Tamilselvan *et al.* demonstrated that micron-sized urchin-shaped SbSI degraded 97% of methyl orange [392], which is comparable to BiOI/MoS$_2$ nanocomposites and higher than BiOI alone [391]. Investigations found that SbSI induced the degradation of methyl orange by forming oxygen radicals through the oxidation of singlet oxygen with photogenerated holes [392]. Wang *et al.* achieved a slightly higher degradation rate of methyl orange in aqueous solution (99% within 1 min when illuminated with a 300 W xenon lamp) through the use of SbSI nanocrystals at room temperature or at 65 °C [393]. However, they attributed the degradation of methyl orange to the formation of singlet oxygen rather than oxygen radicals [393]. For all materials (BiOI/MoS$_2$, BiSI/MoS$_2$, SbSI), it was found that the photocatalysts maintained their activity after several cycles [391].

Defects in bismuth oxyhalides have been found to enhance photocatalytic activity. For example, it was found that oxygen vacancies in BiOBr and BiOCl led to an increase in the product yield for the reduction of CO$_2$ to CO or CH$_4$ [394, 395]. It is believed that these oxygen vacancies act as sites for CO$_2$ adsorption, where they are activated to radical ions that can be reduced with photogenerated electrons. It is also proposed that the oxygen vacancies can trap electrons to prolong their separation from photogenerated holes [394]. These oxygen vacancies were induced in BiOBr by modifying the solvothermal synthesis, in which ethylene glycol was used as the solvent instead of water (which gave stoichiometric BiOBr [394]). For BiOCl, Zhang *et al.* induced oxygen vacancies by illuminating for 5 h with a Xe lamp. But despite the improvements in photocatalytic activity, the product yield of CH$_4$ in BiOBr and BiOCl remained lower than $Cs_3Bi_2I_9$ [384, 394, 395]. Nevertheless, the materials were found to maintain their activity after several cycles, and, in the case of BiOCl, it was found that the oxygen vacancies could be regenerated through illumination [394, 395].

Bismuth oxyhalides have also been investigated in photoelectrochemical cells. BiOI was used in photoelectrochemical cells with an $I_3^-/I^-$ redox couple (0.53 V reduction potential relative to the standard hydrogen electrode) in an acetonitrile-based solution. Platelets of BiOI were grown by successive ionic layer adsorption and reaction (SILAR) and spray pyrolysis [256, 396]. Such cells can be used as photovoltaic devices, but the performance was low, with PCEs of 1% or below, and EQEs reaching only 60% [256, 396]. Possibly a limiting factor was a non-compact structure of the BiOI platelets, or a non-optimal arrangement. Subsequent work on growing BiOI by chemical vapor deposition resulted in more compact films, with PCEs reaching 1.8% and EQEs reaching 80% (refer to Section 4.7) [254]. Beyond photovoltaics, bismuth oxyhalides could be used for photo-assisted water conversion. Bhachu *et al.* synthesized thin films of BiOI, BiOBr and BiOCl by aerosol-assisted chemical vapor deposition (AA-CVD) and performed photoelectrochemical testing in a 0.5 mol $L_{-1}$ $Na_2SO_4$ electrolyte (pH of 6.5), with a Pt counter electrode and Ag/AgCl reference electrode [255], and with 1 sun illumination. BiOI was found to not give any photocurrent at 0 V(RHE), suggesting that it would not reduce $H_+$. However, the photocurrent was ~0.1 mA $cm_{-2}$ at 1.23 V(RHE), suggesting that the BiOI photoelectrochemical cells could oxidize water. BiOBr was also found to be capable of oxidizing water, with a photocurrent of ~0.3 mA $cm_{-2}$ at 1.23 V(RHE). BiOCl showed very little photocatalytic activity. But all three materials were found to be unstable in the electrolyte. For example, the photocurrent from BiOBr at 1 V *vs.* Ag/AgCl (1.6 V(RHE)) was found to decrease by 23% after 1 h, whereas BiOI at the same potential decreased after only 100 s [255]. Moreover, the photocurrents were significantly lower than those achieved with $Cu_2O$, which has a similar band gap. But recent $Cu_2O$ photoelectrodes have used a heterojunction structure, as a well as a surface catalyst to improve the photocatalytic activity [388]. Similar strategies could be investigated for V-VI-VII materials.

*5.3 Radiation detection*

The detection of ionizing radiation (particularly X-rays and γ-rays) is important for a wide range of applications, including security screening, medical diagnostics and the characterization of materials [186, 397–399]. Furthermore, it is important to be able to detect and quantify heavy charged particles, such as α-particles, which are by-products from nuclear reactions [400]. The key properties of radiation detector materials are that they need to have a large: 1) average atomic number ($Z_{avg}$), 2) product of mobility and charge-carrier lifetime (μτ), and 3) resistivity (>$10^9$ Ω cm [397, 398, 401]). A high atomic number is needed because the attenuation coefficient (α) of high energy photons is proportional to $Z_4/E_3$, where $E$ is the photon energy [246, 402, 403]. A large μτ product is needed for the efficient extraction of charge-carriers because the thickness required to completely absorb radiation is often at the millimeter scale [402] or larger. The dependence of the photocurrent ($I$) from a radiation detector to the μτ product is given by the Hecht equation:

$$I = \frac{I_0 \mu \tau V}{L^2} \frac{1 - e^{-L^2/\mu \tau V}}{1 + \frac{Ls}{V\mu}} \tag{6} [398]$$

In Equation 6, $I_0$ is the saturated photocurrent, $V$ the applied bias, $L$ the thickness of the radiation detector, and $s$ the surface recombination velocity. Typically μτ >$10^{-4}$-$10^{-3}$ cm$^2$ V$^{-1}$ [404] is needed. If the μτ product is low, a large applied field is needed to extract the carriers, but this can increase the dark current, which reduces the signal-to-noise ratio. A high resistivity is also needed to maintain a low dark current to achieve a high signal-to-noise ratio (which gives high image contrast), as well as the ability to detect lower doses of radiation (*i.e.*, reduced lowest detectable dose rate). This is especially important for medical diagnostic applications and security screening, in which a lower dose of harmful radiation can be used, reducing the risk of causing cancer in the human subject. The lowest detectable dose rate required for medical applications is 50 mGy$_{air}$ s-1 [399], but there is motivation to be able to detect lower doses down to 0.1 nGy$_{air}$ s-1, which is the background radiation levels in the US [403], and also because using lower doses improves the spatial resolution of the detector [403]. Note that the unit Gy (gray) represents the energy of ionizing radiation absorbed by a particular mass (1 Gy = 1 J kg-1). The dark current can be reduced by increasing the band gap, and at room temperature, the band gap should be between 1.4 and 2.5 eV [399, 401, 405]. The upper limit to the band gap is so that the electron-hole ionization energy is small [401]. Another important parameter in radiation detectors is the sensitivity ($S$), which is defined as $S = (I_{ON} - I_{OFF})/$(dose rate [187]). The sensitivity would therefore depend on $Z$, μτ product, as well as the dark current. The spatial resolution, response speed (which determines the duration of exposure to harmful radiation that is needed), energy resolution, linear dynamic range, uniformity and stability of the detectors are also important parameters [398, 403].

Often the materials used for radiation detectors are single crystals. This is because the penetration depth of radiation (from soft X-rays to hard γ-rays) are on the millimeter to centimeter scale. It is also because single crystals tend to have a lower defect density and larger μτ products than thin films. A common radiation detector is silicon. But silicon has a low $Z$ value and low band gap. Thus, despite its high μτ product, it has low stopping power, low resistivity, low sensitivity and a high lowest detectable dose rate (Table 3). Silicon therefore tends to only be used for soft X-ray ($E$ <10 keV) detection and in portable detectors (*e.g.*, for EDX or XRF [398]). Amorphous selenium (a-Se) has a higher Z value and therefore stronger stopping power for radiation. a-Se is the most common solid-state material for the direct detection of X-rays, but has a low μτ product. High applied biases are therefore needed, which leads to low sensitivities (Table 3 [398]). Other alternatives contain toxic elements ($Cd_xZn_{1-x}Te$ and $HgI_2$ [398]), or require liquid nitrogen cooling (such as high purity Ge [405], which has a small band gap of 0.66 eV and needs cooling to reduce the dark current [398]). Beyond direct-conversion materials, radiation detection can also occur by the use of scintillators to absorb and down-convert the radiation to lower-energy photons that can be detected using standard silicon-based photodiodes [402]. But common scintillators are alkali halides doped with toxic Tl [398, 402], and have given lower resolution for imaging and lower linearity in their response to the radiation dose [405] than direct detectors. Recently, lead-halide perovskites have gained attention for radiation detectors, owing to their composition of high atomic number elements, the high μτ products achievable, their widely tunable band gaps, and their facile processability from solution [397, 399, 400, 402]. As a result, perovskite X-ray detectors can operate at room temperature with improved performance (in terms of sensitivity and lowest detectable dose rate) over a-Se and $Cd_xZn_{1-x}Te$ have been demonstrated (Table 3). Furthermore, lead-halide perovskites have demonstrated promising radiation hardness to protons and γ-rays [398]. Lead-halide perovskites, with high PLQEs to visible light excitation, have also demonstrated reasonable X-ray excited luminescence yields at room temperature, making them suitable for consideration as scintillators [406]. The price of each perovskite single crystal is estimated to be US$0.5–1.0 $cm_{-3}$ based on the price of the precursors [397]. By contrast $Cd_xZn_{1-x}Te$ single crystals cost $3000 $cm_{-3}$, in part due to the low demand for these materials for radiation detectors [398]. Despite these advantages, lead-halide perovskites have limited stability in air [176], and can degrade under continuous irradiation with high flux-density X-rays in air [20] for extended periods of time. In addition, the lead content of perovskite single crystals significantly exceeds the limited imposed by regulations on hazardous elements [403]. Ion migration in lead-halide perovskites could also result in drift in the dark current, which could increase the lowest detectable dose rate [186]. It is therefore important to consider perovskite-inspired materials for radiation detection.

Perovskite-inspired materials have many desirable properties for radiation detectors: the band gaps of most materials are within the ideal 1.4-2.5 eV range, and they are comprised of high atomic number elements (*e.g.*, Bi and I). In particular, many of these materials are stable in air, have low toxicity and can be processed as single crystals using facile fabrication methods. Prior to the work on lead-halide perovskites, $BiI_3$ had already been investigated for radiation detectors [407–410]. $BiI_3$ has a band gap of 1.67±0.09 eV [411], and can be grown as single crystals by the vertical Bridgman method and physical vapor transport [246, 409] (refer to Section 3.7 for details). Owing to the high $Z_{avg}$ of 60.5 and density (5.78 g cm$_{-3}$), the mass absorption coefficient is large (6.746 cm$_2$ g$_{-1}$ at 60 keV). In addition, $BiI_3$ has a low ionization energy (5.8 eV) that is comparable to $Cd_xZn_{1-x}Te$ (5 eV) and lower than a-Se (45 eV). $BiI_3$ can therefore easily convert absorbed X-rays into charges through the photoelectric effect [409]. For device characterization, $BiI_3$ is used in a photoconductor structure owing to its high resistivity ($10_9$-$10_{11}$ $\Omega$ cm), in which its dark current is several orders of magnitude smaller than its photocurrent. That is, $BiI_3$ has two ohmic contacts, usually Au or Pd [407, 409]. These can be applied on the front and back surface of the single crystal (planar) or both on the front surface (coplanar). Sun *et al.* found that the planar structure resulted in 1.8-2.5 times higher sensitivities of ~1.3 × 10$_4$ μC Gy$_{-1air}$ cm$_{-2}$, which are comparable to $MAPbBr_3$ and larger than $Cd_xZn_{1-x}Te$ or a-Se [409] (Table 3). This was attributed to a more uniform electric field in the planar configuration [409]. Dmitriev *et al.* used the planar configuration to measure an electron mobility-lifetime product of 10$_{-5}$ cm$_2$ V$_{-1}$ [407] through photoconductivity measurements.

By contrast, $Cs_2AgBiBr_6$, $MA_3Bi_2I_9$ and $(NH_4)_3Bi_2I_9$ single crystals are grown from solution [186, 187, 403]. Au electrodes were used for $Cs_2AgBiBr_6$ and $MA_3Bi_2I_9$ radiation detectors [186, 403], whereas Ag was painted the $(NH_4)_3Bi_2I_9$ single crystal [187]. The ionization energies for these materials are low: 5.61 eV for $Cs_2AgBiBr_6$ [403] and 5.47 eV for $(NH_4)_3Bi_2I_9$ [187]. As-grown $Cs_2AgBiBr_6$ was reported to have a trap density of $4.54 \times 10^9$ cm$^{-3}$ and mobility of 3.17 cm$^2$ V$^{-1}$ s$^{-1}$, as determined from space-charge limited current measurements. By post-annealing the single crystals at 100 °C for 2 h in $N_2$, the trap density was reduced to $1.74 \times 10^9$ cm$^{-3}$, whereas the mobility increased to 11.81 cm$^2$ V$^{-1}$ s$^{-1}$. This resulted in the μτ product increasing from $3.75 \times 10^{-3}$ cm$^2$ V$^{-1}$ to $6.3 \times 10^{-3}$ cm$^2$ V$^{-1}$, which is larger than a-Se and $BiI_3$ (Table 3). The surface recombination velocity was reduced by washing the single crystals in isopropanol to remove surface defect states, resulting in the resistivity increasing to $10^{11}$ Ω cm. Despite these promising properties, and despite $Cs_2AgBiBr_6$ having a higher $Z_{avg}$ than $MAPbBr_3$, the sensitivity achieved was only 8 μC Gy$_{air}^{-1}$ cm$^{-2}$, which increased to 105 μC Gy$_{air}^{-1}$ cm$^{-2}$ after applying an electric field of 25 V mm$^{-1}$. However, the larger resistivity led to a low dark current (~0.15 nA cm$^{-2}$ at 5 V) and therefore a small lowest detectable dose rate of 0.0597 μC Gy$_{air}$ s$^{-1}$ [403]. A smaller lowest detectable dose rate and significantly larger sensitivity were achieved with $MA_3Bi_2I_9$ single crystals, which have a 0D crystal structure (Table 3 [186]). This may be due to the wide band gap and high resistivity ($10^{11}$ Ω cm in the out-of-plane direction) resulting in a low dark current of only 0.98 nA cm$^{-2}$ at 120 V bias [186]. Notably the activation energy barrier for ion migration was found to be 0.31-0.46 eV [186], comparable to $Cs_2AgBiBr_6$ (0.35 eV) and larger than $MAPbBr_3$ (0.19 eV [187]). A higher activation energy barrier was found for $(NH_4)_3Bi_2I_9$. $(NH_4)_3Bi_2I_9$ is a layered material, and the activation energy barrier was found to be higher in the [001] direction (*i.e.*, between planes; 0.91 eV *cf.* 0.72 eV along the [100] direction [187]). Owing to the layered structure, $(NH_4)_3Bi_2I_9$ exhibited anisotropic transport properties, in which higher μτ products and sensitivities were obtained in the [100] direction ($1.1 \times 10^{-2}$ cm$^2$ V$^{-1}$; $0.8 \times 10^4$ µC Gy$_{-1air}$ cm$^{-2}$) than the [001] direction ($4.0 \times 10^{-3}$ cm$^2$ V$^{-1}$; 803 µC Gy$_{-1air}$ cm$^{-2}$). However, the signal to noise ratio in the [001] direction was found to be larger, leading to a smaller lowest detectable dose rate of 0.055 µGy$_{air}$ s$^{-1}$ (*cf.* 0.210 µGy$_{air}$ s$^{-1}$ in the [100] direction), which was attributed to reduced ion migration [187]. All three materials were found to be stable in ambient air and withstand temperatures >200 °C [186, 187, 403].

**Table 3.** Materials properties of direct X-ray detectors and their performance

| Material | $Z_{avg}$ | µτ (cm$^2$ V$^{-1}$) | Resistivity ((Ω cm) | Sensitivity (µC Gy$_{-1air}$ cm$^{-2}$) | Lowest detectable dose rate (µGy$_{air}$ s$^{-1}$) | Ref |
|---|---|---|---|---|---|---|
| Si | 14 | >1 | $10^4$ | 8 | <8300 | [412–415] |
| a-Se | 34 | $10^{-7}$ | | 20 | 5.5 | [416, 417] |
| Cd$_x$Zn$_{1-x}$Te | 48.2-49.1 | 0.01 | $10^{10}$ | 318 | 50 | [418] |
| MAPbBr$_{3a}$ | 45.1 | 0.012 | ~$10^7$ | $2.1 \times 10^4$ | 0.039 | [419] |
| BiI$_3$ | 60.5 | $10^{-5}$ | $10^9$–$10^{11}$ | $10^4$ | - | [407, 409] |
| Cs$_2$AgBiBr$_6$ | 53.1 | $6.3 \times 10^{-3}$ | $10^9$–$10^{11}$ | 105 | 0.0597 | [403] |
| MA$_3$Bi$_2$I$_9$ | - | ~$10^{-3}$ | $10^{10}$–$10^{11}$ | $1.1 \times 10^4$ | 0.0006 | [186] |
| (NH$_4$)$_3$Bi$_2$I$_9$ | - | $1.1 \times 10^{-2}$ | $10^6 - 10^8$ | $0.8 \times 10^4$ | 0.210 | [187] |

$_a$ MA is CH$_3$NH$_{3+}$

Beyond the direct conversion of X-rays to electrical energy, perovskites have also been investigated as scintillators. A recent work found CsPbBr$_3$ nanocrystals to demonstrate significantly higher sensitivity than Bi$_4$Ge$_3$O$_{12}$ (one of the standard scintillator materials), and have a sensitivity to X-rays on the same order of magnitude as CsI:Tl [420]. Recently, (C$_8$H$_{17}$NH$_3$)$_2$SnBr$_4$ was demonstrated as a lead-free perovskite scintillator. These had a PLQE of 98%. It was suggested that the mechanism for X-ray induced emission (radioluminescence) was photoelectric ionization, followed by relaxation to the band-edge to form excitons that recombine radiatively. The radioluminescence spectrum was found to be almost the same as the PL spectrum. The emission peak was centered at 596 nm, which is detectable by standard CCDs. The tin-based perovskite was made into a composite with PMMA. The radioluminescence was found to be linear with the X-ray dose down to $10^4$ µGy s$^{-1}$, which is a sufficiently low dose to be safely used in X-ray imaging. The spatial resolution achieved with the perovskite scintillator/CCD system was 200 μm, which was lower than reported for lead-based perovskite systems. The tin-based perovskite/PMMA composite also demonstrated no decrease in radioluminescence intensity after cycled X-ray exposure over 800 s [421].

**Table 4.** Properties of direct γ-ray and α-particle detectors

| Material | $Z_{avg}$ | μτ (cm$_2$ V$_{-1}$)$_a$ | Resistivity (Ω cm) | Energy resolution (%) | Ref |
|---|---|---|---|---|---|
| Gamma ray detector | | | | | |
| HPGe | 32 | >1 | $10_2$–$10_3$ | 0.2 | [422, 423] |
| $Cd_xZn_{1-x}Te$ | 48.2-49.1 | 0.004–0.01 | $10_{10}$ | 0.5 | [424, 425] |
| $CsPbBr_3$ | 65.9 | $1.34 \times 10_{-3}$ | $10_{11}$ | 3.8–3.9 | [426] |
| $Sb:BiI_3$ | 60.5 | $10_{-4}$ | $10_8$–$10_{10}$ | 2.2 | [404] |
| Alpha particle detector | | | | | |
| $Cd_xZn_{1-x}Te$ | 48.2-49.1 | $10_{-3}$–$10_{-2}$ | $10_{10}$ | ~1% | [422, 427, 428] |
| $CsPbBr_3$ | 65.9 | $9.5 \times 10_{-4}$ | $10_9$ | 15% | [400] |
| $BiI_3$ | 60.5 | - | $10_{10}$ | 40.2% | [404, 408, 410] |
| $Sb:BiI_3$ | 60.5 | - | $10_9$ | 33 – 66% | [246] |
| $Cs_3Bi_2I_9$ | 57.7 | $5.4 \times 10_{-5}$ | $9.4 \times 10_{12}$ | - | [405] |
| $Cs_3Sb_2I_9$ | 53.1 | $1.1 \times 10_{-5}$ | $5.2 \times 10_{11}$ | - | [405] |
| $Rb_3Bi_2I_9$ | 53.9 | $1.7 \times 10_{-6}$ | $3.2 \times 10_{11}$ | - | [405] |
| $Rb_3Sb_2I_9$ | 49.3 | $4.5 \times 10_{-6}$ | $8.5 \times 10_{10}$ | - | [405] |

$_a$ the highest μτ product is shown for anisotropic materials

In γ-ray detection, the energy resolution of the detector is an important parameter. This is because the detector needs to accurately identify the characteristic spectrum of specific radionucleotides. Although high energy resolution can be achieved with high purity Ge and $Cd_xZn_{1-x}Te$ (Table 4), these materials are limited by their need to operate at cryogenic temperatures or high cost, respectively, as discussed above. Recently, He *et al.* achieved high sensitivity γ-ray detectors using $CsPbBr_3$ single crystals, which demonstrated an energy resolution of 3.8-3.9% (FWHM relative to the energy of the peak) for the γ-ray peak [426]. In perovskite-inspired materials, there has been extensive work on $BiI_3$ for γ-ray detection. The total attenuation length for photoelectric absorption (the main process by which γ-rays are converted to electron-hole pairs) in $BiI_3$ is smaller than $Cd_xZn_{1-x}Te$ (Figure 25b) due to the higher $Z_{avg}$ for BiI$_3$ [405] (Table 4). But despite the high $Z_{avg}$ and reasonable µτ products demonstrated in X-ray detectors, BiI$_3$ capable of detecting moderate energy γ-rays (>100 keV) have only been recently demonstrated. The limited performance has been attributed to the formation of voids and defects in BiI$_3$ during growth by the vertical Bridgeman method [404]. The density of iodine vacancies was found to be reduced through Sb doping, which led to improved resistivities (from $1.45 \times 10_8$ Ω cm for BiI$_3$ to $2.63 \times 10_9$ Ω cm for Sb:BiI$_3$) and a four-order-of-magnitude reduction in the leakage current in BiI$_3$ single crystals [246]. The higher-quality Sb:BiI$_3$ crystals demonstrated an energy resolution of 2.2% for $_{137}$Cs (662 keV) radiation [404], substantially lower than reported for CsPbBr$_3$ [426].

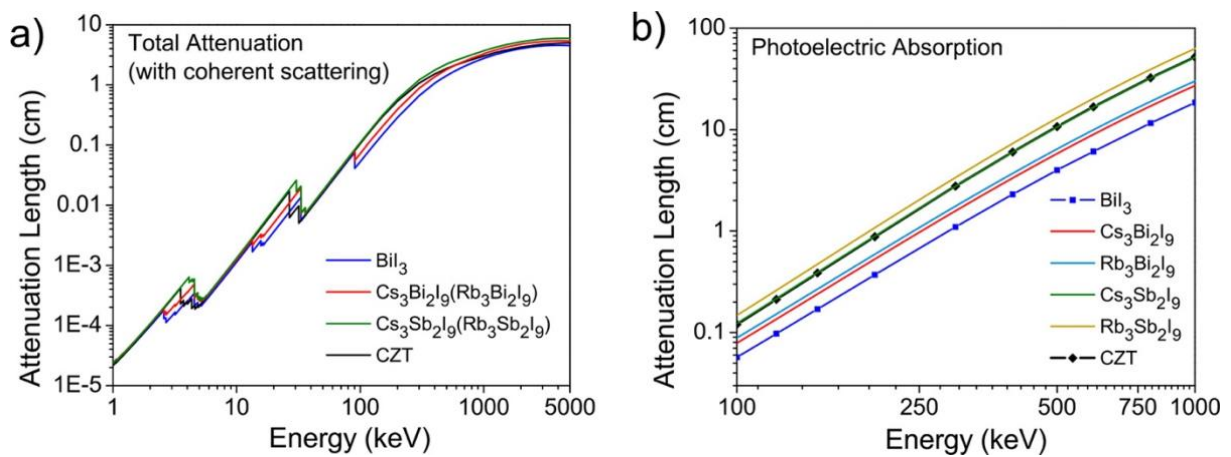

**Figure 25.** Attenuation length of perovskite-inspired materials compared to Cd$_x$Zn$_{1-x}$Te (CZT) under (a) total attenuation and (b) photoelectric absorption. Reprinted with permission from Ref. [405]. Copyright 2018 American Chemical Society.

Beyond X-rays and γ-rays, the detection of α-particles is important because these are the byproducts of nuclear reactions and nuclear decays [400]. µτ particles are also used in the measurement of composition, such as in Rutherford backscattering spectrometry. The energy resolution of the α-particle detector is therefore an important factor. In testing α-particle detectors, 5.5 MeV α −particles from an $_{241}$Am source are commonly used [246, 400, 405, 410, 428]. Current work on perovskite and perovskite-inspired materials for $\alpha$-particle detectors have not demonstrated energy resolutions or $\mu\tau$ products matching industry-standard $Cd_xZn_{1-x}Te$ (Table 4). In fact, the defect perovskites ($A_3B_2I_9$ materials) are not able to resolve the $\alpha$-particle peak, and have low $\mu\tau$ products. This was attributed to the low mobilities, which limit the charge collection efficiency [405]. By contrast, Sb:BiI$_3$ single crystals were found to have electron mobilities of $1000\pm200$ cm$_2$ V$_{-1}$ s$_{-1}$, which are similar to the mobilities of $Cd_xZn_{1-x}Te$ [246]. Sb:BiI$_3$ demonstrated an energy resolution of 66% to $\alpha$-particles, which improved to 33% after applying an electric field of 532 V cm$_{-1}$ for 8 h, and the improvement in resolution was attributed to reduced leakage currents [246]. CsPbBr$_3$ single crystals were also found to be capable of resolving the $\alpha$-particle peak, demonstrating an energy resolution of 15%. An important factor was the high mobilities of 63 cm$_2$ V$_{-1}$ s$_{-1}$ for electrons and 49 cm$_2$ V$_{-1}$ s$_{-1}$ for holes, leading to reasonable $\mu\tau$ products reaching up to 10$_{-3}$ cm$_2$ V$_{-1}$ [400]. Future efforts to improve the performance of defect perovskites for $\alpha$-particle detection should focus on improving the mobility [405]. While the mobilities could be limited due to strong-electron phonon coupling, defect perovskite X-ray detectors have achieved $\mu\tau$ products reaching 10$_{-2}$ cm$_2$ V$_{-1}$ s$_{-1}$ (Table 3), and improved performance may be expected from these materials for $\alpha$-particle detection.

*5.4 Electronic devices*

Two critical electronic devices are thin film transistors (TFTs) and memristors. The operating principles of these devices can be found in review articles, such as Ref. [402, 429]. TFTs are ubiquitously used in electronic circuits, in which they are used as switches, for amplification, as well as for the fabrication of logic devices (particularly the Complementary Metal Oxide Semiconductor or CMOS). Memristors are actively explored for next-generation terabyte-scale non-volatile memory storage that can be operated with low power. Historically, the materials considered for these applications have been silicon or oxides [402, 429]. But more recently, lead-halide perovskites have been actively explored for these applications with promising results [402, 429]. As with all other electronic applications, it will be important to reduce the lead content used in these devices, as well as improve the stability under operation.

In TFTs, tin-based perovskites have been widely explored as a lead-free alternative. In fact, the first organic-inorganic perovskite TFT was a 2D tin-based perovskite (($C_6H_5C_2H_4NH_3$)$_2$SnI$_4$ or (PEA)$_2$SnI$_4$), reported by Mitzi *et al.* well before lead-halide perovskites were actively considered for photovoltaics. This TFT demonstrated *p*-channel character with a field-effect mobility of 0.6 cm$^2$ V$^{-1}$ s$^{-1}$ [430]. Significantly higher electron mobilities have recently been reported in lead-based perovskites, with field-effect mobilities exceeding 10 cm$^2$ V$^{-1}$ s$^{-1}$ demonstrated at room temperature [431], which is comparable to the electron field-effect mobility of strontium titanate perovskites [429]. We note that an electron mobility of 396 cm$^2$ V$^{-1}$ s$^{-1}$ has been reported from MAPbI$_3$ TFTs at room temperature [432], but there is uncertainty on whether this was overestimated. Recently, tin-based perovskite TFTs have been demonstrated with hole field-effect mobilities exceeding those of lead-based perovskites [431]. This was achieved by Matshushima *et al.* using the same material previously reported by Mitzi *et al.* Critical improvements were: 1) the passivation of the surface of PEA$_2$SnI$_4$ with self-assembled monolayers of NH$_3$I-SAM, and 2) the development of the top-gate, top-source/drain device structure (Figure 26a) to ensure both surfaces are passivated. These led to minimal hysteresis in the transfer curves (Figure 26b), with the hole field-effect mobilities reaching 15 cm$^2$ V$^{-1}$ s$^{-1}$ [433] and on/off ratios reaching 10$^6$. *p*-type TFTs are especially important because CMOS requires both *n*- and *p*-type TFTs, and fewer options for high-performing *p*-type TFTs are currently available. The hole mobility of MASnI$_3$ single crystals has been found to be at least an order of magnitude larger than the field-effect mobilities achieved thus far (refer to Section 3), and it is expected that the in-plane mobility of 2D $(PEA)_2SnI_4$ is comparably high [433]. Thus, future improvements in the hole field-effect mobility could be expected. However, $(PEA)_2SnI_4$ has low stability, and TFTs have been found to degrade quickly, even when they are in high vacuum or inside a glovebox [434]. Improved stability was achieved by replacing PEA with the π-conjugated oligothiphene ligand 4Tm (4Tm = 2-(3"',4'-dimethyl-[2,2':5',2":5",2"'-quaterthiophen]-5-yl)ethan-1-ammonium). $(4Tm)_2SnI_4$ perovskite TFTs were found to retain their performance after storage in air for a month. TFTs were made using a bottom gate, top-source/drain structure, which gave a hole field-effect mobility of 2.32 $cm^2$ $V^{-1}$ $s^{-1}$ and an on/off ratio of $10^5$–$10^6$. While the field-effect mobility is lower than that reported by Matshushima *et al.*, a different structure was used; $(PEA)_2SnI_4$ TFTs made in the same structure as the $(4Tm)_2SnI_4$ TFTs had a field-effect mobility of 0.15 $cm^2$ $V^{-1}$ $s^{-1}$, with an on/off ratio of $10^4$–$10^5$ [434].

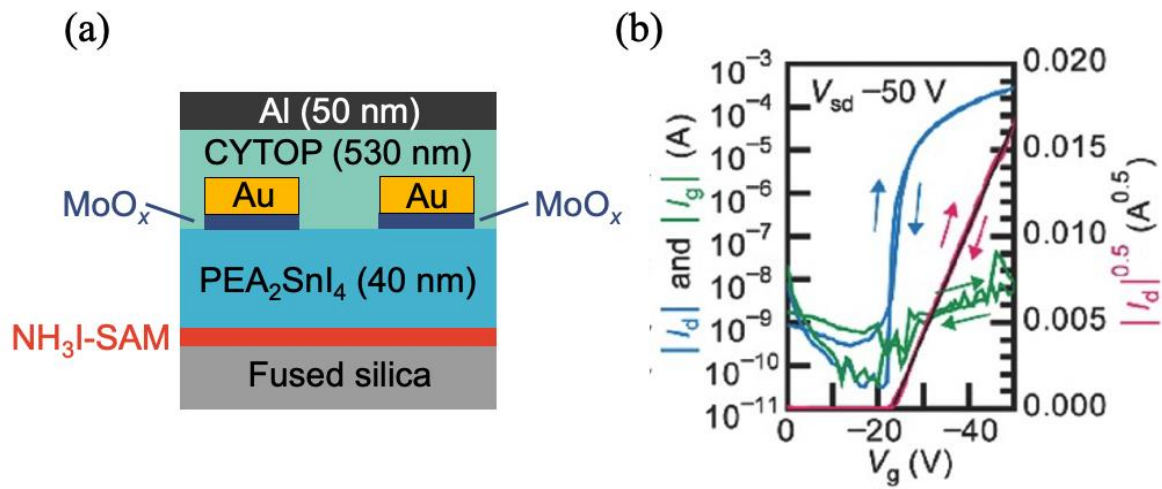

**Figure 26.** $(PEA)_2SnI_4$ perovskite TFTs: (a) Device structure and (b) transfer curves. Reprinted with permission from Matshushima *et al.* [433]. Copyright 2016 by Wiley.

Memristors are devices with variable resistance states, which depend on the applied voltage and current history [435]. Such devices were first proposed as a fourth fundamental circuit element in 1971 (along with resistors, inductors and capacitors) [436], and realized in 2008 [437]. These variable resistance states can be used for high-density memory storage, logical computations, as well as for neuromorphic computing [435]. Such devices switch between resistance states through the application of an electric field. As such, memristors have low power consumption and retain their resistance states after the electric field has been removed, which contrasts to common commercial non-volatile memory (*e.g.*, NAND flash [438]). The key requirements for memristors are 1) rapid switching between the high and low resistance states, 2) low operating voltage (for set/reset and read), 3) large ratio in resistance of the on and off states (on/off ratio), 4) large endurance (number of switching cycles in which the resistance states are maintained), and 5) long retention time [402] of the resistance states after power has been removed. Thin film memristors are typically comprised of an insulator sandwiched between two degenerate contacts (*e.g.*, metal-insulator-metal structure). Wide band gap oxides are commonly used for the insulator layer, such as $TaO_x$, $HfO_2$ and $SrRuO_3$, which have achieved on/off ratios ranging from 10 to $10_3$, endurances ranging from $10_3$–$10_{10}$ cycles, and retention times between $10_4$ s and 10 y [402]. Lead-halide perovskites ($MAPbI_3$, $CsPbBr_3$ and 2D perovskites, among others) have recently been demonstrated to exhibit resistive switching, with high on/off ratios reaching $10_7$ and rapid switching between resistance states (640 µs) but lower endurances of up to 3000 cycles. Retention times up to $10_5$ s have been measured, but the set/reset voltages are lower than their oxide counterparts (which commonly have set/reset voltages with absolute values exceeding 1 V) [402]. But as with many other optoelectronic device applications, the need eliminate toxic lead from memristors has motivated the exploration of perovskite-inspired materials for resistive switching [435, 438–440].

The wide band gaps of many perovskite-inspired materials leads to a low dark resistance, which is advantageous for achieving a high resistance in the off state. The ability to process these materials using low temperature, facile methods is also advantageous and makes the materials compatible with flexible substrates [438]. At the same time, these materials have low formation energies for vacancies (especially halides), which can act as conductive filaments in resistive switching [439]. This ionic vacancy conduction mechanism (known as valence change memory) has been identified in $MASnBr_3$, $CsSnI_3$ and $Cs_3Sb_2Br_9$ memristors [435, 438, 439] using inert Au electrodes. For $MASnBr_3$ and $Cs_3Sb_2Br_9$, the conducting filaments are attributed to bromide vacancies [435, 439]. When a bias exceeding the set voltage is applied, $Br_-$ accumulate at the anode, while bromide vacancies form and accumulate at the cathode [435, 439]. The $Br_-$ may be oxidized to Br atoms that are stored in the inert electrode [435]. The bromide vacancies have a low activation energy barrier to migration (0.22–0.57 eV), and these vacancies can form a conductive channel bridging the two electrodes, resulting in the memristor switching from the high to low resistance state. Composition and Raman measurements have shown that the segregation of Br occurs. Applying a negative reset voltage causes the Br atoms to diffuse away from the electrode they were stored at and recombine with the halide vacancies, causing the conductive filaments to rupture and the memristor return to the high resistance state [435, 439]. The conductive filaments for $CsSnI_3$ has been attributed to tin vacancies, which form and are removed in a similar manner to bromide vacancies [438]. Valence change memory has also been suggested as the mechanism for resistive switching in BiOI memristors, which are believed to form conductive filaments through oxygen vacancies [441].

An alternative to valence change memory for the formation of conductive filaments is through electrochemical metallization (ECM), in which a reactive metal anode (often Ag) is oxidized when an electric field is applied. The metal cations are the conductive filament and migrate through the solid film electrolyte to the cathode, where they are reduced and electrocrystallize to form the metal again. To reset the device from the low to high resistance state, a negative bias is applied, which results in the dissolution of the conducting filaments through Joule-heating assisted dissolution [438]. This ECM mechanism was found to occur in $CsSnI_3$ and $Cs_3Cu_2I_5$ memristors with Ag top electrodes. In both devices, a layer of polymethyl methacrylate (PMMA) was spin coated over the perovskite-inspired materials to improve its stability in ambient air, as well as preventing the reaction between Ag and the iodide-based thin film, which is important for improving the endurance of the devices [438, 440]. The PMMA layer did not act as a barrier to silver dissolution and migration, allowing resistive switching to occur. Han *et al.* compared $CsSnI_3$ memristors using Ag *vs.* Au electrodes to compare the ECM and VCM mechanisms. They found the memristors operating based on ECM to give improved on/off ratios (>1000) and low set/reset voltages of 0.13 V and -0.08 V (with an electroforming voltage of 0.36 V), whereas VCM memristors gave improved retention ($10_4$ s [438]).

To summarize the devices discussed in Section 5, a comparison of the state-of-the-art performance of perovskite-inspired materials across all applications is given in Table 5.

**Table 5.** State-of-the-art performance of perovskite-inspired materials in the applications covered in Section 5

| Application | Active material | Performance properties | Performance values | Ref. |
|---|---|---|---|---|
| Phosphor for white-light LEDs | $Cs_2(Ag_{0.6}Na_{0.4})InCl_6$ with 0.04% $Bi_{3+}$ | PLQE CIE coordinates Color temperature | 86±5% (0.396,0.448) 4054 | [277] |
| LED | FPMAI-MAPb$_{0.6}$Sn$_{0.4}$I$_3$ | PLQE | - | [367] |

| | | EQE | 5.0 | |
| | | EL wavelength | 917 | |
| | | EL FWHM | 80 nm | |
| Photocatalysis | $Cs_3Sb_2Br_9$ nanocrystals | Precursor | $CO_2$ (g) | [385] |
| | | Light source | 1 sun AM 1.5G | |
| | | Product | CO | |
| | | Yield | 516 µmol $g^{-1}$ (5 h) | |
| X-ray detector | $MA_3Bi_2I_9$ single crystals | µτ | ~$10^{-3}$ $cm^2$ $V^{-1}$ | [186] |
| | | Sensitivity | $1.1 \times 10^4$ | |
| | | Lowest detectable dose rate | 0.0006 | |
| Gamma-ray detector | Sb:$BiI_3$ single crystals | µτ | $10^{-4}$ | [404] |
| | | Energy resolution | 2.2% (662 keV) | |
| α-particle detector | Sb:$BiI_3$ single crystals | Energy resolution | 33 – 66% (5.5 MeV) | [246] |
| Thin film transistor | $(C_6H_5C_2H_4NH_3)_2SnI_4$ thin films | Hole mobility | 12±1 (max. 15) $cm^2$ $V^{-1}$ $s^{-1}$ | [433] |
| | | Threshold voltage | -22±2 | |
| | | $I_{ON}/I_{OFF}$ | $(1.9±2.1) \times 10^6$ | |
| Memristor | $CsSnI_3$ (orthorhombic phase; Ag top electrode) | SET voltage | 0.13 V | [438] |
| | | RESET voltage | -0.08 V | |
| | | On/off ratio | >$10^3$ | |
| | | Endurance | >600 cycles | |
| | | Retention time | $> 7 \times 10^3$ s | |

# 6   Conclusions and Outlook

Lead-halide perovskites have developed at an unprecedented rate in photovoltaics due to an ideal set of optoelectronic properties. Among these properties is its tolerance to point defects, which allows perovskite devices to achieve high efficiencies despite being processed at low temperature using simple fabrication methods that give rise to high densities of point defects. This contrasts to traditional semiconductors, such as silicon and III-V compounds, and has spurred efforts to identify 'perovskite-inspired materials' that could replicate the exceptional optoelectronic properties of LHPs, but which are free from the toxicity burden of lead and which are also more thermally and environmentally stable. The effort to develop PIMs is a multi-disciplinary challenge involving theorists, experimentalists and device engineers.

For theorists, the primary challenges have been: 1) unveiling the origins of the tolerance of LHPs to point defects, and 2) developing improved approaches to design PIMs [442]. As discussed in Section 2.2, defect tolerance is a phenomenon that rarely originates from any single material property. Indeed, there are several factors which can contribute to the defect tolerance of a material, each with different levels of efficacy and rarity. To achieve the remarkable defect tolerance of LHPs, successful PIMs will likely be required to replicate the wealth of 'defect-tolerant' material properties exhibited by LHPs (atypical band structure, strong dielectric screening, anharmonic carrier capture *etc.*). Through deeper understanding of defect-tolerant properties and their inter-relationships, as well as extending the materials discovery abstraction pyramid (Figure 2) to the explicit treatment of crystal defects, the accuracy of predictions for next-generation, high-performance PIMs will be improved immensely.

For materials design, recent years have witnessed an explosion in the popularity of data mining (materials informatics) and machine learning procedures, which promise to revolutionize the materials discovery process [34, 41, 442]. An inspiring, prototypical example is the use of text mining and natural language processing to extract previously-hidden insights from the vast dataset of materials science literature. In 2019, Ceder *et al.* exemplified this concept by retrieving nearly 20 000 "codified synthesis recipes" from over 50 000 paragraphs in the recent solid-state literature [443]. In addition to pure data mining, the integration of physics-based machine learning (ML) models into the high-throughput screening process could significantly augment the power and efficiency of materials discovery. For instance, by implementing accurate ML models of material stability and chemical descriptors as a pre-screening step, before the application of expensive electronic structure theory (*i.e.* DFT), the computational load could be considerably reduced, thereby accelerating the process and affording the use of expanded input compositional space [442]. Moreover, ML techniques may be leveraged to uncover subtle chemical trends in successful candidate materials (*e.g.* defect-tolerant material properties), which may not be initially apparent to the 'human eye' [41]. Indeed, the ability to physically interpret trained ML models and distinguish causation from correlation is an ongoing area of research [444].

For materials scientists, chemists, physicists and engineers, one of the main challenges has been to develop suitable fabrication methods for each class of materials. A wide range of methods have been investigated, from solution processing to chemical and physical vapor deposition (which can be used at scale), through to 'hybrid' methods that combine both solution- and vapor-based processing. A series of dedicated efforts has led to pinhole-free thin films being achieved across a wide range of materials, with methods to control the grain size and preferred orientation demonstrated, which all have led to improvements in the performance of the materials in devices. Furthermore, it has been widely shown across many families of PIMs that the materials are more stable thermally and in air compared to LHPs (*e.g.*, $Cs_2AgBiBr_6$ and BiOI). There are a handful of exceptions, namely tin- and germanium-based perovskites, as well as InI (owing to In being unstable in the +1 oxidation state). But the incorporation of additives and developing 2D structures have been utilized to enhance their ambient stability. Critically, defect tolerance has been demonstrated computationally across several materials, such as BiOI, $CuSbS_2$ and $CuSbSe_2$.

One of the challenges across many PIMs is their wide band gaps, which are not suitable for single-junction photovoltaic devices. Although these materials may be used as top-cells in tandems, recent work has shown these materials to also hold significant promise in applications beyond PV. For example, in light-emission applications, photocatalysis and radiation detectors, wide band gaps of 2 eV or larger are advantageous, which compounded with the stability of these materials, makes them of practical interest for these applications. For example, inorganic double perovskites have been demonstrated as efficient, stable white-light emitters, which will be important for solid-state lighting. Many PIMs have suitable band positions to reduce $H_+$ or $CO_2$, and $Cs_3Sb_2Br_9$ nanocrystals were demonstrated to more efficiently reduce $CO_2$ than $CsPbBr_3$. In radiation detectors, the wide band gap of PIMs results in low dark current, which is important for them to achieve a high signal-to-noise ratio, with a low lowest detectable dose rate, down to 0.6 $nGy_{air}$ $s^{-1}$ in $MA_3Bi_2I_9$. This is smaller than the lowest dose detectable by $MAPbBr_3$ as well as standard X-ray detectors, such as a-Se and CZT, and will allow lower and safer doses of X-rays to be used in security equipment or in medical screening. Furthermore, the composition of PIMs of high atomic number elements allows these materials to have high stopping power for radiation, which in turn results in sensitivities to X-rays being comparable to $MAPbBr_3$ and orders of magnitude larger than conventional Si, a-Se and CZT materials. The high resistivities of PIMs owing to their wide band gap is also advantageous for achieving a high resistance 'off' state in memristors, and on/off ratios comparable to or larger than conventional oxide-based materials have been reported. The exploration of PIMs in these electronic applications is still in the nascent stages, but there are already very promising signs of the potential of these low-toxicity materials to give stable, efficient performance.

However, an important challenge faced by the field is the limited mobility of some PIMs. Many PIMs either have a 0D crystal structure (*e.g.*, $Cs_3Bi_2I_9$), or a 0D electronic structure (*e.g.*, $Cs_2AgBiBr_6$), which limits mobilities and drift/diffusion lengths. Furthermore, these lower-dimensional materials have higher exciton binding energies, as well as soft lattices and strong electron-phonon coupling. These factors further restrict mobilities due to the formation of excitons (which need to be split to free carriers), as well as polarons or self-trapped excitons. This limits the current densities and efficiencies in PVs, as well as the $\mu\tau$ product of PIM single crystals, which are lower than in standard radiation detector materials. The main exception to the low mobilities in PIMs are tin-based perovskites, which have disperse bands. Indeed, this has resulted in these materials achieving field-effect mobilities comparable to LHPs in *p*-type TFTs.

Many PIMs also have crystal structures based on low dimensional networks (e.g. 1D and 2D connectivity), which leads to certain crystallographic orientations having higher mobility. This anisotropy in carrier transport necessitates careful control over the preferred orientation of the materials in thin films, such that: 1) the high mobility directions connect both electrodes in devices, while 2) also maintaining a compact morphology that does not lead to shunting. For example, BiOI has a layered structure, and a {012} preferred orientation is desirable to connect the top and bottom electrodes in vertically-structured PV devices. But this open structure leads to low shunt resistances, which limits open-circuit voltages. New device structures or strategies that allow high mobility and high shunt resistance to be simultaneously achieved are needed.

Beyond these materials and fabrication challenges, further efforts are needed to identify the cause of defect tolerance. While some $ns^2$ compounds have been found to be defect tolerant, others have been found to have deep traps, *e.g.*, $BiI_3$ and $SbI_3$. A deeper understanding of defect tolerance will be necessary to develop increasingly refined, accurate, and comprehensive design principles, and thus more efficient, targeted screening procedures. On the computational front, recent advances in theory and technology have enabled both the qualitative and quantitative descriptions of bulk and defect-related material properties from *ab-initio* modelling. Overcoming the challenges associated with complex defect calculations (Section 2.3.5) will facilitate accurate prediction and validation of promising candidate materials, alongside enhanced understanding of experimental results and defect behavior.

Ultimately, for future research efforts in materials discovery to be successful, consensus and consistency with selection metrics will be immensely beneficial, allowing greater transferability and comparability between studies. For similar reasons, alongside the reduction of duplicate work, the facilitation of sharing both data and methodologies should be vehemently encouraged. Frameworks which automate this process (*e.g.* AiiDA [102] for materials modelling), maintaining data provenance and reproducibility, will be especially advantageous in this regard. This effort should not be limited to only computational data, but also experimental data (*e.g.*, optical and transport measurements) and synthesis details. This will help to refine computational models, as well as providing 'training' datasets for more general ML models. Overall, the future of PIMs research will require an interlinked, synergistic approach between theory and experiment in order to identify further promising materials.

**Acknowledgements**

Yi-Teng Huang acknowledges funding from Education Ministry of Taiwan Government as well as Downing College Cambridge. Seán R. Kavanagh acknowledges funding from the EPSRC Centre for Doctoral Training in Advanced Characterisation of Materials (CDT-ACM)( EP/S023259/1), as well as the use of VESTA [445] (for crystal structure visualization) and Freepik.com (for vector image templates) in the preparation of Figure 2. D.O.S. acknowledges support from the European Research Council, ERC (Grant 758345). This work was supported by a National Research Foundation of Korea (NRF) grant funded by the Korean government (MSIT) (No. 2018R1C1B6008728). R.L.Z.H. acknowledges support from the Royal Academy

**ORCID IDs**

Yi-Teng Huang https://orcid.org/0000-0002-4576-2338

Seán R. Kavanagh https://orcid.org/0000-0003-4577-9647

David O. Scanlon https://orcid.org/0000-0001-9174-8601

Aron Walsh https://orcid.org/0000-0001-5460-7033

Robert L. Z. Hoye https://orcid.org/0000-0002-7675-0065

(2017).